\documentclass{emulateapj} 

\shorttitle{Modeling Nuclear IR SED of AGNs}
\shortauthors{Lira et al.}

\begin{document}

\title{Modeling the Nuclear Infrared Spectral Energy Distribution\\ 
of Type II Active Galactic Nuclei}

\author{Paulina Lira, Liza Videla}
\affil{Departamento de Astronom\'ia, Universidad de Chile.}

\author{Yanling Wu} 
\affil{Infrared Processing and Analysis Center, California Institute of Technology.}

\author{Almudena Alonso-Herrero} 
\affil{Instituto de F\'isica de Catabria, CSIC-UC, 39005 Santander, Spain}

\author{David M. Alexander, and Martin Ward}
\affil{Department of Physics, Durham University.}

\begin{abstract}

We present results from model fitting to the Spectral Energy Distribution
(SED) of a homogeneous sample of Seyfert II galaxies drawn from the $12\mu$m
Galaxy Sample. Imaging and nuclear flux measurements are presented in an
accompanying paper (Videla et al., 2013). Here we add IRS {\emph{Spitzer}}
observations to further constrain the SEDs after careful subtraction of a
starburst component. We use the library of CLUMPY torus models from Nenkova et
al.~(2008ab) and also test the two-phase models recently produced by Stalevski
et al.~(2012). We find that photometric and spectroscopic observations in the
mid-IR ($\lambda \ga 5\mu$m) are crucial to properly constrain the best-fit
torus models. About half of our sources show clear near-IR excess of their
SEDs above the best fit models. This problem can be less severe when using the
Stalevski et al.~(2012) models. It is not clear what is the nature of this
emission since best fitted black body temperatures are very high ($\sim
1700-2500$ K) and the Type II classification of our sources would correspond
to a small probability to peer directly into the hottest regions of the
torus. Crucially, the derived torus parameters when using CLUMPY models are
quite robust,, independently of whether the sources require an additional
black body component or not. Our findings suggest that tori are characterized
by $\mathcal{N}_0 \ga 5$, $\sigma \ga 40$, $\tau \la 25$, $\angle i \ga 40$
degrees, $Y \la 50$ and $A_v^{\rm los} \sim 100-300$, where $\mathcal{N}_0$ is
the number of clouds in the equatorial plane of the torus, $\sigma$ is the
characteristic opening angle of the cloud distribution, $\tau$ is the opacity
of a single cloud, $\angle i$ is the line-of-sight orientation of the torus,
$Y$ is the ratio of the inner to the outer radii, and $A_v^{\rm los}$ is the
total opacity along the line-of-sight. From these we can determine that
typical torus sizes and masses of $0.1-5.0$ pc and $10^{4-6} M_\odot$.  We
find tentative evidence that those nuclei with a detected Hidden Broad Line
Regions are characterized by lower levels of extinction than those without
one. Finally, we find no correlation between the torus properties and the
presence of circumnuclear or more global star-formation.

\end{abstract}


\section{Introduction}

Evidence for the presence of circumnuclear obscuration in Active Galactic
Nuclei (AGN) is undeniable. Obscuration has been determined from the excess
absorption in X-ray, UV and optical wavelengths
(\citep{lawrence82,mass-hesse93,turner97,malkan98,risaliti99}; see also
further references in Videla et al., 2012; hereafter Paper I). However,
whether this obscuration has similar properties in all sources or presents a
wide variety of properties that vary from source to source is not yet
determined.

From a statistical view-point there is strong evidence that the properties of
the obscuring material change as a function of luminosity, and possibly,
redshift (Barger et al., 2005; Hopkins, Richards \& Hernquist, 2007; Gilli,
Comastri \& Hasinger, 2007). Still, because of the higher spatial resolution,
only local intermediate-luminosity ($\la 10^{46}$ ergs/s) sources can be
studied in enough detail to disentangle the dust emission coming from the
region close to the active nuclei from that coming from the remaining host
galaxy in the crucial near-IR domain. Also, the number counts in distant
samples are dominated by Seyfert luminosity-class sources, and therefore their
local properties might be a good representation of the higher-redshift
counterparts.

All AGN, whether obscured or not, are thought to harbor a central engine that
emits because of the release of gravitational potential energy in an accretion
disk surrounding a supermassive Black Hole. The Unified Model, proposed more
than 25 years ago, is still strongly advocated to claim that, in fact, any AGN
might look like a Type I source (i.e., unobscured) or Type II source (i.e.,
obscured), depending on the orientation of the obscuring material, shaped as a
torus, with respect to our line of sight (Antonucci 1993). If one assumes that
all tori share similar physical traits, then the Unified Model imposes some
strong constraints on the observables, such as an expected correlation between
the torus orientation angle and the optical depth towards the central source,
among others. More recently, a new interpretation has been proposed, where
Type I and Type II objects are preferentially drawn from the ends of the
distribution of torus covering factor (Elitzur 2012).

The early simple picture of a homogeneous torus (e.g., Pier \& Krolik,
1992,1993; Granato \& Danese, 1994; Efstathiou \& Rowan-Robinson, 1995), which
implied some clear differences in the properties between Type I and Type II
sources, has been superseded by more complex models, where a clumpy media,
with an overall geometry that still resembles that of a torus, give a much
more realistic representation of the obscuring region (Nenkova et al., 2002;
H\"onig et al., 2006; Nenkova et al., 2008ab; Stalevski et al.~2012). Many
efforts have already been carried out to fit the Spectral Energy Distributions
(SEDs) of Type I and Type II AGN using clumpy distributions of the dusty
medium (e.g., Nikutta et al.~2009, Mor et al.~2009, Ramos-Almeida et al.~2009;
Alonso-Herrero et al.~2011 (hereafter AH11); Mullaney et al.~2011).

These new prescriptions imply that the correlation between the different
observables should be regarded as statistical ones, where individual sources
might not be a good representation of the median of a sample. In the same way,
the classification as Type I or Type II, also becomes a probabilistic
problem. Therefore, samples as large and homogenous as possible have to be
gathered in order to draw significant results about the torus properties.

With this idea in mind we have gathered near and mid-IR observations for a
sample of Seyfert II galaxies to determine and fit their SEDs and in this way
derive physical and structural parameters of their tori. This paper is
organized as follows: Section 2 presents the imaging and spectroscopic
observations and their treatment previous to the fitting process; Section 3
presents the clumpy models and the fitting procedure; Section 4 presents the
obtained results and discusses the best-fit parameters; Section 5 looks into
the possible correlations between the results and other characteristics of our
sources, such as hydrogen column determined from X-ray data or the presence of
a Hidden Broad Line Region (HBLR) inferred from spectropolarimetric
observations.

\section{Observations}

The sample included in this work is formed by all Seyfert II galaxies in the
southern hemisphere found in the Extended 12 $\mu$m Galaxy Sample
\citep{rush93}, comprising 48 Seyfert II galaxies\footnote{This research has
  made use of the NASA/IPAC Extragalactic Database (NED) which is operated by
  the Jet Propulsion Laboratory, California Institute of Technology, under
  contract with the National Aeronautics and Space Administration.}.

The most important characteristic of the 12 $\mu$m Galaxy Sample for this work
is that it includes a fairly unbiased sample of nearby (z$\leq$0.07) active
galaxies that have been observed in many spectral regimes. It is selected in
the mid-infrared (MIR), minimizing possible biases: it includes elliptical,
lenticulars and spirals galaxies (allowing to avoid systematic errors in the
decomposition process of the surface brightness profiles, presented in Paper
I); it includes a wide range in galaxy inclinations and most importantly, it
includes a wide range of obscuration properties of the nuclear source
(allowing a more general test of the nuclear emission models and probing a
range of hydrogen columns of $\sim10^{22}-10^{25}$ cm$^{-2}$).

The Type classification of the objects was obtained from existing catalogs of
active galaxies \citep{veron-cetty91,hewitt91}. Some objects have been
re-classified as modern observations have become available. In order to
corroborate the classification of the targets 38 objects were observed with
the RC spectrograph on Blanco Telescope in CTIO, in two runs in August 2007
and February 2008 (see below and Paper I).

\subsection{Infrared Imaging and Optical Spectroscopy}

Details on the imaging observations for our sample are found in Paper I. We
presented the near and mid-IR SEDs for 40 Type II Seyfert galaxies drawn from
the Extended 12 $\mu$m Galaxy Sample \citep{rush93}, six of which had been
determined previously by Alonso-Herrero et al.~(2003). A detailed account on
the data reduction and nuclear flux determination are given in Paper I.

Additionally, optical long slit data were obtained in order to determine the
nature of the nuclear ionizing source and the host stellar population and are
also presented in Paper I. This analysis has shown that 3 objects from the
original sample have HII nuclei (MCG\,+0-29-23, NGC\,6810 and Mrk\,897) and
therefore have been discarded from any subsequent analysis in this work
leaving a total sample of 37 Type II Seyfert nuclei.

\subsection{Spectroscopic Observations}

\begin{figure}
  \begin{center}
    \includegraphics[scale=0.6,angle=0]{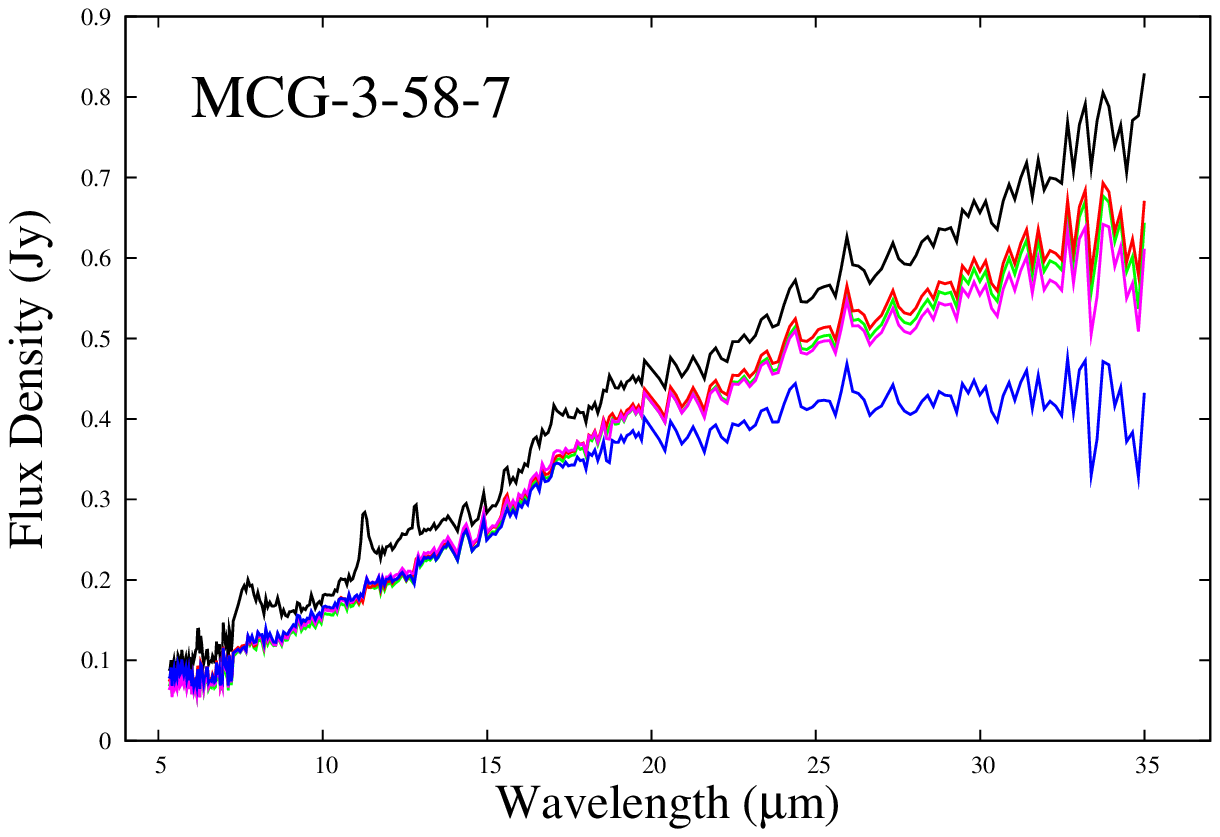}
    \includegraphics[scale=0.6,angle=-0]{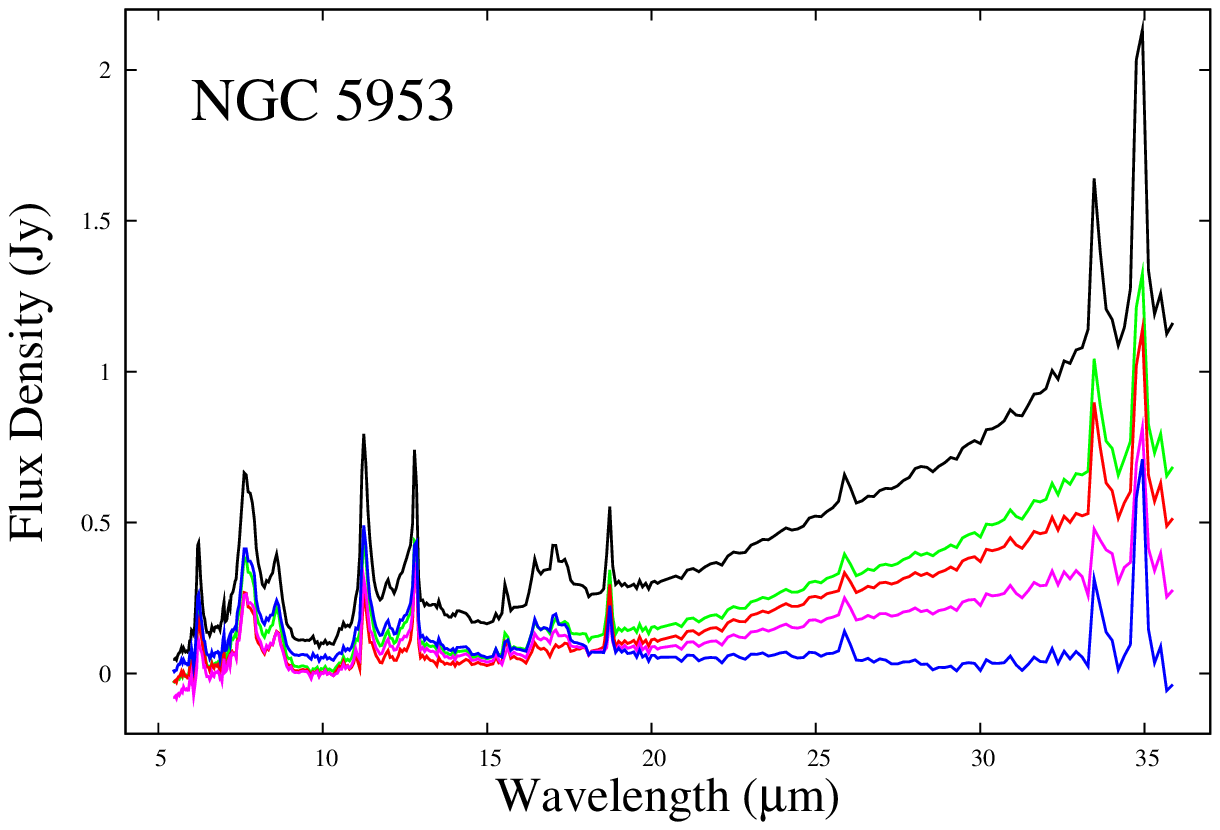}
 \end{center}
  \figcaption[]{Two extreme examples of subtraction of the emission from star
    forming regions using the templates published by Smith et al.~(2007). From
    top to bottom in each panel the curves represent the original
    {\emph{Spitzer}}-IRS data obtained using a small extraction aperture (\S
    2.2.1) followed by the subtracted spectra using the four templates for
    star forming emission. The top panel presents the data for MCG-3-58-7,
    with very a small star forming component. The bottom panel presents the
    data for NGC\,5953, which shows strong PAH features even after the
    subtraction attempts. This object has a very strong circumnuclear star
    forming region which clearly is contaminating the \emph{Spitzer}
    observations.\label{example_subt}}
\end{figure}

\subsubsection{Spitzer Data}

Most of the galaxies in our sample have been observed using the Infrared
Spectrograph (IRS) on board {\emph{Spitzer}}. The IRS provides moderate
resolution spectroscopy from 5.2 to 38.0 $\mu$m. It is composed of four
separate modules, with two modules providing R$\sim$60-120 spectral resolution
over 5.2-38.0 $\mu$m (the Short-Low or SL, from 5.2 to 14.5$\mu$m; and the
Long-Low or LL, from 14.0 to 38.0 $\mu$m).

The wide wavelength coverage of IRS spectra can allow a good representation of
the mid-IR continuum of Seyfert galaxies. In contrast, $\sim10\mu$m MIR
ground-based spectroscopy is normally dominated by Polycyclic Aromatic
Hydrocarbure (PAH) features and the 9.7$\mu$m silicate absorption present in
that regime, resulting in very little free continuum to be observed.

Results using {\emph{Spitzer}} IRS data for nearby Seyfert galaxies have
already been published by several authors (Buchanan et al., 2006; Deo et al.,
2007; Mel\'endez et al., 2008; Wu et al., 2009; Thompson et al., 2009;
Tommasin et al., 2010). In more detail, \citet{wu09} published the spectra for
103 AGN from the $12\mu$m Galaxy Sample, of which 44 are in this sample (the
remaining seven lack {\emph{Spitzer}}/IRS SL spectra). Most of the 44
observations correspond to Program ID 3269 and were obtained between 2004 and
2005, a good match for the timing of our photometry obtained between 2002 and
2004 (see Paper I). The exceptions are IC\,5063 and NGC\,4941 (PID 30572,
observed in 2006 and 2007, respectively), and F\,05189-2524 (as part of the
IRS calibration campaign).

For those objects observed in mapping mode (all sources in the 3269 program),
sky subtraction was performed by differentiating the on- and off-source
observations of the same order in each module. In order to isolate the AGN
emission from the stellar components, and since rather large apertures were
used to obtain the 1-dimensional spectra presented by \citet{wu09}, here we
have reprocessed the data in order to have the smallest possible aperture,
$2\times2$ pixels, or $3.6\times3.6$ arcsec$^2$ on the sky for the SL module.
For data obtained with the IRS staring mode (IC\,5063 NGC\,4941 and
F\,0518-2524), the reduction was done in the following manner. Individual
pointings to each nod position of the slit were co-added using median
averaging. Then on and off source images were subtracted to remove the
contribution of the sky emission. Spectra from the final 2-D images were
extracted with a point source extraction mode, which scaled the extraction
aperture with wavelength to recover the same fraction of the diffraction
limited instrumental PSF (for details see Wu et al., 2009). Therefore, no
"small'' aperture spectra are available for these 3 sources. All the spectra
were flux calibrated using the IRS standard star $\alpha$ Lac, for which an
accurate template was available.

\begin{figure}
  \begin{center}
    \includegraphics[scale=0.65]{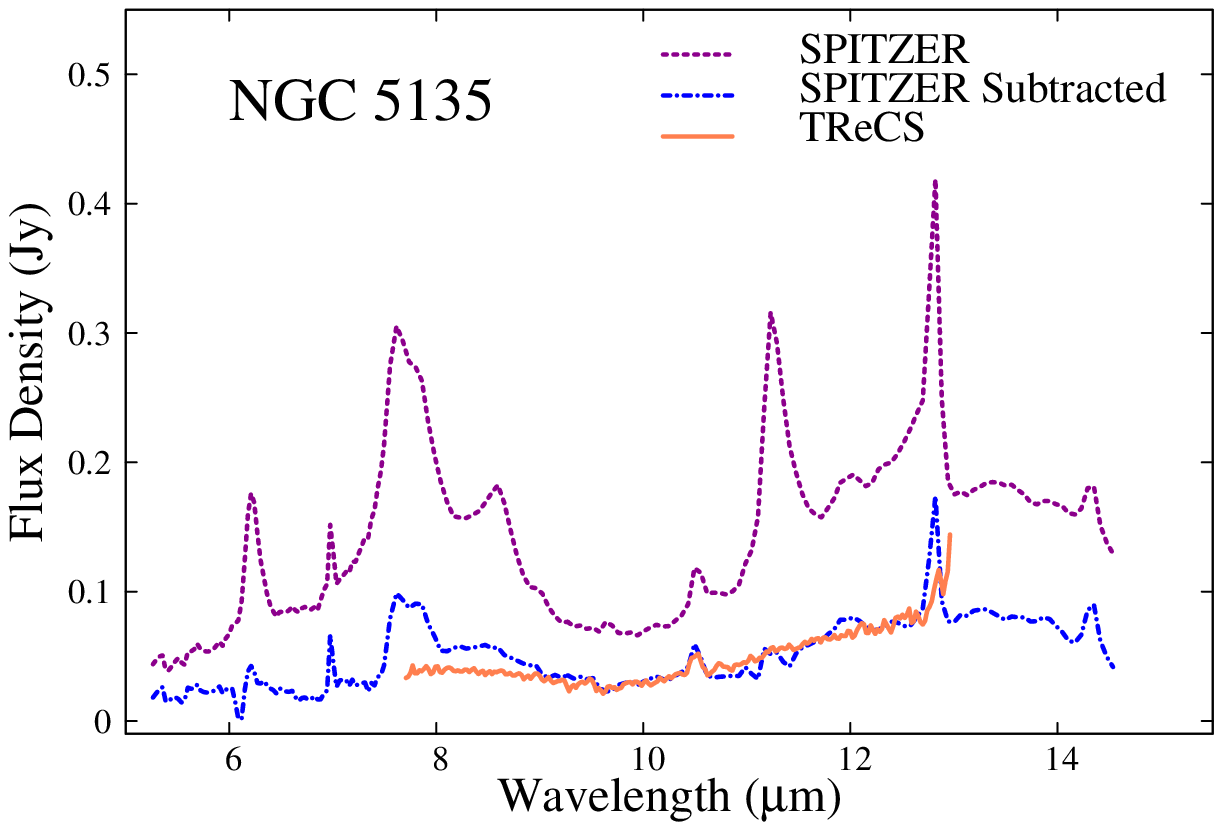}
    \includegraphics[scale=0.65]{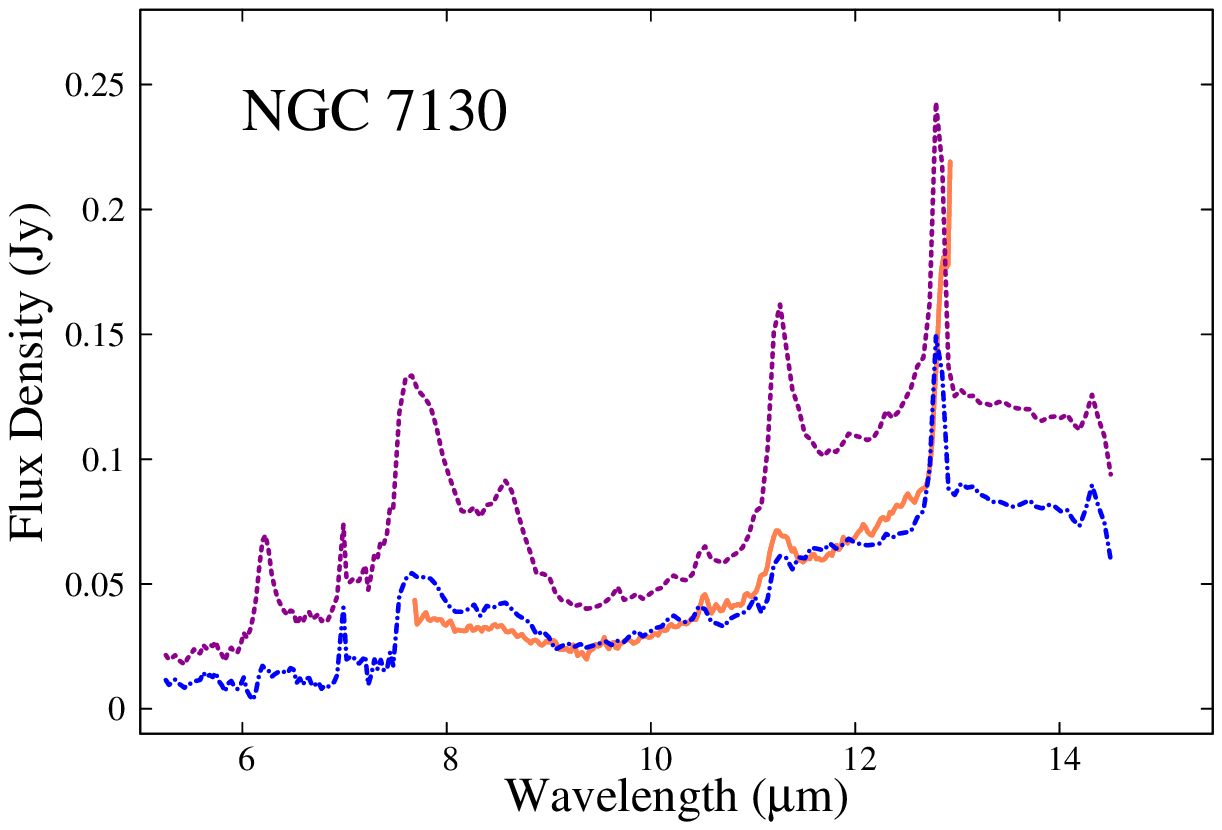}
 \end{center}
  \figcaption[]{Examples of star-formation corrected {\emph{Spitzer}}-IRS
    spectra compared with high-resolution ground-based, $\sim 8-13$ $\mu$m
    T-ReCS data for NGC\,5135 and NGC\,7130, as published by D\'iaz-Santos et
    al.~(2010). The \emph{Spitzer} spectra are presented before and after the
    subtraction of the starburst template and are shown with dotted lines. The
    subtracted IRS data have been scaled to match the T-ReCS data at
    $\sim10\mu$m. \label{trecs_examples}}
\end{figure}

\subsubsection{Star Formation in the {\emph{Spitzer}}-IRS Spectra}\label{clean_spec}

As the apertures used to extract the {\emph{Spitzer}}-IRS are larger than the
spatial resolution of our photometry and it is likely to include extended
emission coming from star formation located close to the nucleus, the spectra
needs to be modeled to subtract a star-forming component before including the
spectra in the nuclear SED of the galaxies in our sample. The presence of
polycyclic aromatic hydrocarbons (PAH), which dominate the MIR spectra of star
forming regions, will allow to isolate the AGN emission.

\citet{smith07} used low-resolution {\emph{Spitzer}}-IRS spectra of a sample
of nearby galaxies spanning a large range in star formation properties to
construct 4 templates in order to account for the differences of the PAH
feature strengths and continuum properties. These 4 templates were used to
subtract the star formation component from the MIR spectra of our
galaxies. Each template was scaled independently ensuring an optimal
subtraction of the PAH features but without oversubtracting the continuum
(i.e., without yielding negative fluxes). As can be seen in Fig.~18 of
\citet{smith07} the templates differ the most for $\lambda > 20 \mu$m. This
region was excluded during the model fitting of our SEDs because the lack
of photometric data at $\lambda > 10 \mu$m makes it impossible to determine
which template describes best the continuum emission for each particular
galaxy for wavelengths longer than 20 $\mu$m.

In Fig.~\ref{example_subt} extreme examples of the subtraction process are
shown. The resulting spectra for MCG\,-3-58-7 have no PAH features, and the
subtraction only modifies the continuum for $\lambda \gtrsim 20 \mu$m. The
second example, NGC\,5953, shows strong PAH residuals indicating that some
emission from star forming regions is still present in the subtracted
spectra. As discussed in Paper I, this object has a strong off-nuclear
starburst region which is contaminating the {\emph{Spitzer}} data. Still, for
the best subtraction a large fraction of the PAH emission for $\lambda \gtrsim
10\mu$m is gone, while the 7.7$\mu$m feature leaves strong residuals. Also, again
it can be seen that the main difference between the four different subtracted
spectra starts at $\lambda \gtrsim 20\mu$m. Only one object in our sample,
MCG\,-2-8-39, did not require any correction for the presence of a star
forming component since no evidence for the presence of PAH emission was found
in its spectrum.

\begin{deluxetable*}{ccl}
\tabletypesize{\footnotesize}
\tablecolumns{3}
\tablecaption{Parameter values for torus models.}
\tablehead{
\colhead{Par} & \colhead{Values} & \colhead{Description}}
\startdata
\multicolumn{3}{c}{CLUMPY Models (as described in Nenkova et al.~(2008b))}\\ 
\multicolumn{3}{c}{}\\
\hline
$\tau$                  &5, 10, 20, 30, 40, 60,  & V-band optical depth of individual clouds\\
                        &80, 100, 150& \\
$\mathcal{N}_0$         &1-15 in steps of 1      & Number of clouds at torus equator    \\
$\sigma$                &15-70 in steps of 5     & Torus half-opening angle (0 at equator)  \\
$q$                     &0-3 in steps of 0.5     & Radial cloud distribution exponent   \\
$Y$                     &5, 10-100 in steps of 10& Torus thickness ($=R_{out}/R_{in}$)  \\
$\angle i$              &0-90 in steps of 10     & Viewing angle (0 at polar axis)    \\
\hline
\multicolumn{3}{c}{}\\
\multicolumn{3}{c}{2pC Models (as described in Stalevski et al.~(2012))}\\
\multicolumn{3}{c}{}\\
\hline
$\tau_{9.7}$              &0.1, 1, 5, 10, 15, 20  & Total 9.7$\mu$m optical depth at equator\\
$R_{in}$                  & 0.5                  & Torus inner radius in pc\\
$R_{out}$                 & 15                    & Torus outer radius in pc\\
$p$                     & 0, 1                  & Radial clump distribution exponent\\
$q$                     & 0, 2, 4, 6            & Polar clump distribution exponent\\
$\Theta$                & 50                    & Torus half-opening angle (0 at equator)  \\
Inclination, $i$        & 0, 40-90 in steps of 10& Viewing angle (0 at polar axis)    \\
Filling factor$\dag$    & 0.15, 0.25            & Frequency of clumps\\
Contrast                & 100                   & Fraction of dust in clumps\\
Clump size$\dag$        & 0.15, 1.2             & in pc\\
Resolution$\dag$        & $1\!\!\times\!\!1\!\!\times\!\!1$, $8\!\!\times\!\!8\!\!\times\!\!8$ & number of grid cells per clump\\ 
\enddata

\tablecomments{$\dag$ available parameter combinations are: filling factor of
  0.15, size clump of 0.15 pc and resolution of
  $1\!\!\times\!\!1\!\!\times\!\!1$; or filling factor of 0.25, size clump of
  1.2 pc and resolution of $8\!\!\times\!\!8\!\!\times\!\!8$.}
\label{par_nenk}
\end{deluxetable*}

We would like to asses the efficiency of our starburst component subtraction
procedure. Using high-resolution ground-based T-ReCS data, D\'iaz-Santos et
al.\ (2010) have studied the characteristics of the nuclear MIR emission in
nearby AGN. Two objects presented by D\'iaz-Santos et al.\ (2010), NGC\,5135
and NGC\,7130, are also common to our sample.

Fig.~\ref{trecs_examples} compares our subtracted {\emph{Spitzer}}-IRS
observations with those obtained from the ground. The {\emph{Spitzer}} spectra
have been scaled so that the subtracted spectrum has a common density flux at
$\sim10\mu$m with the ground-based data.

Fig.~\ref{trecs_examples} shows that a good match is achieved between the
{\emph{Spitzer}} subtracted spectra and the T-REcS data, particularly for the
regions dominated by the continuum. PAH residuals can be seen for both objects
at the position of the 7.7 and 8.6$\mu$m features, even though the $\sim 12-13
\mu$m region presents a very clean subtraction. We can tentatively conclude
that whenever a good subtraction of the PAH features is achieved the resulting
spectrum can provide a good representation of the emission from the active
nucleus. In fact, to avoid being too sensitive to a poor subtraction, whenever
PAH residuals are seen, we use only a few spectral windows of the
{\emph{Spitzer}} observations to constrain our model fitting (\S4.1). It has
to be stressed, however, that PAH and dust emission are not the only
components present in the MIR spectra of starbursts. Silicate absorption can
be seen towards starburst nuclei with a mean flux ratio of the local continuum
to the base of the line of $0.2\pm0.1$ (Brandl et al., 2006).

In particular, NGC\,1144, NGC\,5135, NGC\,5953, NGC\,7130, NGC\,7496 and
NGC\,7582, show a strong starburst component before the template subtraction,
and significant structure afterwards. For NGC\,1144 a clean PAH subtraction is
achieved, but because of the intrinsic silicate absorption towards starburst
nuclei just mentioned it is impossible to asses whether the residual dip seen
at $\sim 10 \mu$m is due to the presence of a dusty torus or not. NGC\,7582
shows some PAH residuals at the position of the 7.7 $\mu$m PAH feature after
the template subtraction, and therefore, again it is not simple to asses the
presence of the residual $10 \mu$m feature. However, results from small
aperture, ground-based spectroscopy of NGC\,7582 shows that in this case the
presence of the silicate absorption is related to the nuclear source (AH11).

\subsubsection{Other Spectroscopic Data}

NGC\,1068, was observed with the {\emph{ISO}} satellite in 2001, and its
spectrum was published by Le Floc'h et al., (2001), already divided into AGN
and starburst components. Our photometry for this galaxy comes from an earlier
date (1998 and 1999), but IR variability of this nucleus was of the order of
0.1 magnitudes in the $K$-band during that period (Taranova \& Shenavrin,
2006)

NGC\,7172 was observed using the TRecS spectrograph mounted on the Gemini
South telescope (Roche et al., 2007). Therefore this spectrum presents a
narrower spectral coverage than the space-born observations. Because of the
high spatial resolution, this spectrum isolates the AGN emission from the
active nucleus.

\section{Modeling the nuclear IR SEDs}

\subsection{Clumpy Torus Model}

As was suggested in many previous works, a clumpy structure of the
distribution of gas and dust in the torus is more realistic than a continuous
one. The fundamental difference between clumpy and continuous density
distributions is that radiation can propagate freely between different regions
of a medium populated by optically thick clouds when it is clumpy, implying
that cold dust may exist near the nucleus, and dust directly illuminated by
the central source may exist far from it. The difficulties in modeling such an
environment plus the time consuming calculations (and technical limitations)
prevented the developing of such approaching until recently.

\citet{nenkova02} were the first in studying a clumpy distribution for the gas
and dust in the torus. In \citet{nenkova08a} they presented the general
formalism for handling this clumpy media. The resulting SEDs (hereafter, the
CLUMPY models) were presented in \citet{nenkova08b}, with an erratum on their
calculations in \citet{nenkova10}. They assume that all the clouds are
identical and characterize each one by its size (which should be much smaller
than the separation between clouds), its opacity and its spatial distribution
(the angular distribution of the clouds). Also the different dust temperatures
in the illuminated surface and the dark side of the cloud are accounted
for. The size distribution of the dust grains is that described by MRN,
composed by the standard Galactic mixture (53\% of silicate and 47\% of
graphite). Scattering is taken into account, dominating at short wavelengths
($\lambda\lesssim 1~\mu$m). They argue that the illuminating spectrum makes no
difference at wavelengths dominated by dust emission, $\lambda \gtrsim
2-3\mu$m, but at shorter wavelengths the AGN scattered radiation dominates.

Through the clumpy treatment \citet{nenkova02} can naturally explain the
rather low and diverse dust temperatures found close to the nucleus of
NGC\,1068 \citep{jaffe04,schartmann05}. Furthermore, it is found that the
X-ray attenuating column density is widely scattered around the column density
that characterizes the IR emission, because the IR flux is collected from the
entire observed area (averaging over all the clouds), while the X-ray opacity
is calculated from one particular line of sight.

The parameters for the CLUMPY models are the optical opacity of each cloud
($\tau_V$), the number of clouds through the torus equator ($\mathcal{N}_0$),
the angular and radial distribution of the clouds ($\sigma$, measured from the
torus equator, and $r^q$, respectively), $Y = \mathrm{R}_{out} /
\mathrm{R}_{in}$, and the angle between the axis of symmetry of the system and
the line of sight ($\angle i$). The values for each parameter are shown in
Table \ref{par_nenk}.

\begin{deluxetable*}{lccccccccccc}
\tablecolumns{12}
\tablecaption{Compiled Data for our Sample \label{gralinfo}}
\tablehead{
\colhead{Galaxy}&\colhead{Class}  &\colhead{HBLR} &\colhead{BD}     &\colhead{$F^c_{\rm OIII}$} &\colhead{$F^c_{\rm 2-10 keV}$}&\colhead{$L^{\rm bol}_{\rm OIII}$}&\colhead{$L^{\rm bol}_{\rm 2-10 keV}$}&\colhead{log($N_H$)}&\colhead{$F_{5-8GHz}$}&\colhead{$R_L$}         &\colhead{$\alpha$} \\
                &                 &               &                 &\multicolumn{2}{c}{10$^{-12}$ ergs/seg/cm$^{2}$}    &\colhead{ergs/s}            &\colhead{ergs/s}              &\colhead{cm$^{-2}$}  &\colhead{mJy}      &                        &\\
\colhead{(1)\dag}&\colhead{(2)\ddag}&\colhead{(3)\S}&\colhead{(4)\ddag}&\colhead{(5)$\ast$}    &\colhead{(6)$\ast$}      &\colhead{(7)}               &\colhead{(8)}                 &\colhead{(9)$\ast$} &\colhead{(10)$\o$} &\colhead{(11)$\natural$}&\colhead{(12)$\diamondsuit$}
}
\startdata
%#Name		Class_AGN  PBL Bal Dec F_OIII		F_RX	L_bol  L_bol   log_NH     F_GHz R_L
%1		2	  3	     4	 5 	 6       7      8       9          10    11	  12
NGC\,34		&Sy2-HII   & N	& 13.7&1.0-1.3	      & ---   & 44.2 & ---  &$>$24~$\wr$&14.5 &4.7    & 0.97$\pm$1.26\\
F\,00198-7926~N &Sy2	   & N	& --- &0.1-0.4        & ---   & 44.9 & ---  &$>$24~$\wr$&---  &---    & 0.10$\pm$0.15$\langle$\\
F\,00198-7926~S &Sy2	   & N	& --- &0.1-0.4        & ---   & 44.9 & ---  &$>$24~$\wr$&---  &---    & 5.02$\pm$0.16 \\
F\,00521-7054	&Sy2	   & N	& 8.3:&0.6-1.3	      & ---   & 45.5 & ---  &  ---	&---  &---    & 1.69$\pm$0.28 \\
ESO\,541-IG012	&Sy2	   &--- & 6.1:&0.5	      & ---   & 44.9 &  --- &  ---  	&0.8  &0.4    & 2.28$\pm$0.62 \\    
F\,01475-0740	&Sy2	   & Y	& 7.1 &0.8-0.9 	      &0.8    & 44.0 & 42.7 &  21.6  	&130  &70     & 0.34$\pm$0.09$\langle$\\
NGC\,1068	&Sy2	   & Y	& --- &98-190-280     & ---   & 45.3 & ---  &$>$25~$\wr$&1342 &1.2    & 5.77$\pm$0.86 \\
NGC\,1144	&Sy2	   & N	& --- &0.1-0.3-0.5    & ---   & 43.9 & 44.9 &  23.8  	&2.3  &3.4    & -0.67$\pm$0.24$\langle$\\
MCG\,-2-8-39	&Sy2	   & Y	& 4.2 &0.1-0.2-0.7    & 2.7   & 44.2 & 43.9 &  23.5  	&$<$0.3&$<$0.2& -1.00$\pm$0.01$\langle$\\
NGC\,1194	&Sy2	   &---	& 6.6 &0.3            & 10.9  & 43.1 & 43.8 &  24.0     &0.9  &2.6    & 5.92$\pm$0.43 \\
NGC\,1320 	&Sy2	   & N	& 3.8 &0.3-0.5-0.6    & ---   & 43.0 & ---  &$>$24~$\wr$&1.0  &2.1    & 2.86$\pm$0.17 \\
F\,04385-0828	&Sy2	   & Y	& 4.0:&0.04-0.09      &2.4-6.0& 42.5 & 43.3 &  ---	&6.0  &132    & 3.43$\pm$0.64 \\
ESO\,33-G2	&Sy2	   &--- & 5.0: &0.6	      & ---   & 43.8 & ---  &  22.0  	&---  &---    & 2.86$\pm$0.68 \\
F\,05189-2524	&Sy2	   & Y	& --- &0.8-1.2-1.3    &4.3-6.4& 45.1 & 44.6 &  22.8  	&6.9  &1.2    & 2.57$\pm$0.66 \\
%MCG\,+0-29-23	&HII	   &---	& 8.1 &0.2-0.6        & ---   & 43.9 & ---  &  --- 	&$<$0.3&$<$0.4& 0.02$\pm$0.11 \langle\\
NGC\,3660	&Sy2	   & N	& 3.3:&0.6-1.0        & 2.3   & 43.4 & 42.9 &  20.5  	&0.8  &0.9    & -0.32$\pm$0.37$\langle$\\
NGC\,4388	&Sy2-SB    & Y	& 4.0 &1.1-2.8-4.5-4.8& 43.0  & 43.9 & 44.0 &  23.4  	&34.6 &4.9    & 4.18$\pm$2.17 \\
NGC\,4501   	&Sy2       & N	& 3.7:&0.06-0.07      & 0.02  & 41.8 & 40.2 &  22.2  	&2.6  &95     & -0.93$\pm$0.11$\langle$\\
TOL\,1238-364	&Sy2	   & Y	& 4.0 &1.2-2.8-3.4    & ---   & 44.0 & ---  &$>$24~$\wr$&2.3  &0.4    & -0.46$\pm$0.02$\langle$\\
NGC\,4941	&Sy2	   & N	& 4.6 &1.1-3.6-4.6    & 3.0   & 43.0 & 41.8 &  23.7  	&9.0  &2.7    & --            \\
NGC\,4968	&Sy2	   &--- & 4.3 &0.4-0.7 	      & ---   & 43.1 & ---  &$>$24~$\wr$&2.1  &3.3    & 5.32$\pm$0.01 \\
MCG\,-3-34-64	&Sy1.8     & Y	& --- &4.2-5.2        & 4.0   & 44.8 & 43.6 &  23.6  	&42.2 &2.2    & -0.14$\pm$0.19$\langle$\\
NGC\,5135	&Sy2-HII   & N	& 5.6 &1.4-2.3-2.4    & ---   & 44.2 & ---  &$>$24~$\wr$&$<$2.3&$<$0.4& 3.16$\pm$2.08 \\
NGC\,5506	&Sy1.5-1.9 & N	& 4.4 &1.8-1.9-5.5-5.7&58-108 & 43.6 & 44.1 &  22.5-20.4&67.6 &11.9   & 2.83$\pm$0.80 \\
NGC\,5953	&LINER-Sy2 &---	& 4.2 &0.2-0.7 	      & ---   & 42.6 & ---  &  ---	&1.1  &3.1    & -1.68$\pm$1.53\\
NGC\,5995	&Sy2-HII   & Y	& --- &6.6-18.1       & ---   & 45.9 & 42.5 &  21.9	&2.4  &---    & 1.65$\pm$0.10 \\
F\,15480-0344	&Sy2	   & Y	& 4.6:&2.2-2.6-5.0    & ---   & 45.3 & ---  &$>$24~$\wr$&12.4 &0.6    & 0.50$\pm$0.23$\langle$\\
%NGC\,6810	&HII       &---	& 7.4:&0.6            & ---   & 42.8 & ---  &  ---	&---  &---    & -1.37$\pm$0.21\langle\\
NGC\,6890	&Sy2	   & N	& 5.0 &0.5-0.6-0.7    & ---   & 43.0 & ---  &  ---	&0.5  &0.8    & 1.89$\pm$0.92 \\
IC\,5063	&Sy2	   & Y	& --- &3.5-6.5	      &12-30  & 44.4 & 44.0 &  23.4	&506  &31.8   & -0.37$\pm$4.00\\
%MRK\,897	&Sy2	   &---	& 4.3 &0.04-0.04      & ---   & 42.8 & ---  &  ---	&3.5  &96     & 0.08$\pm$0.54\langle\\\
NGC\,7130	&LINER     & N	& 7.9 &2.1-4.5-6.0    & ---   & 44.7 & ---  &$>$24~$\wr$&18.1 &1.1    & -0.08$\pm$0.33$\langle$\\
NGC\,7172	&Sy2-HII   & N	& --- &0.04-0.07      & 21    & 41.9 & 43.7 &  22.9	&4.7  &259    & --            \\
MCG\,-3-58-7	&Sy2	   & Y	& 4.7 &0.7-1.5        & 2.3   & 44.7 & 43.9 &  23.4	&0.5  &0.1    & 2.42$\pm$0.35 \\
NGC\,7496	&Sy2-HII   & N 	& 5.1 &0.1-0.1-0.3-0.5& ---   & 42.0 & ---  &  22.7	&3.8  &32.2   & 0.39$\pm$0.24$\langle$\\
NGC\,7582	&Sy2	   & N	& --- &1.6-2.8-3.8    &4.0-27.2& 43.3& 42.3 &  23.1	&51.8 &14.2   & 2.66$\pm$0.31 \\
NGC\,7590	&Sy2-HII   & N 	& 5.9 &0.2-0.2        &1.2-1.14& 42.0& 41.8 &  $<$21 	&$<$0.2&$<$2.2& --            \\
NGC\,7674	&Sy1-HII   & Y	& --- &1.2-1.6-1.7-1.9& ---   & 44.8 & ---  &$>$25~$\wr$&12.8 &1.9    & 3.04$\pm$0.54 \\
CGCG\,381-051	&HII       &---	& --- &0.2	      & ---   & 43.6 & ---  &  ---	&0.6  &2.4    & -0.27$\pm$0.20$\langle$\\
\enddata
\tablecomments{
\dag: For alternative galaxy names see Paper I.
\ddag: Spectral Class and Balmer Decrements (BD) from Paper I; unreliable
BD values are labeled with ':'.
\S: Compilation of HBLR taken from Shu et al.\ (2007);
$\ast$: Data from 
Bassani et al.\ (1999),
Brightman \& Nandra (2008), 
Greenhill, Tilak \& Madejski (2008), 
Noguchi et al.\ (2010), 
Panessa \& Bassani (2002), 
Sazonov et al.\ (2007), 
Shu et al.\ (2007), 
Tran (2003), 
Winter et al.\ (2010);
only direct X-ray flux components are quoted; 
$\wr$ indicates Compton Thick sources;
$\o$: Data from Thean et al.\ (2000) and Gallimote et al.\ (2006) at 8.4 and 5
GHz, respectively; radio flux for IC\,5063 at 5 GHz was obtained from Gregory et al.\ (1994);
$\natural$: $R_L \equiv f_B/f_{5GHZ}$ (see text for details);
$\diamondsuit$: near-IR slopes from Paper I; values of $\alpha < 1$ at a 1 sigma
level are flagged with a $\langle$.}

\end{deluxetable*}

Essentially, the characteristics of the emerging emission depend on whether
the IR photons originate on the side of a cloud that is directly illuminated
by the AGN, or the region that is heated by the radiation emitted by other
clouds, the dark side. The optical depth will determine how different the SEDs
of these individual clouds are. The number of clouds of the whole torus
population and their geometrical distribution will determine how many clouds
are in the shadows of other clouds, while our orientation with respect to the
torus will dictate how much of directly illuminated or dark parts of clouds we
see.

\subsection{Two-Phase Torus Models}

Very recently Stalevski et al.~(2012) have proposed two-phase models where
dust might not only be found in clumps or clouds, but also in a diffuse medium
filling the space between the clumps (hereafter, the 2pC models). The diffuse
medium is controled by two parameters: the filling factor and contrast (see
Table \ref{par_nenk}). The filling factor determines the statistical frequency
of clumps, with a value $\gtrsim 0.25$ yielding enough clumps to form an
interconnected, sponge-like structure. The contrast parameter determines what
fraction of the dust is found in the clouds, with a value of 1 corresponding
to a smooth distribution and a value $>1000$ having effectively all the dust
in the clumps.

Stalevski et al.~(2012) claim that the largest differences when this diffuse
medium is included can be seen as a larger flux output in the
near-IR. 

\subsection{Fitting Procedure}

\subsubsection{$\chi^2$ Test}

We compare our observed SEDs with models where the torus is described as a
clumpy distribution of gas and dust as proposed by
\citet{nenkova08a,nenkova08b}, and which also includes a diffuse component as
proposed by Stalevski et al.~(2012).

In order to include the {\emph{Spitzer}} data in the modeling we had
to take into account the scaling of the spectra, the spectral windows
to be used, and the associated errors.

For those objects where we had a N-band flux measurement, the spectra was
scaled to match the photometric value. For those objects without a
measurement, the average of the scaling factors found for the objects with
N-band photometry was used. This was found to be $\sim 0.06\pm0.03$, with the
error corresponding to the standard deviation of the determined scaling
factors. Photometric upper limits are also shown in the SEDs but they were not
used during the fitting process.

To avoid using {\emph{Spitzer}} spectra with a poor subtraction of the
starburst component, we masked out strong PAH residuals during our fitting
procedure, Also, for very noisy spectral data, the windows were averaged into
"photometric'' points, assuming the corresponding error of the N-band
measurement. For well subtracted, high signal-to-noise spectra, the whole
spectrum was used during the model fitting. We assumed a 5\%\ error for each
spectroscopic point (Wu et al.~2009) of the {\emph{Spitzer}} observations.

We programmed a simple $\chi^2$ routine that calculates the scaling factor
needed shift the clumpy torus models to match the observed SEDs. The best fits
correspond to the smaller $\chi^2$ values. Notice that given our adopted
scaling for the {\emph{Spitzer}} data, the determined $\chi^2$ do not
necessarily represent the actual goodness of the fit but allows to determine
which is the best possible model.

Some fits obtained using the CLUMPY library, however, are at odds with the
expected results. For example, the canonical Seyfert II galaxy NGC\,1068,
which has a well constrained SED, including a large number of IR photometric
observations and a well determined MIR spectrum, is best fitted by models
presenting a small value of the torus angle, this is, models with face-on
orientations. This is not supported by a large body of evidence on the
geometry of the central region in this object. In fact, based on this
evidence, AH11 restricted the torus angle to values corresponding to high
inclination only.

It can be concluded, that the shape of the SED alone might not be sufficient
to distinguish between models in the very large CLUMPY library. Whether or not
a similar situation would be found when using the 2pC models is not clear due
to the much smaller set of available models.

\subsubsection{$L^{\rm bol}$ Estimates: Breaking the Degeneracy}

\begin{figure}
  \begin{center}
    \includegraphics[scale=0.45,trim=50 20 0 200]{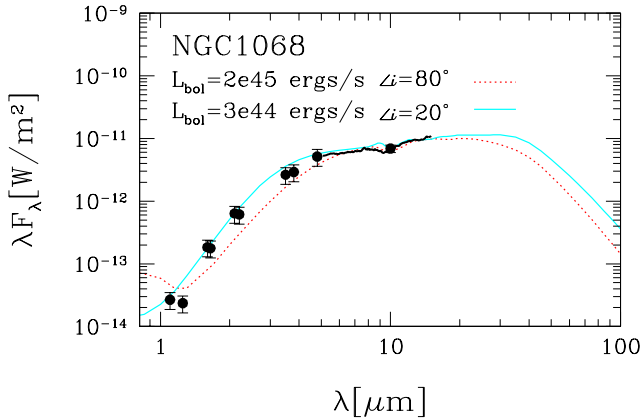}
  \end{center}
  \figcaption{Fits to the Spectral Energy Distribution of NGC\,1068 with and
    without (dashed and continuous lines, respectively) a $L^{\rm bol}$ prior
    (see \S 3.2.2 for details). \label{ngc1068}}

\end{figure}

In order to break the possible degeneracy between different CLUMPY models that
present similarly shaped SEDs in the near and mid-IR regime, we have
introduced another restriction. Different physical parameters or geometry of
the CLUMPY models, scaled to the same flux level, will necessarily imply
different bolometric luminosities of the central AGN. Therefore, we can use
the 'observed' bolometric luminosities to discriminate between competing
models. This is similar to the approach followed by Mor et al.~(2009) when
modeling {\emph{Spitzer}} observations of 26 luminous quasars.

Determining the bolometric luminosity of Type II sources is not
straightforward. The two most promising luminosity proxies are the [OIII] line
emission arising from the Narrow Line Region (NLR), and the 2-10 keV emission
in the X-rays. However, both methods present important drawbacks. [OIII] fluxes
can be difficult to correct for slit losses and aperture effects for very
extended NLRs, while determining an accurate extinction correction can be
problematic when the Balmer decrement is affected by absorption line features
from the underlying stellar population. Probably of less importance would be
taking into account the fraction of UV flux intercepted by the putative torus,
and determining the contamination by a starburst component. At the same time,
the 2-10 keV emission is strictly a proxy only for Compton-thin sources
(i.e., those with $N_H < 10^{24}$ cm$^{-2}$), and determining accurate
unabsorbed 2-10 keV fluxes can be difficult for highly absorbed sources ($N_H
\sim 10^{23-34}$ cm$^{-2}$) with average quality data. See the work by LaMassa
et al.~(2010) for more discussion on the subject of luminosity proxies.

We have compiled extinction and absorption corrected [OIII] and 2-10
keV fluxes for our sample (Table \ref{gralinfo}). [OIII]
measurements are available for all of our sources, and it can be seen
that they generally agree within a factor of 3 or better. X-ray fluxes
are only available for about 40\%\ of the sample, mainly due to the
fraction of Compton-Thick sources.

Using the expressions derived by Marconi et al.~(2004) we have
estimated the bolometric luminosities from the 2-10 keV X-ray
observations (see Table \ref{gralinfo}). For the [OIII] data we
have followed the work of Lamastra et al.~(2009) who compared
extinction corrected [OIII] luminosities and 2-10 keV fluxes for a
sample of Type II Seyfert galaxies to find $L_{\rm 2-10 keV} \approx 10
L_{\rm OIII}$. Adopting the Marconi et al.~(2004) bolometric
correction for the X-ray data, Lamastra et al.~(2009) then tabulated 3
different bolometric corrections for $\log(L_{\rm OIII})$ in the
ranges 38-40, 40-42, and 42-44 ergs/s. Instead, we have combined the
[OIII] vs 2-10 keV correlation given above with the bolometric
correction found in Marconi et al.~(2004) to find the cubic
expression:

\begin{eqnarray}
\log L_{\rm OIII}/L_\sun = 9.45 - 0.76 \mathcal{L} + 0.012 \mathcal{L}^2 - 0.0015 \mathcal{L}^3
\end{eqnarray}

with $\mathcal{L} = \log L^{\rm bol}/L_\sun -12$. The results are reported in
Table \ref{gralinfo} where we give the average value of $L^{\rm bol}$ when
more than one [OIII] flux is available. From objects with multiple [OIII]
observations we find a {\em rms\/} of $0.3\pm0.05$ in fractional flux. This
{\em rms\/} value will be used as the standard deviation of the [OIII] flux
distribution in what follows.

Table \ref{gralinfo} shows that when both estimates of the bolometric
luminosities are available, in about 50\%\ of the cases the values agree
within a factor of 5 or better. However, in the other 50\%\ of the cases there
are important disagreements. It is significant that in all but two cases
(NGC\,1144 and NGC\,7172) $L^{\rm bol}_{\rm OIII} > L^{\rm bol}_{\rm 2-10
  keV}$, maybe due to an insufficient absorption correction of the X-ray flux
since several of these sources have large absorbing columns. Because of this,
and the complete availability of [OIII] fluxes for our sample, we will use
$L^{\rm bol}_{\rm OIII}$ as a prior for the bolometric luminosity of our
sources.

From the models, the bolometric luminosity of the AGN illuminating the
torus is found as $L^{\rm bol}_{\rm model} = \Theta \times 4 \pi d^2$,
where $d$ is the distance to the galaxy, and $\Theta$ is the scaling
factor needed to shift the model to the observed SED data points.

We use $L^{\rm bol}_{\rm model}$ to restrict the best fit models by applying a
Bayesian approach. We compute prob(model$|$SED) $\propto$
prob(SED$|$model)$\times$prob(model$|$Lbol), where we assume prob(SED$|$model)
$\propto \exp -(\chi^2/2)$ (i.e., the likelihood that the SED has been
obtained from the model) and prob(model$|$Lbol) $\propto \exp(-(L^{\rm
  bol}_{\rm OIII}-L^{\rm bol}_{\rm model})^2/2\sigma_{\rm OIII}^2)$ as prior,
with $\sigma_{\rm OIII} = 0.3 \times L^{\rm bol}_{\rm OIII}$.

Figure \ref{ngc1068} shows the case study for NGC\,1068. When restricting
$L^{\rm bol}_{\rm model}$ to the value implied by the observed [OIII] fluxes, the
correct picture emerges. When applied to the entire sample, we find that about
$\sim$20-25\%\ of sources change significantly their best-fit model results
when applying the $L^{\rm bol}$ restriction.

\subsubsection{Fitting Sources with near-IR Excess}

\begin{figure}
  \begin{center}
    \includegraphics[scale=0.42,trim=0 0 0 150]{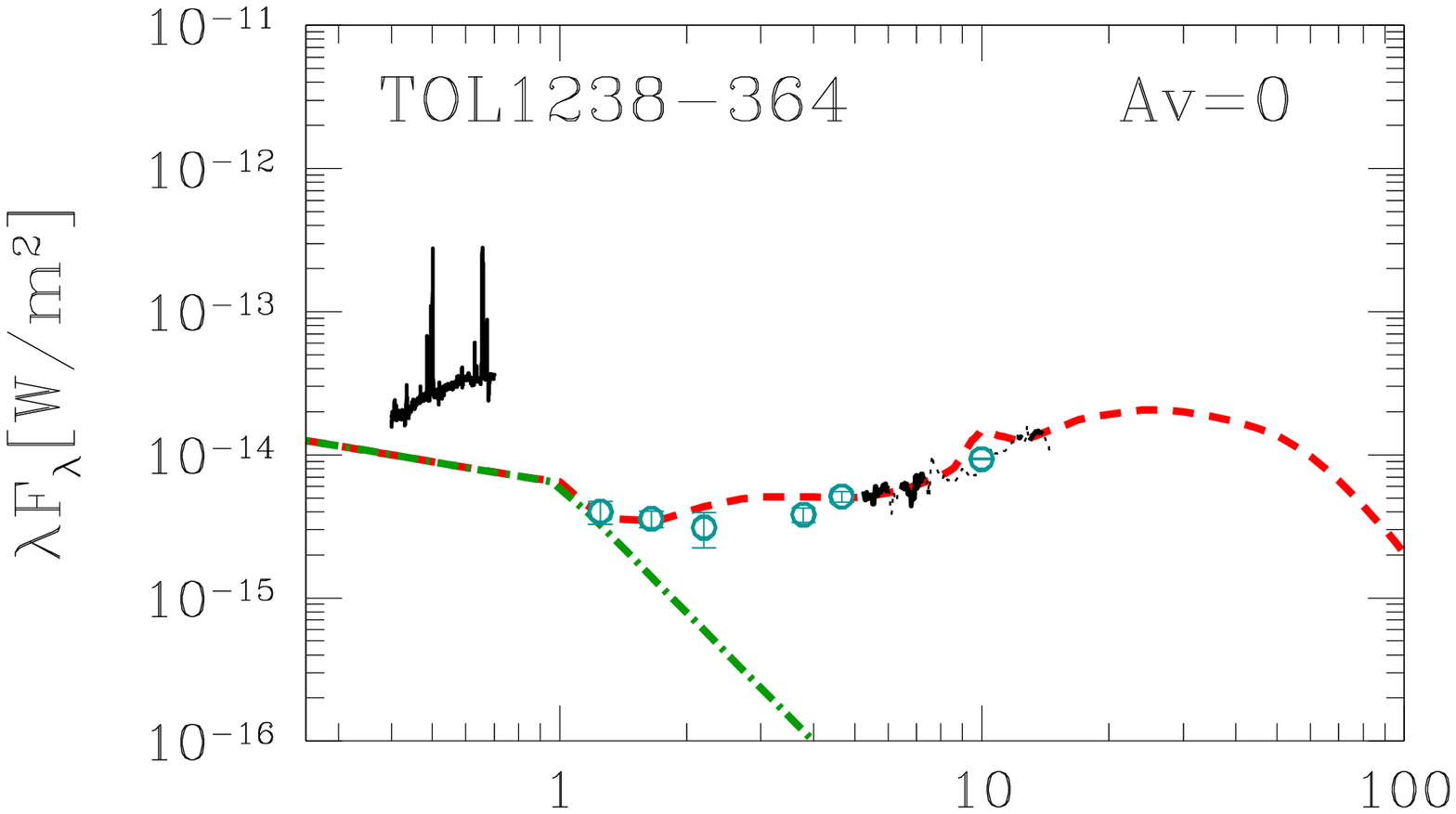}\\
    \includegraphics[scale=0.42,trim=0 0 0 250]{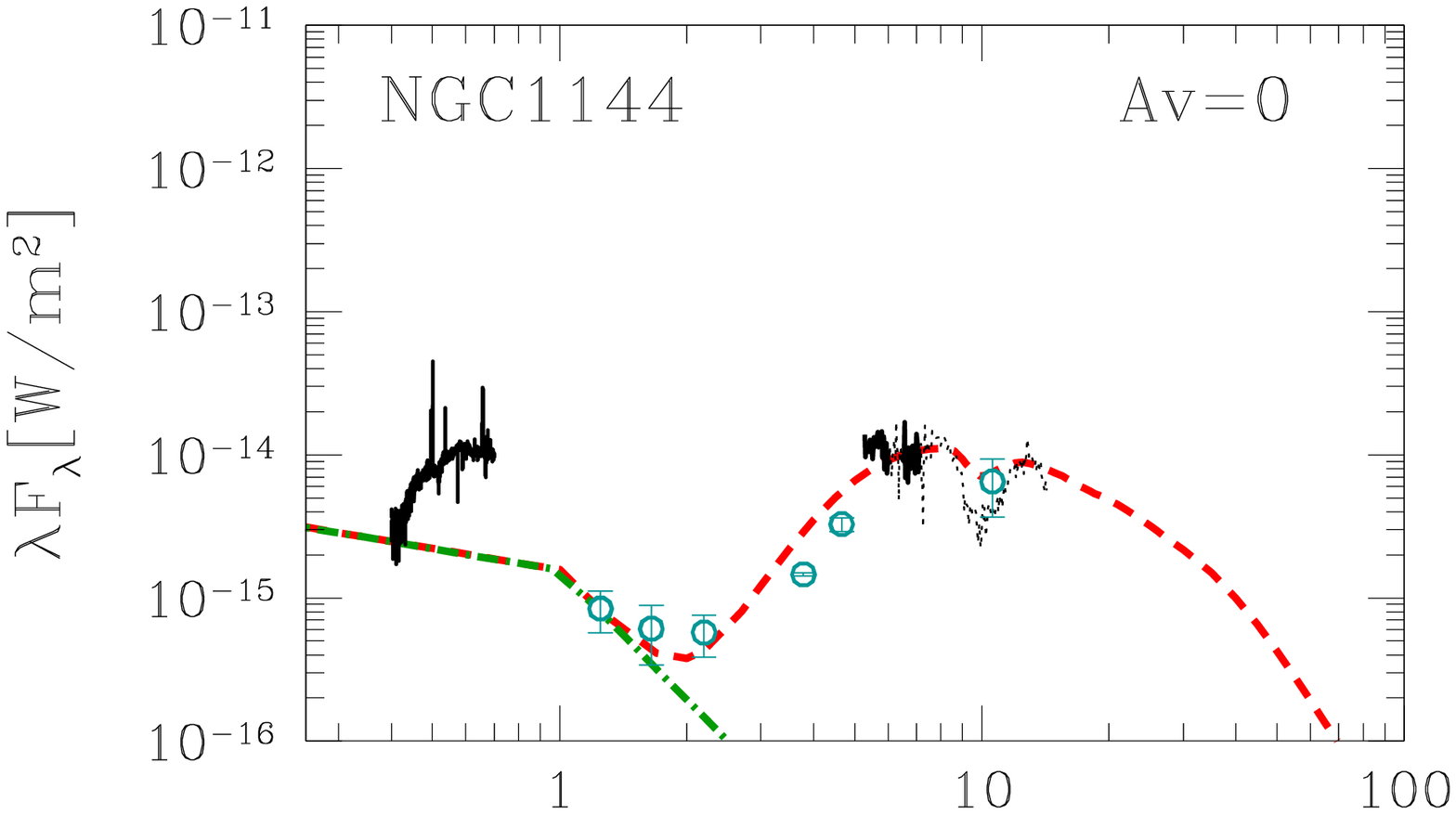}\\
    \includegraphics[scale=0.42,trim=0 0 0 250]{IRAS01475-0740.tor.4m.eps}\\
    \includegraphics[scale=0.42,trim=0 0 0 250]{NGC7130.pow.4m.eps}
  \end{center}
  \figcaption{Example nuclear infrared Spectral Energy Distributions and
    optical spectra from Paper I for 4 of our sources with near-IR excess. The
    best model fit to the infrared SED is shown together with the accretion
    disk component after extinction has been applied to prevent it from
    appearing in the optical regime (dashed and dash-dotted lines,
    respectively). The amount of extinction applied to the accretion disk is
    shown at the top left corner of each panel. Notice that only TOL\,1238-364
    was observed under photometric conditions (Paper I), implying that the
    absolute flux calibration of the optical spectra for NGC1144,
    IRAS01475-0740 and NGC7130 might be underestimated.
    \label{4mspectra}}
\end{figure}

In many of our sources a near-IR excess is observed. These SEDs turn upwards
for $\lambda\! \la\! 3\mu$m, this is, they present an index $\alpha\! <\! 1$
in the near-IR at a 1 sigma level (for $\lambda f_{\lambda} \propto
\lambda^{\alpha}$). These objects are flagged in Table \ref{gralinfo}. Notice
that MCG\,+0-29-23, NGC\,6810 and MRK\,897, which are classified as
star-forming nuclei and therefore are not included in the present analysis,
also show a near-IR excess (Paper I). Also, NGC\,34 and NGC\,5953 have
$\alpha\!  <\! 1$ (although NGC\,34 is not at a 1 sigma level), but as
argued below (see also \S 4.1), it is very likely that these sources are
contaminated by a starburst component and therefore are not flagged in Table
\ref{gralinfo} as presenting a 'genuine' near-IR excess.

In Paper I several possibilities to explain the presence of the observed
near-IR excess were proposed: contribution from a compact nuclear starburst or
a nuclear stellar cluster, emission from the accretion disk, emission from a
compact jet, or emission from a very hot dust component that survives within
the sublimation radius of the torus.

The first scenario, the contribution from a compact starburst, was not
supported by the analysis of diagnostic-diagrams from emission line ratios and
the analysis of the observed stellar continuum (see Paper I). Strong PAH
residuals in the {\emph{Spitzer} data not necessarily imply a strong nuclear
  starburst component because of the large apertures. However, the nuclear
  optical spectra presented in Paper I suggests that NGC\,34 might have a
  strong nuclear starburst, while NGC\,5953 has a very strong circumnuclear
  star-forming region.

A second scenario is the presence of a luminous nuclear stellar cluster. While
nuclear clusters are found in most galaxy nuclei and are characterized by a
light weighted stellar age of $\sim 10^{8-9}$ years (Walcher et al., 2006),
their luminosities (absolute magnitude $z_{\rm AB}=-13$ for the most luminous
cases; Cote et al.~2006) are not sufficient to make a significant contribution
to the observed near-IR fluxes in our sample.

The last three scenarios would require the leaking of the radiation from the
central region through the clumpy obscuring torus. Since the nature of a
clumpy medium would allow for certain lines of sight to peer directly at the
central engine, even for highly obscured sources, in what follows we will
study these scenarios in more detail. As 12/31 sources show evidence for a
near-IR excess, we can postulate that the clumpy structure leaves, on average,
$\sim 40\%$ of the lines of sight free of absorption for Type II sources.

The first working hypothesis is that the presence of a near-IR excess could be
due to central disk emission piercing through the torus for the particular
line-of-sight we have of the system. To test this hypothesis SED fitting was
done using the whole library of CLUMPY models, which besides the pure torus
emission also have 'SED+AGN' models, in which a torus is combined with the
emission from the central nucleus. The advantage of using these models to
simply adding an arbitrary power-law component is that the normalization of
the AGN emission, which in turn illuminates the obscuring torus, is treated
consistently with that of the emission from the torus itself.

A consistency check to the best fit 'SED+AGN' models is to test whether the
accretion disk emission is expected to be seen in the optical, since our
sample is characterized by a strong optical stellar continuum and the lack of
a power-law component. We find that only three sources require a moderate
amount of redenning ($A_V \la 1$) to prevent disk emission from showing in the
optical observed spectra (see Fig.~\ref{4mspectra}). This extinction could
correspond to lines of sight at a grazing angle to some clouds in the torus,
or to extinction introduced at larger scales. In principle, therefore, it is
possible to have a disk component appearing at near-IR wavelengths without a
disagreement with the optical observed continuum.

A second consistency check is to examine the observed hydrogen column
densities. A direct view of the accretion disk should correspond to objects
presenting small X-ray inferred hydrogen columns, since these two observations
would probe the line-of-sight to the central region of the active nucleus. Of
the 12 sources identified as having a near-IR excess, one lacks X-ray
observations (CGCG\,381-051). Of the remaining, 8 have fairly large absorbing
columns ($\log(N_H) \ga 22.5$ cm$^{-2}$; see Table \ref{gralinfo}), while
IRAS\,01475-0740, NGC\,3660 and NGC\,4501 have $\log(N_H) = 21.6, 20.5$ and
22.2 cm$^{-2}$, respectively.

This result could be explained in the context of a clumpy torus if the
probability of the power-law emitting region being obscured would be lower by
$\sim 70\%$ (8/11) than that of the X-ray emitting region. In fact, in the
canonical picture of the central engine of an AGN, the X-ray emitting region
probably does not extend further than a few gravitational radii from the
central Black Hole, while the optically emission coming from the accretion
disk should reside hundreds of gravitational radii away, if a classical
Shakura-Sunyaev $\alpha$-disk is adopted \citep{shakura76}.

However, this argument should also work in the opposite sense: those sources
with small values of $N_H$ should show a high probability to also exhibit an
accretion disk component in the optical and near-IR. From Table \ref{gralinfo}
we find that, besides IRAS\,01475-0740, NGC\,3660 and NGC\,4501, 4 other
galaxies which do not show a near-IR excess have $N_H \la 22$ cm$^{-2}$:
ESO\,33-G2, NGC\,5506, NGC\,5995 and NGC\,7590. In summary, out of 7 sources
with low $N_H$ columns 3 have near-IR excess and 4 do not. Hence, there is a
$\sim 50\%$ chance to have a direct view towards the accretion disk when there
seems to be unimpaired access of the innermost region towards the central black
hole. This result comes from small number statistics but still argues against
our null hypothesis.

Moreover, our sample is composed entirely by objects with Type II
classification from the absence of broad emission lines in their optical
spectra, this is, their BLR is to a large degree completely obscured to us in
direct light. But from the analysis above, in $\sim 40\%$ of the sources we
might have a direct view of the accretion disk. As before, this would suggest
that the probability of seeing a larger structure is smaller than that of
seeing more compact regions. This strongly argues against the interpretation
of the near-IR excess as emission from the accretion disk. It is still
feasible that in some of our sources the accretion disk component is indeed
seen in direct emission, but it seems very unlikely that this can explain the
large fraction of sources with near-IR excess in our sample. Moreover, as it
will be further discussed in \S 4.3, other works have already claimed the
presence of an extra near-IR component in the SEDs of Type I AGN, where the
accretion disk and the innermost torus is readily visible. This is, the
component is required {\em besides} the disk and torus emission, as modeled
by CLUMPY. This extra component has so far been accounted for using a
black-body spectrum.

Consequently we conducted a fitting process using the torus CLUMPY 'SED'
models plus a black-body component with a free scaling parameter (as opposite
to the previous 'SED+AGN' modeling where both components were jointly scaled),
representing emission from a hot component such as hot carbonaceous dust
grains surviving within the sublimation radius of the silicate dust. We tested
temperatures in the 1000--2500 K range using steps of 100 K.

The models proposed by Stalevski et al.~(2012) might offer a different
solution. Hot dust located in the diffuse component might be found further out
in the torus and therefore increase the chance of making a contribution to the
near-IR even for Type II sources. Because of the much smaller library of 2pC
models currently available, and to test whether the treatment of the diffuse
component detailed in Stalevski et al.~(2012) can account for the near-IR
excess emission, we conducted a simple $\chi^2$ minimization without adding
extra components.

\subsubsection{Fitting Radio-loud Sources}

We have compiled nuclear radio measurements for most of our sources from Thean
et al.\ (2000) and Gallimore et al.\ (2006) at 8.4 and 5 GHz, respectively
(see Table \ref{gralinfo}). We computed a radio-loudness parameter as
$R_L=F_{5 GHz}/F_{B}$ as defined by Kellermann et al.\ (1989). Thean et
al.\ (2000) fluxes were taken to 5 GHz assuming $S_{\nu} \propto \nu^{-0.2}$,
implying a correction factor of 1.1 for these measurements. Since we do not
have a direct measurement of the nuclear $B$ magnitudes, we have used the
$L^{\rm bol}_{\rm OIII}$ values found in Table \ref{gralinfo} to determine $B$
fluxes assuming the relation defined by Marconi et al.\ (2004). Radio-loud AGN
are defined as those with $R_L \ga 30$. We can see that 5 of our sources
fulfill this criteria, while most are in the Radio-quiet regime ($R_L <<10$).

As detailed in Paper I, 3/12 of the sources with near-IR excess are classified
as radio-loud: IRAS\,01475-0740, NGC\,4501, and NGC\,7496. We can try to
estimate whether Synchrotron emission is responsible for the near-IR excess
using a correlation between optical and radio flux determined for unbeamed
low-power radio-loud AGN derived by Chiaberge et al.~(1999) after
interpolating to the J-band wavelength.

The predicted fluxes for our sources are extremely small, between 5 and 7
orders of magnitude below the observed fluxes, and clearly will not explain
the observed near-IR excess, unless strong beaming takes place. We cannot
rule-out this last possibility. Radio variability would uncover beamed
sources.

We conducted the fitting process for the 3 candidate beamed radio sources
using the torus CLUMPY 'SED' models plus a power-law with a free scaling
parameter. The power-law would correspond to a beamed jet which can have
strong near-IR emission during a 'high-state' (Bonning et al., 2009).

We used the calculated near-IR slopes as a first guess and explored 10
values around $\alpha$ in steps of 0.1.

\begin{figure*}
  \begin{center}
    \includegraphics[scale=0.35,trim=50 100 45 160]{NGC34.pow.eps}
    \includegraphics[scale=0.35,trim=50 100 45 160]{IRAS00198-7926_N.tor.eps}
    \includegraphics[scale=0.35,trim=50 100 45 160]{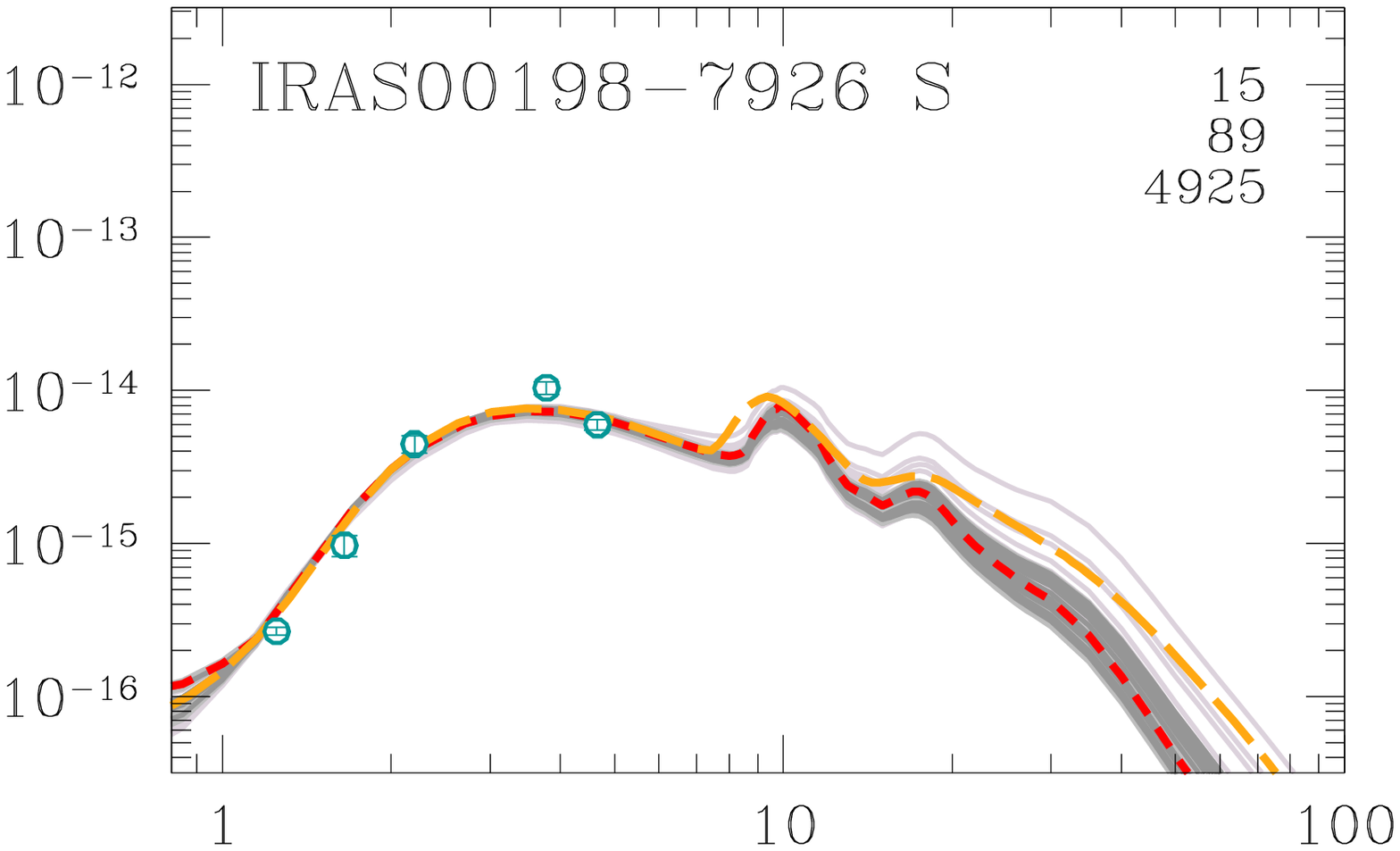}\\
    \includegraphics[scale=0.35,trim=50 100 45 160]{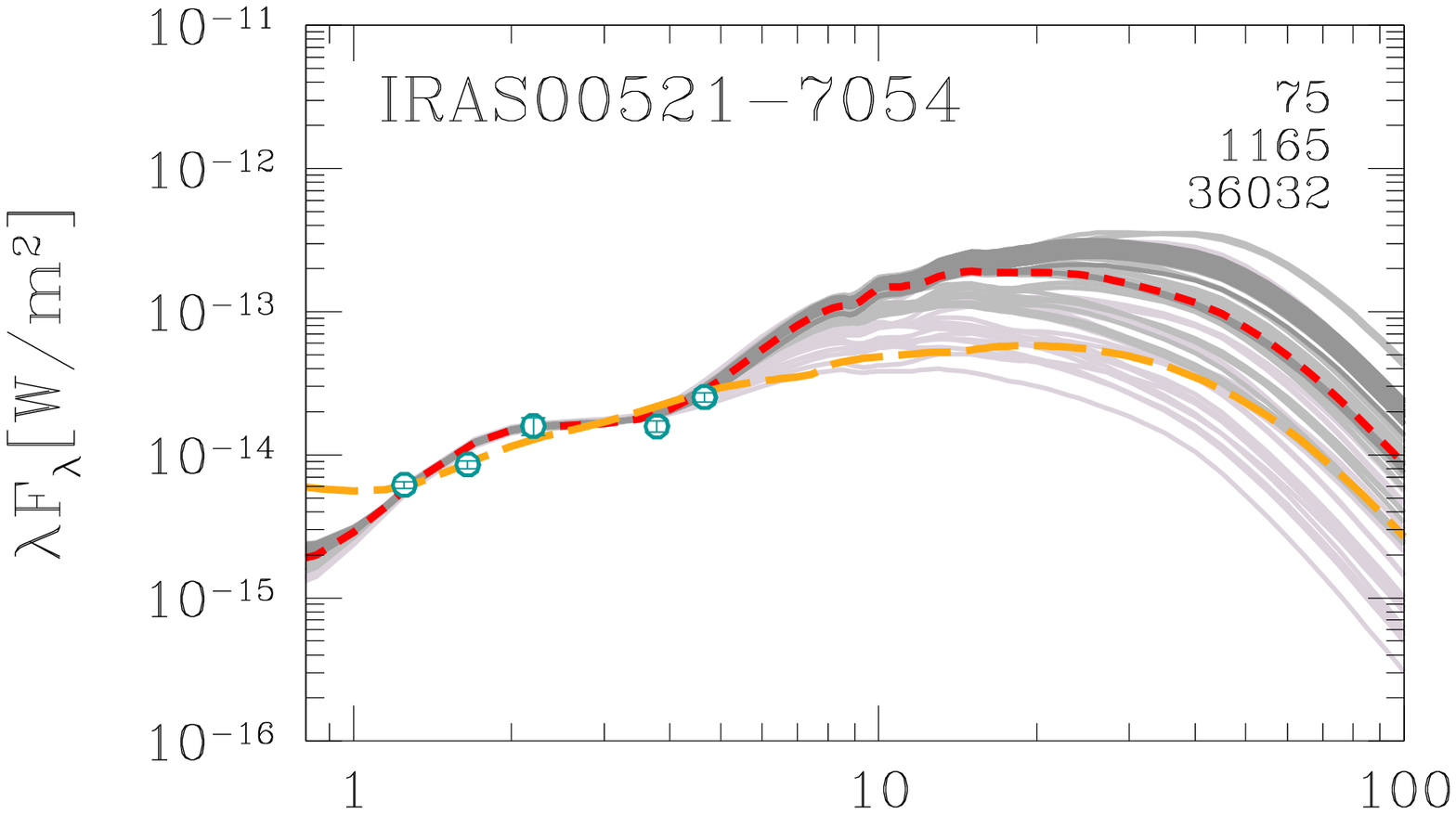}
    \includegraphics[scale=0.35,trim=50 100 45 160]{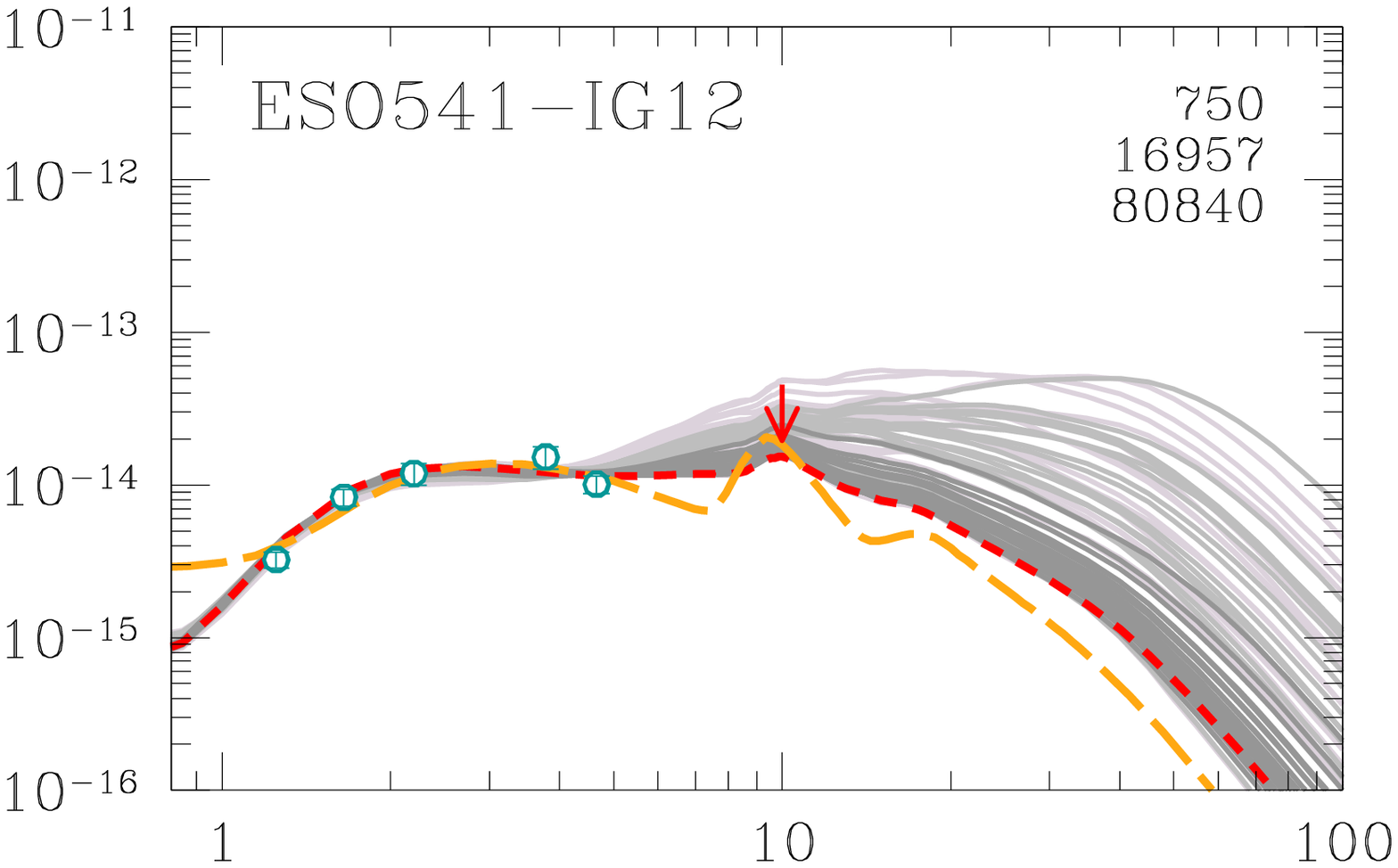}
    \includegraphics[scale=0.35,trim=50 100 45 160]{IRAS01475-0740.tor.eps}\\
    \includegraphics[scale=0.35,trim=50 100 45 160]{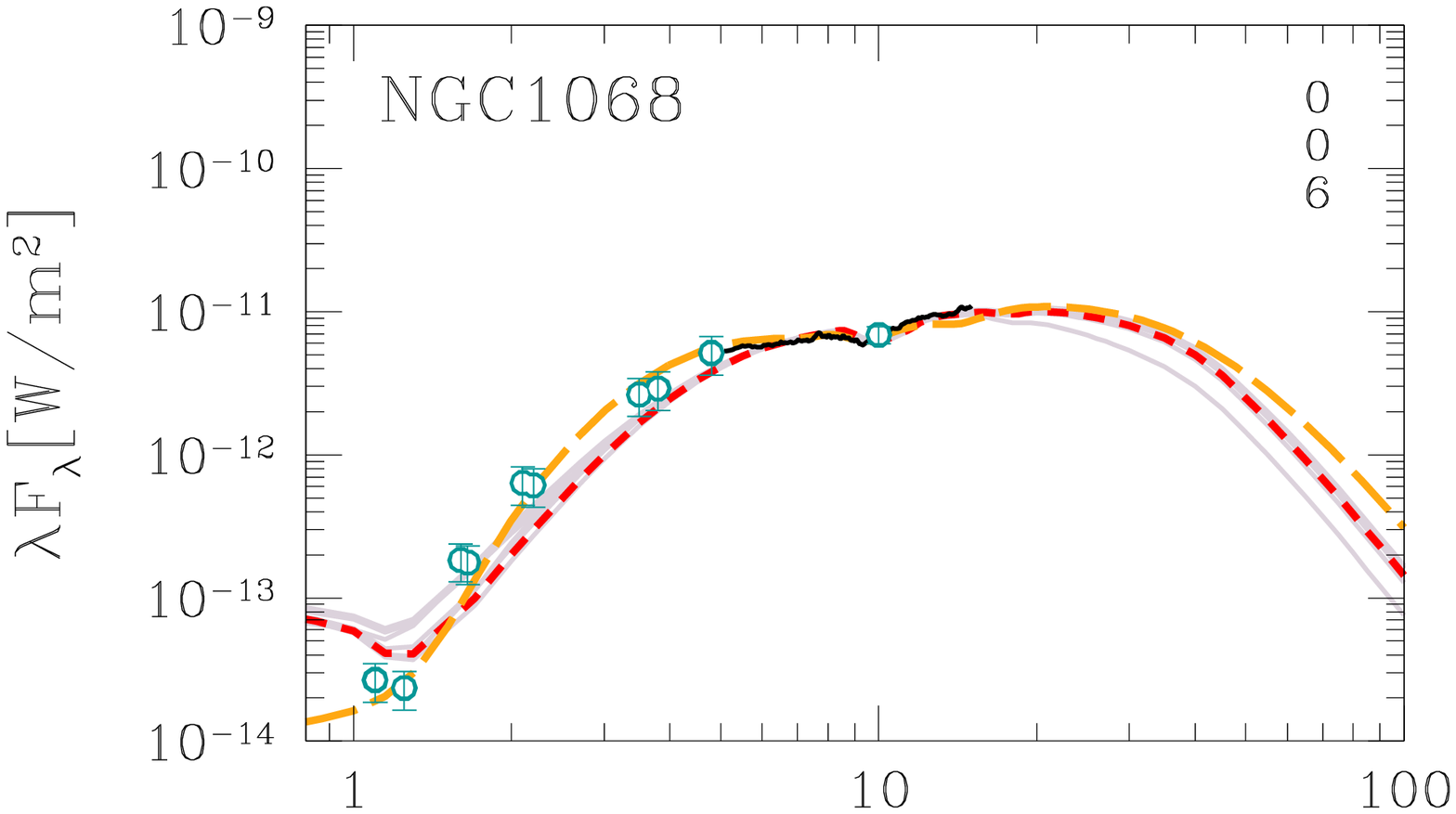}
    \includegraphics[scale=0.35,trim=50 100 45 160]{NGC1144.pow.eps}
    \includegraphics[scale=0.35,trim=50 100 45 160]{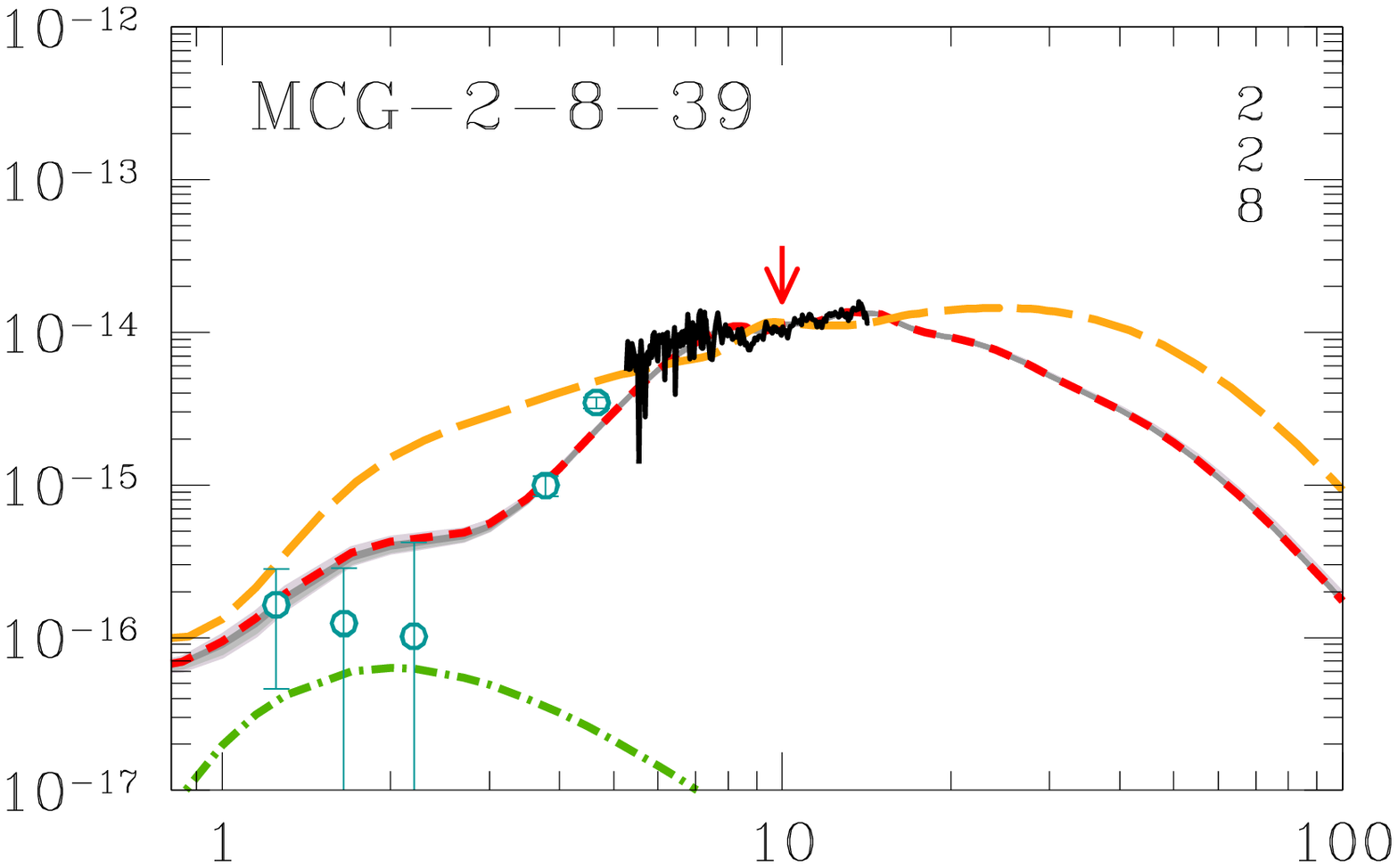}\\
    \includegraphics[scale=0.35,trim=50 100 45 160]{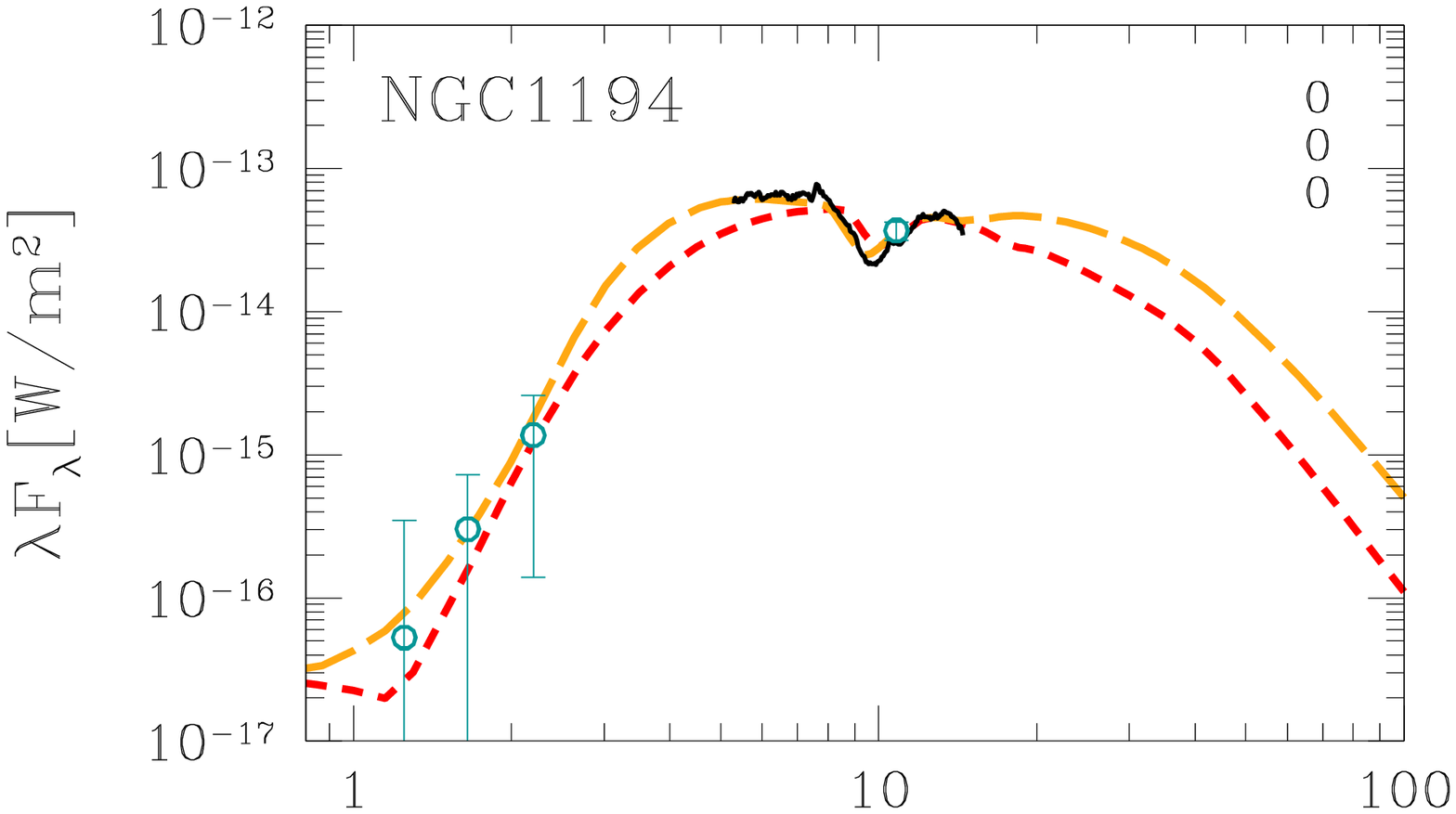}
    \includegraphics[scale=0.35,trim=50 100 45 160]{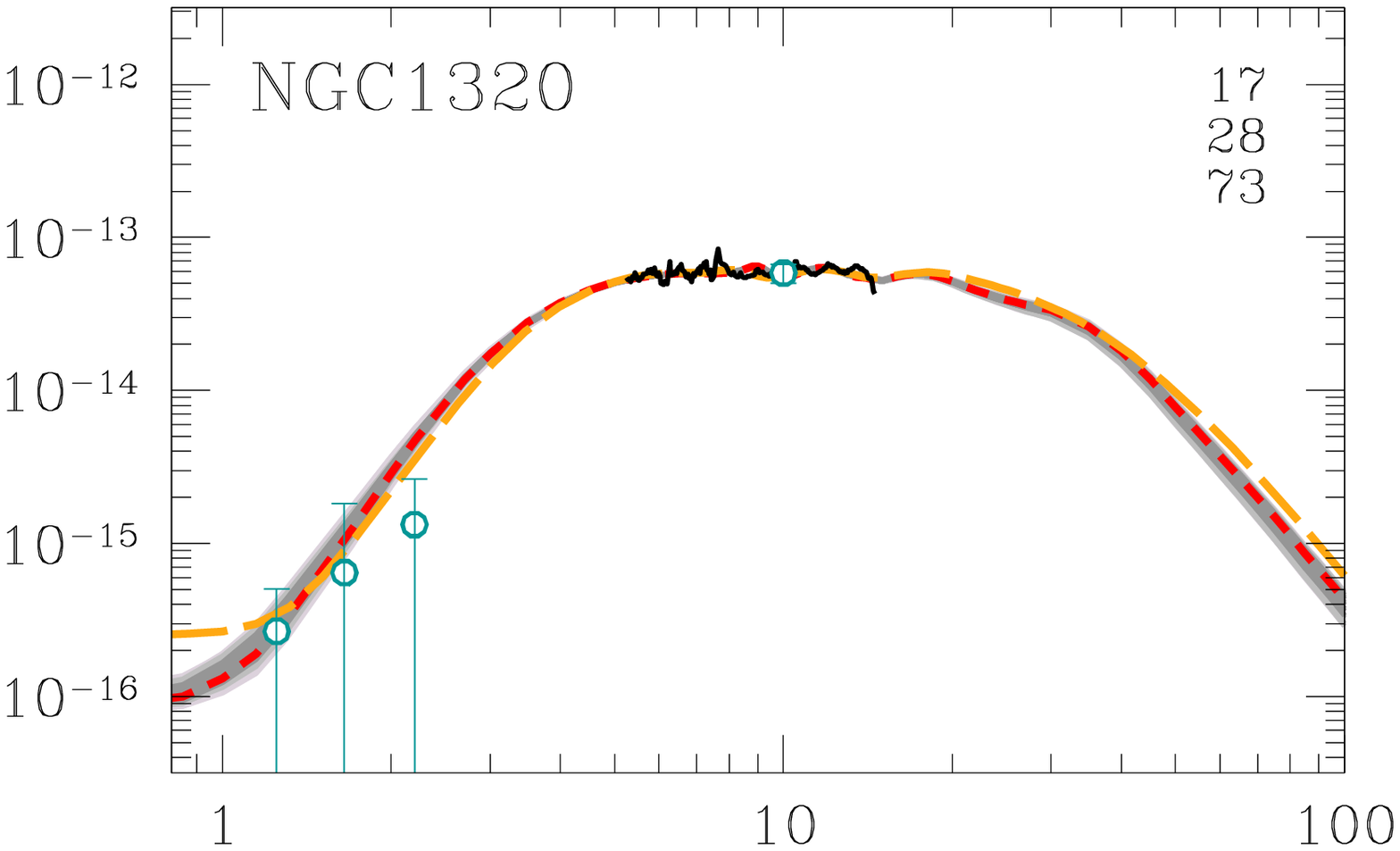}
    \includegraphics[scale=0.35,trim=50 100 45 160]{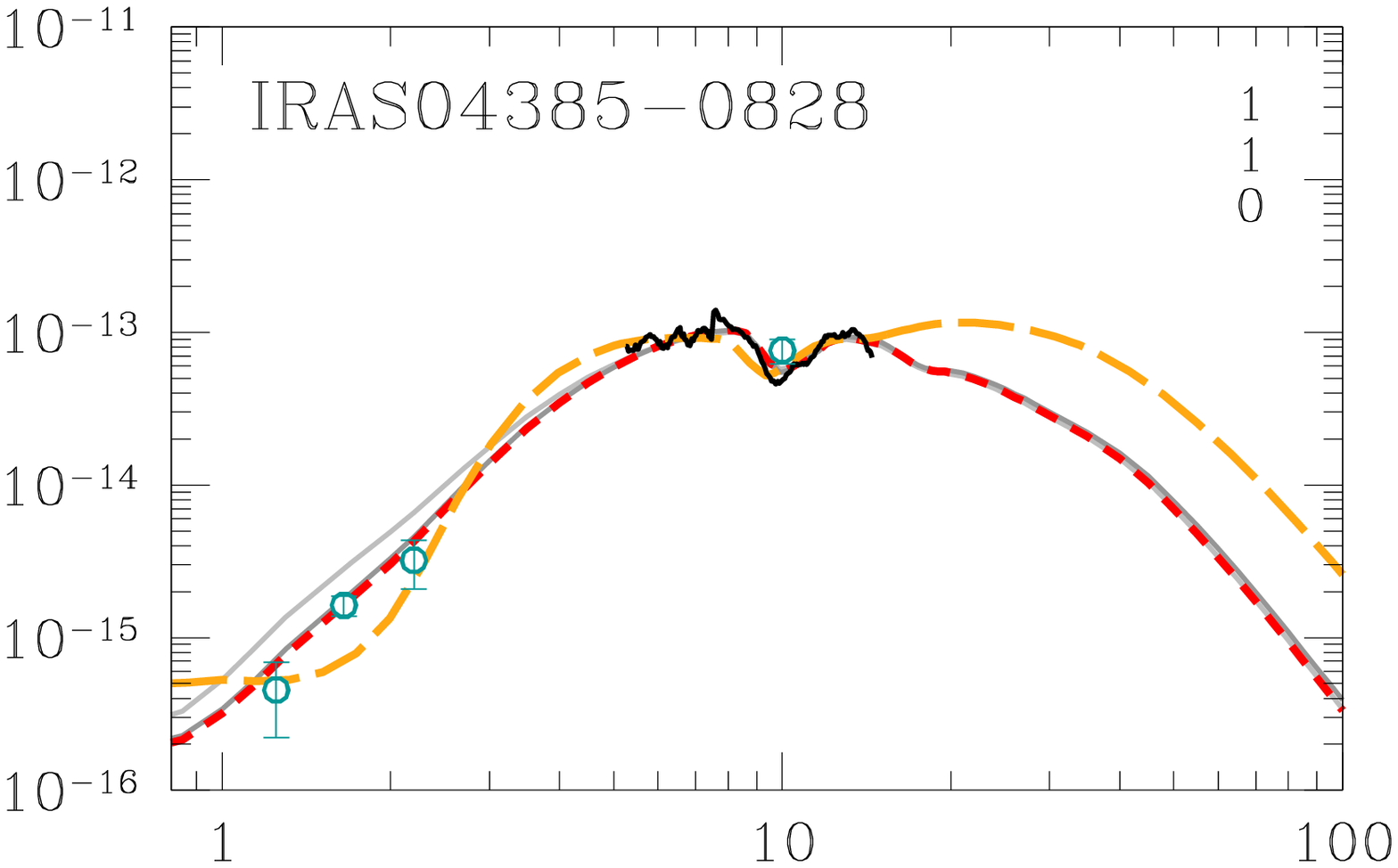}\\
    \includegraphics[scale=0.35,trim=50 100 45 160]{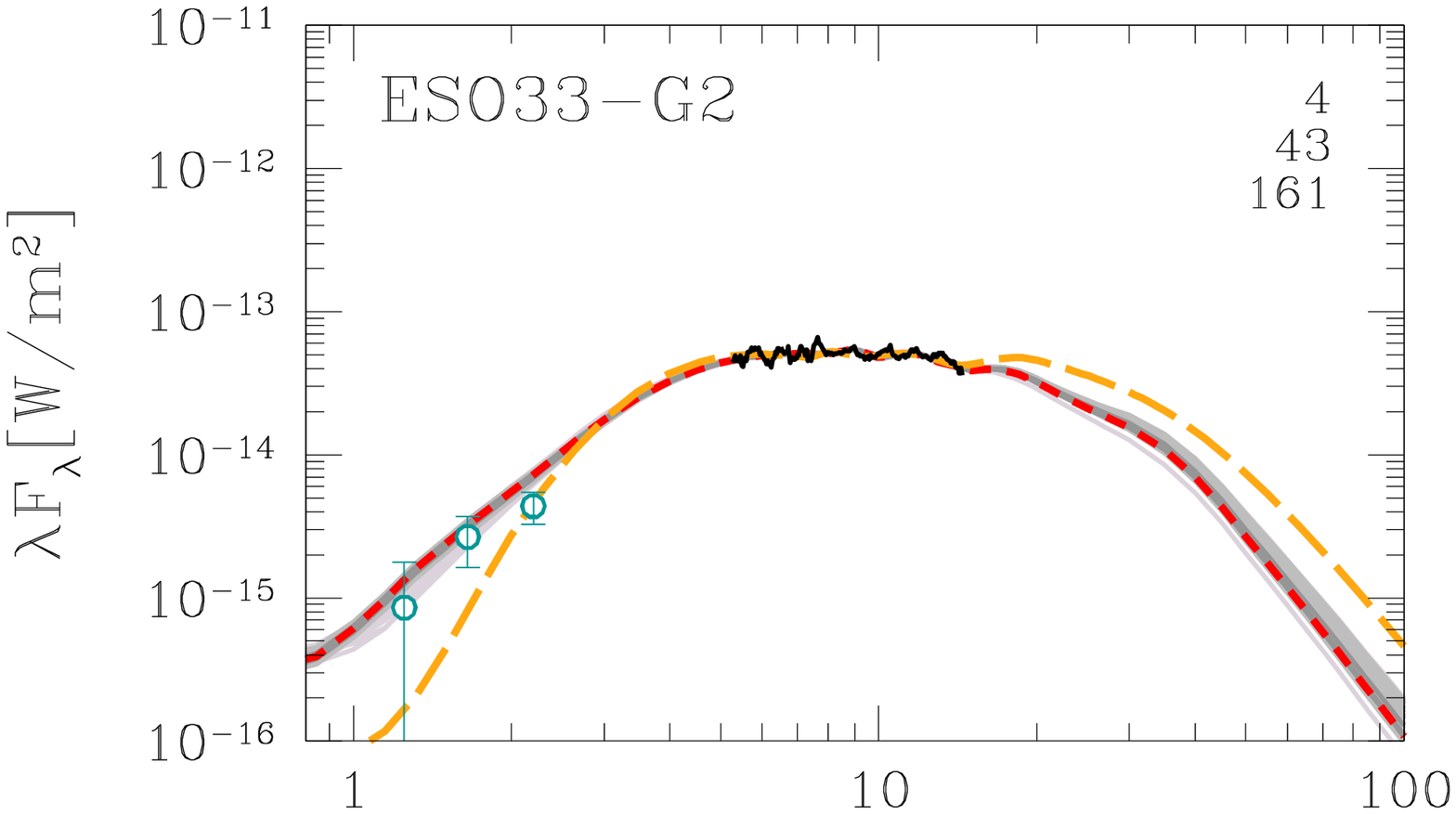}
    \includegraphics[scale=0.35,trim=50 100 45 160]{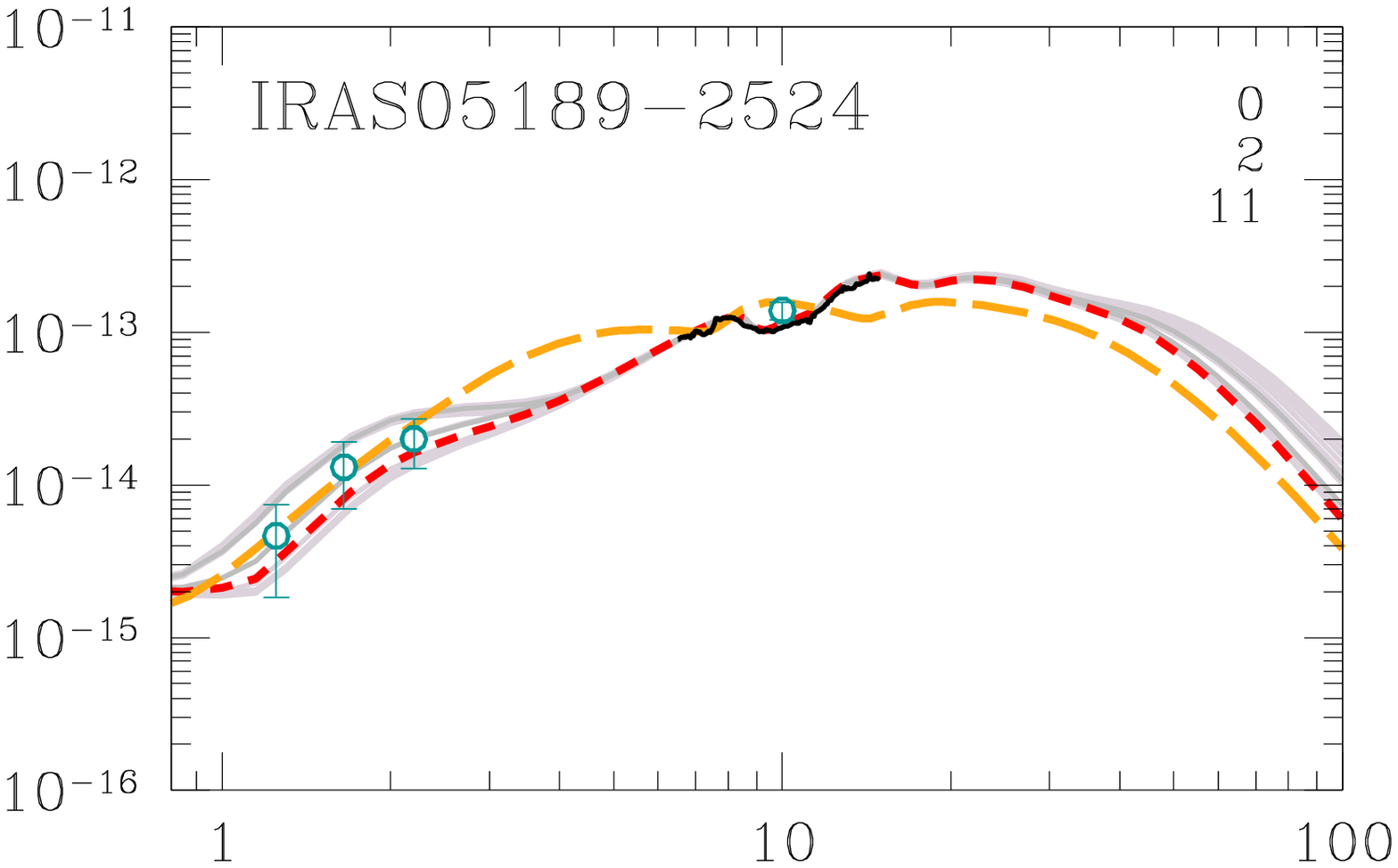}
    \includegraphics[scale=0.35,trim=50 100 45 160]{NGC3660.pow.eps}\\
    \includegraphics[scale=0.35,trim=50  40 45 160]{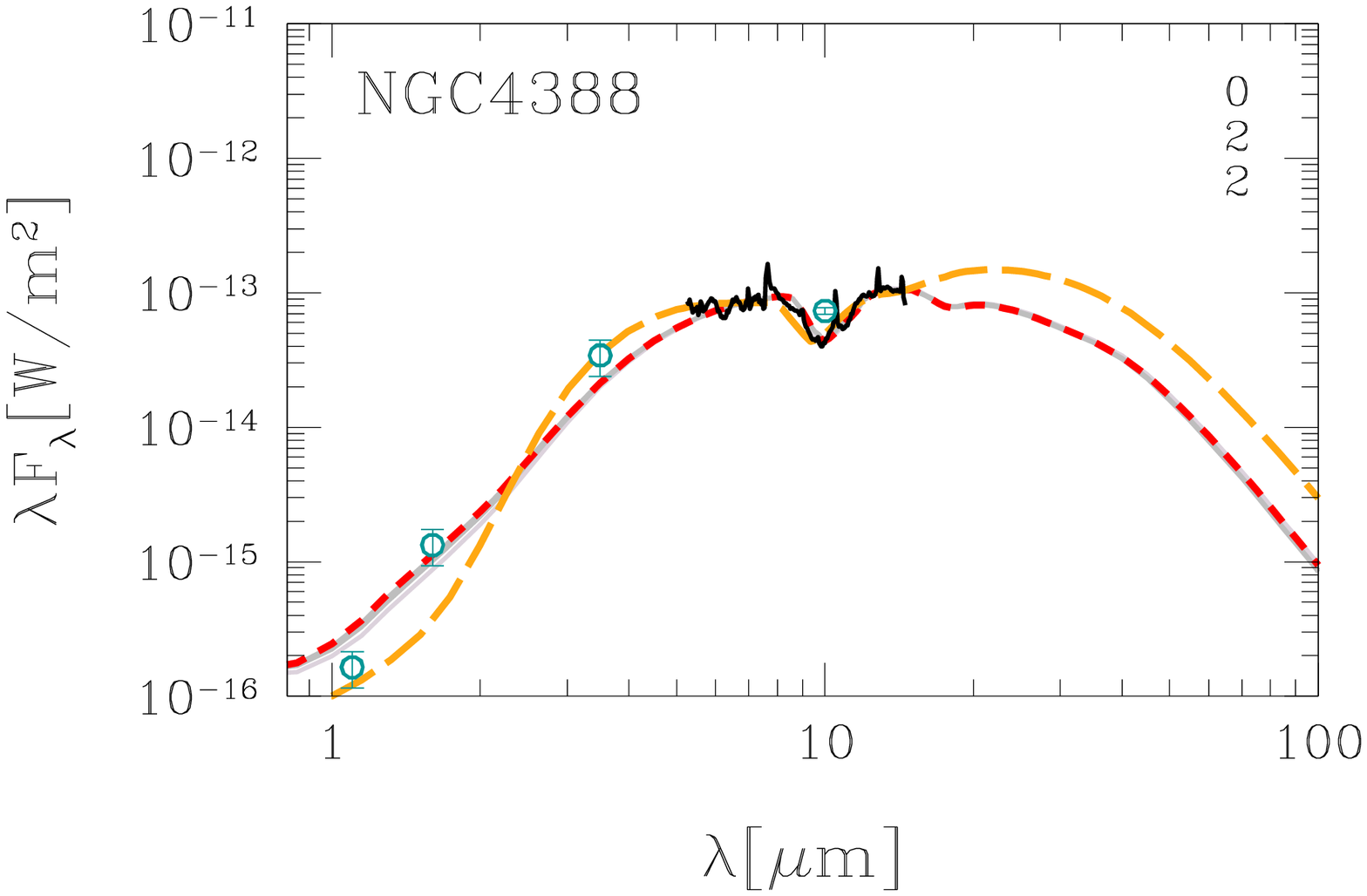}
    \includegraphics[scale=0.35,trim=50  40 45 160]{NGC4501.pow.eps}
    \includegraphics[scale=0.35,trim=50  40 45 160]{TOL1238-364.pow.eps}
%trim=l b r t  	l from the left, b from the bottom, r from the right, and t  from the top. 
  \end{center}
  \figcaption[SEDs]{Spectral Energy Distribution model fits determined for our
    sample. Photometric points are shown with circles and upper limits with
    arrows. Mid-IR spectroscopic observations are shown with a thin black line
    for those data ranges used during the fitting and with a thin dashed black
    line for ranges that were masked out. When the spectroscopic observations
    were instead binned into "photometric'' points, these are shown with
    triangles. The best-fit 2pC model is shown with a thick long-dashed
    line. The best-fit CLUMPY model is shown with a thick dashed line. Up to
    20 CLUMPY models in the top$-50\%$, $50-10\%$ and $10-1$\% from the best
    fit probability value are shown using dark-gray, medium-gray and
    light-gray thin-continuous lines (see text for further details). The
    number of actual models in each of these probability ranges are shown in
    the top-right corner of each panel.\label{seds}}
\end{figure*}

\begin{figure*}
    \includegraphics[scale=0.35,trim=50 100 45 160]{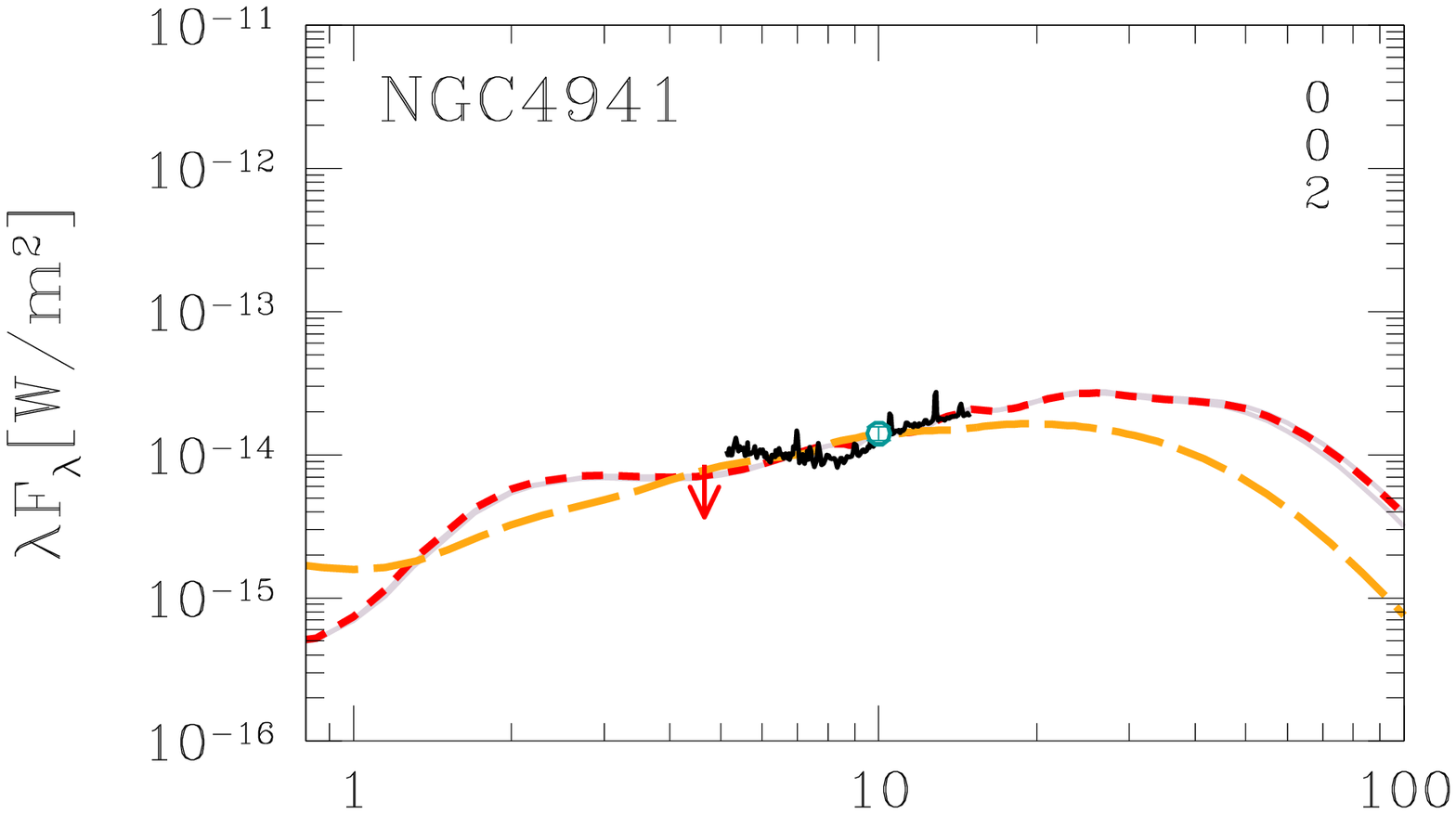}
    \includegraphics[scale=0.35,trim=50 100 45 160]{NGC4968.tor.eps}
    \includegraphics[scale=0.35,trim=50 100 45 160]{MCG-3-34-64.tor.eps}\\
    \includegraphics[scale=0.35,trim=50 100 45 160]{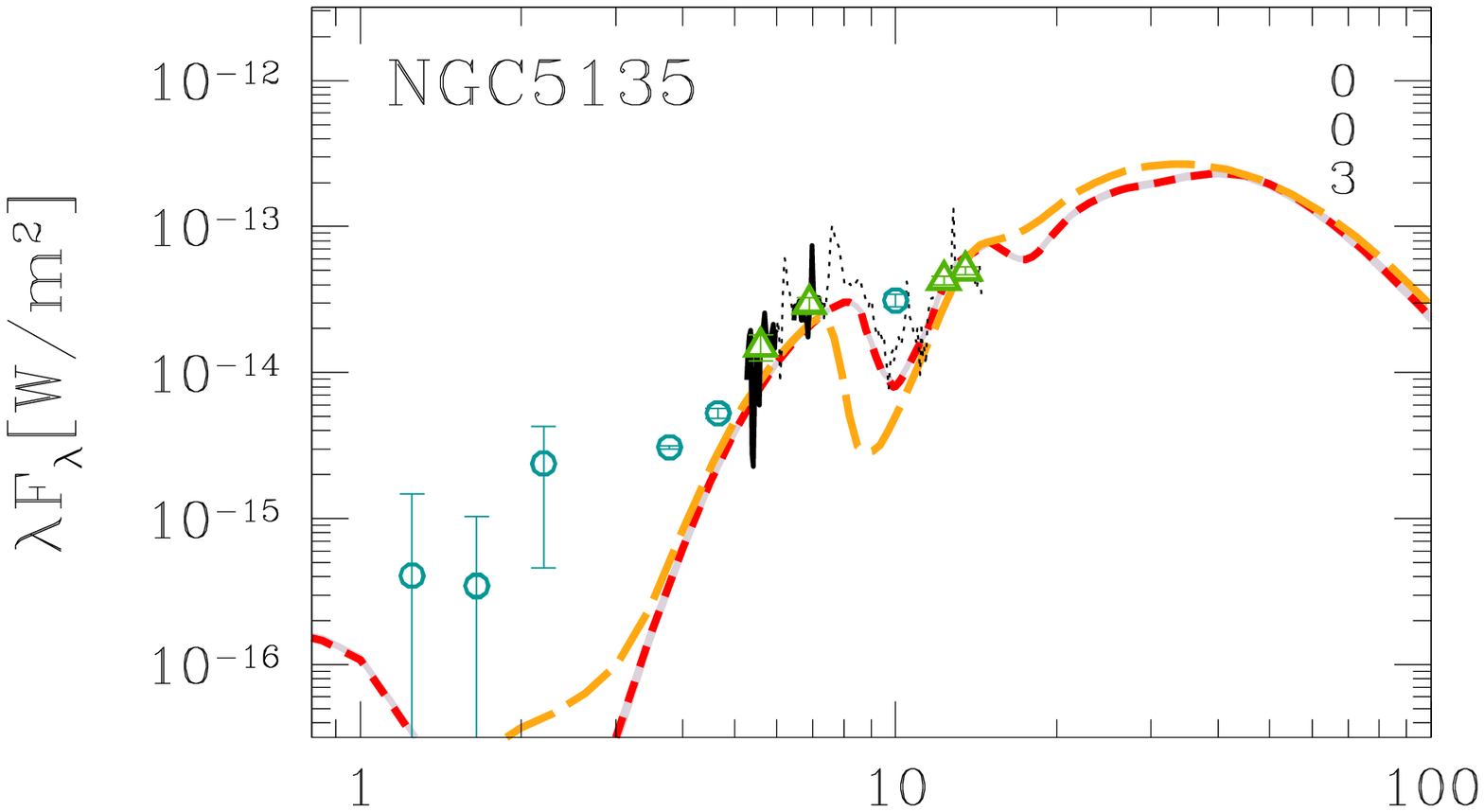}
    \includegraphics[scale=0.35,trim=50 100 45 160]{NGC5506.tor.eps}
    \includegraphics[scale=0.35,trim=50 100 45 160]{NGC5953.pow.eps}\\
    \includegraphics[scale=0.35,trim=50 100 45 160]{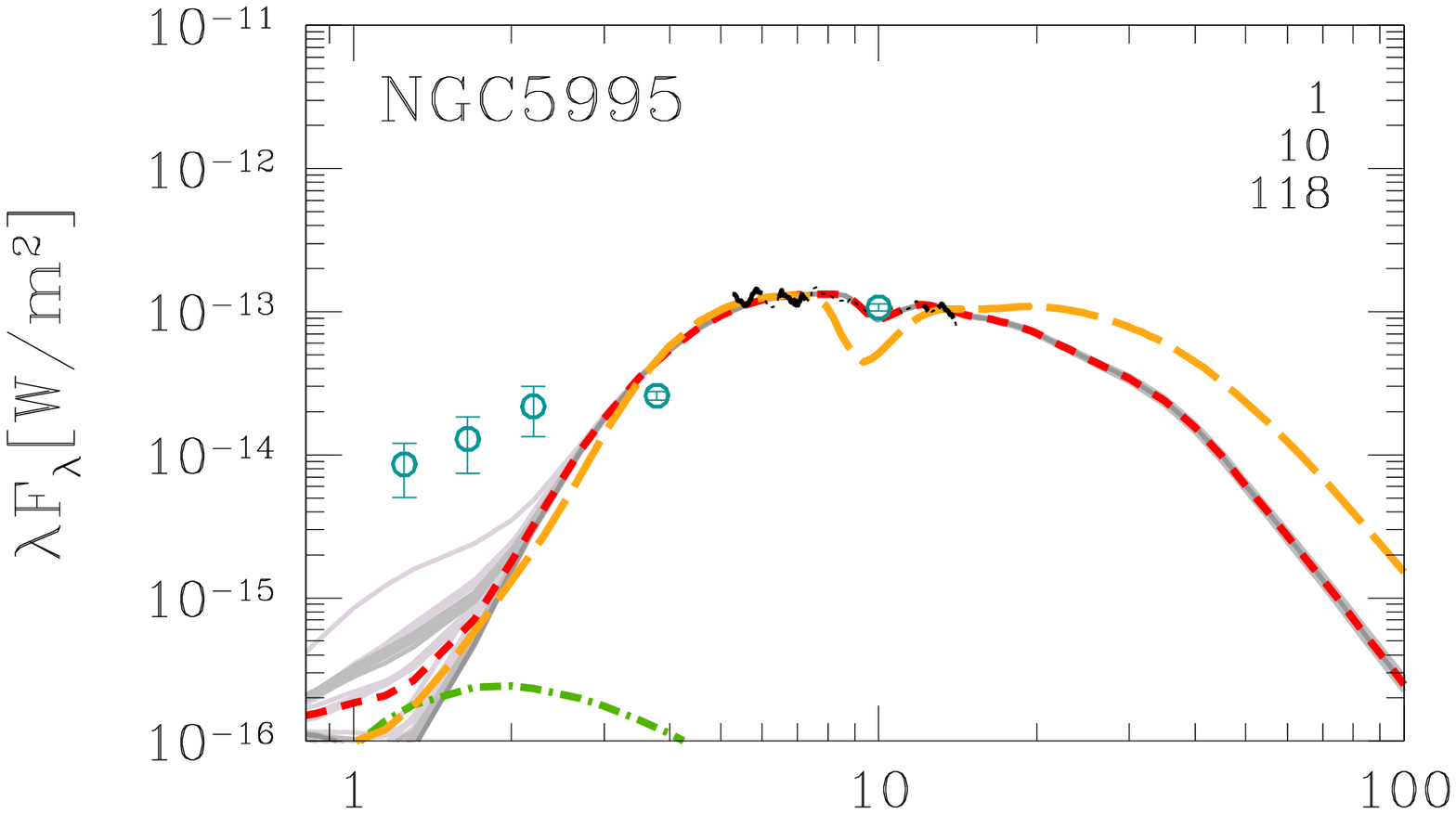}
    \includegraphics[scale=0.35,trim=50 100 45 160]{IRAS15480-0344.pow.eps}
    \includegraphics[scale=0.35,trim=50 100 45 160]{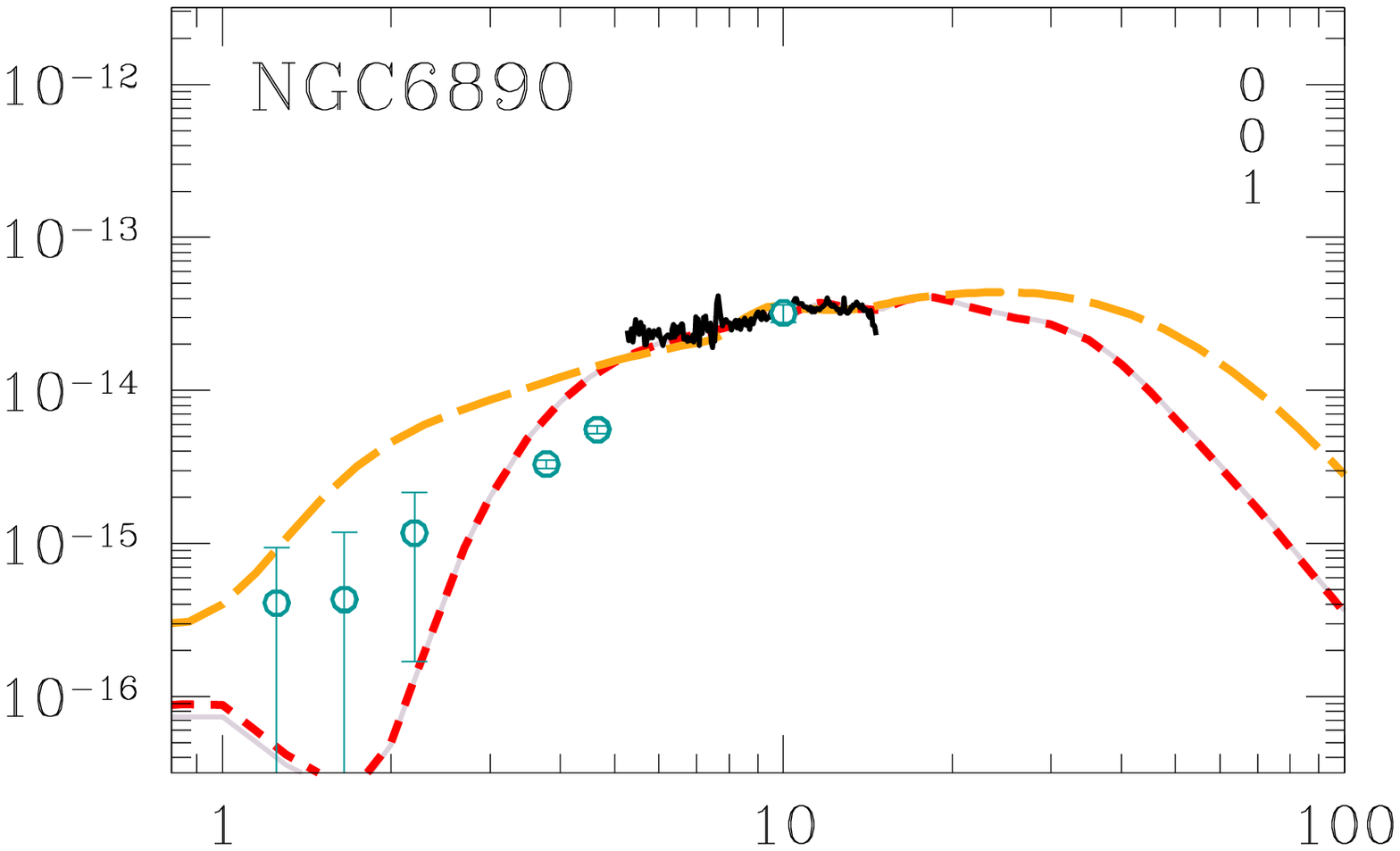}\\
    \includegraphics[scale=0.35,trim=50 100 45 160]{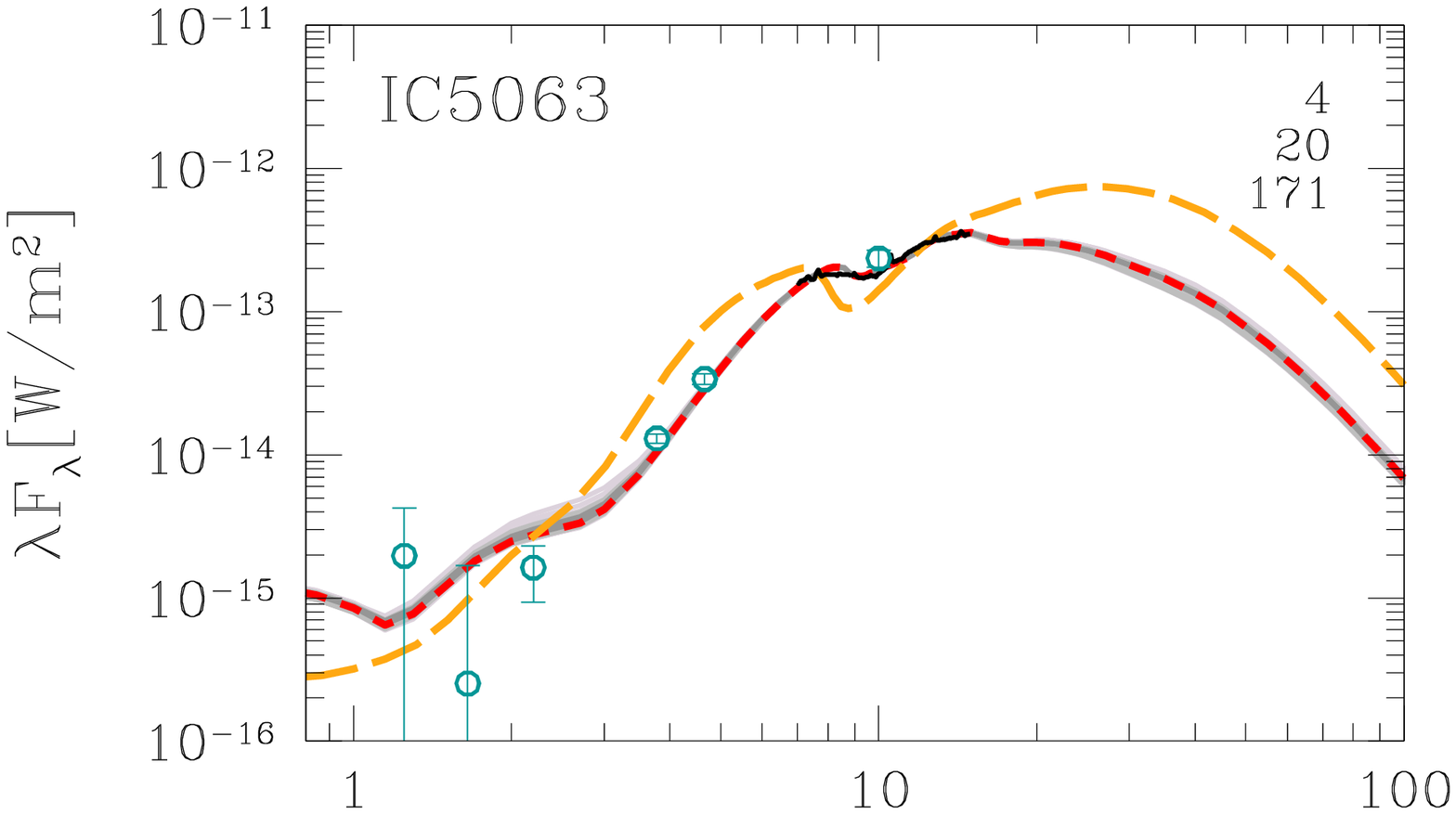}
    \includegraphics[scale=0.35,trim=50 100 45 160]{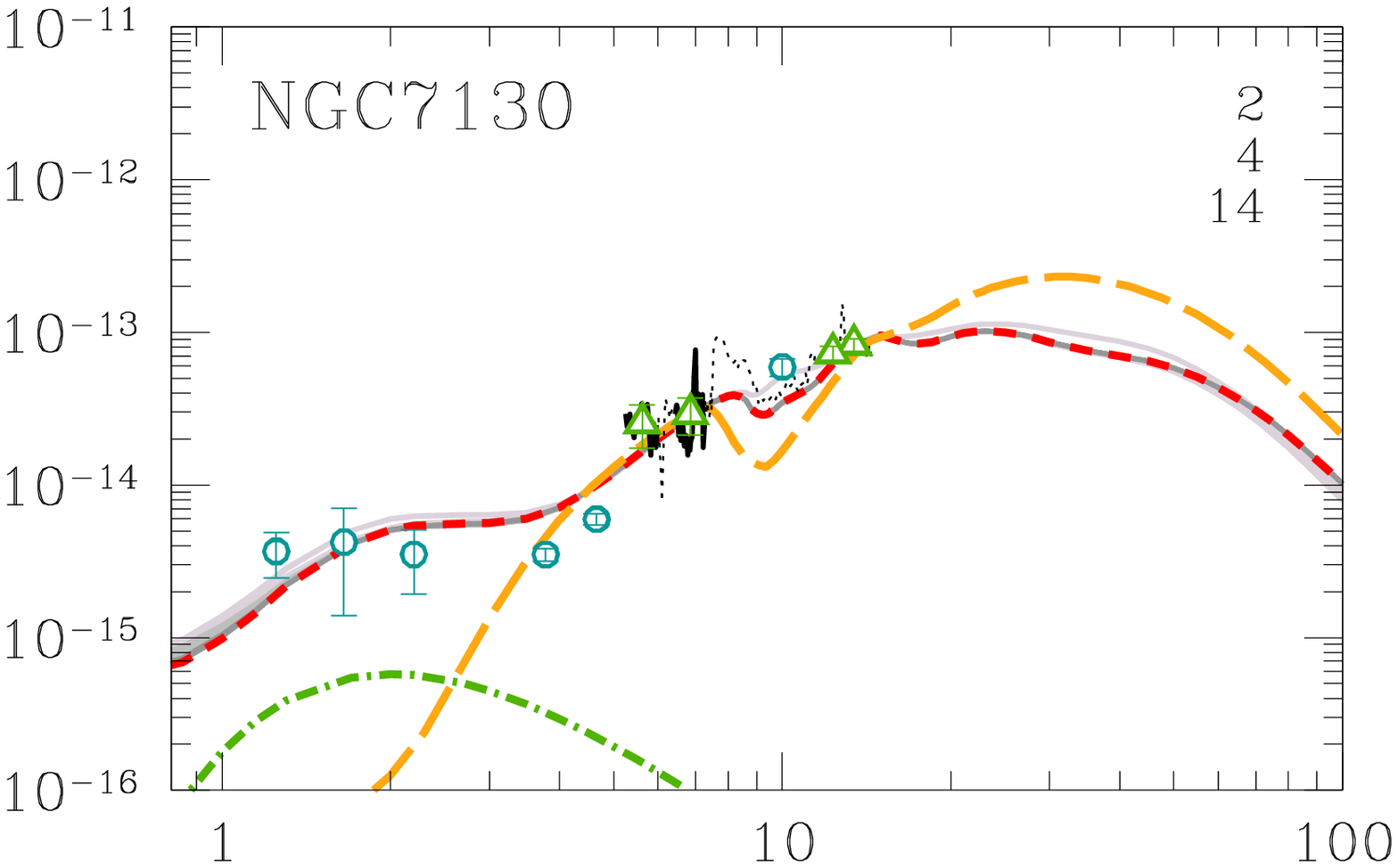}
    \includegraphics[scale=0.35,trim=50 100 45 160]{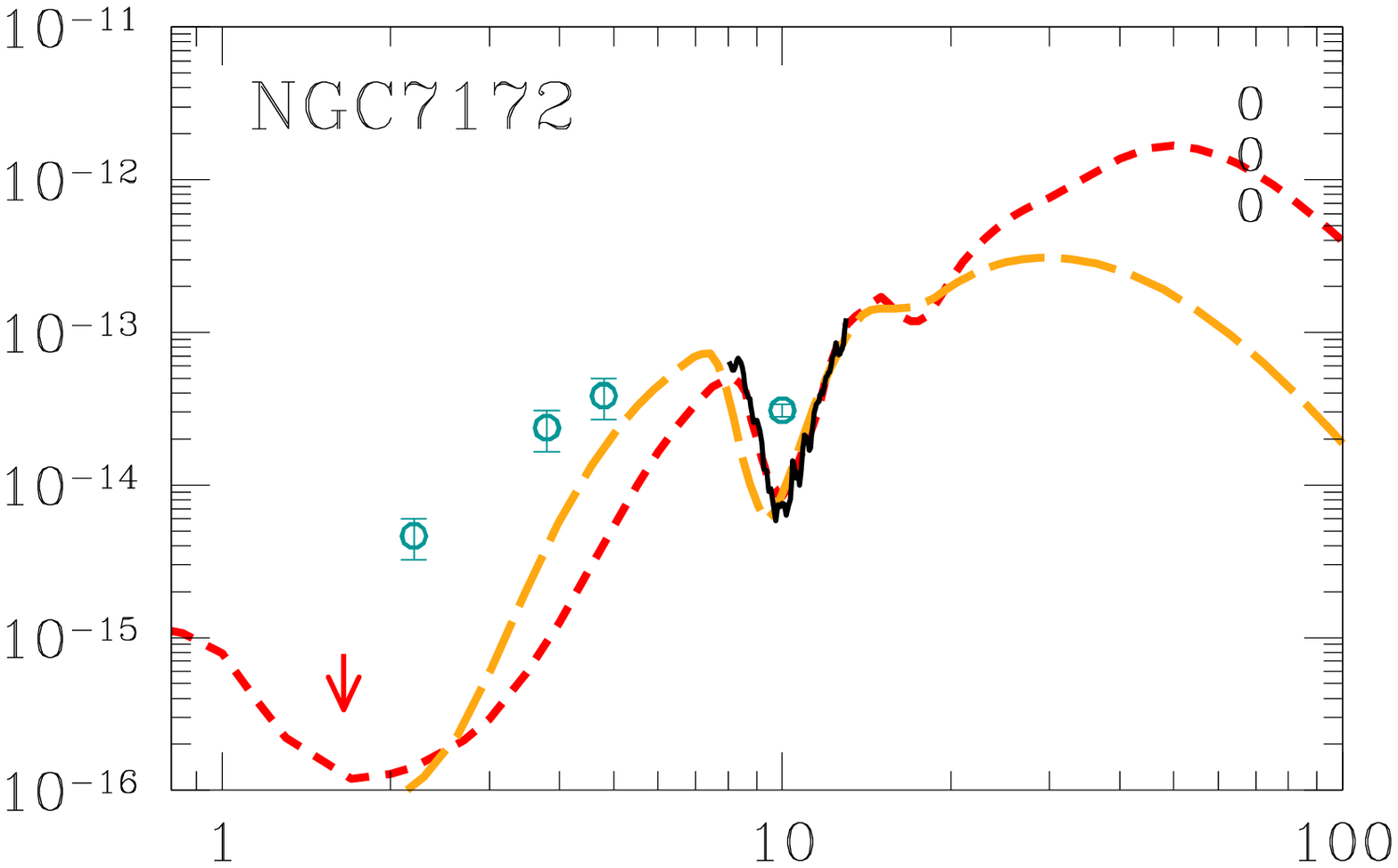}\\
    \includegraphics[scale=0.35,trim=50 100 45 160]{MCG-3-58-7.tor.eps}
    \includegraphics[scale=0.35,trim=50 100 45 160]{NGC7496.pow.eps}
    \includegraphics[scale=0.35,trim=50 100 45 160]{NGC7582.tor.eps}\\
    \includegraphics[scale=0.35,trim=50  40 45 160]{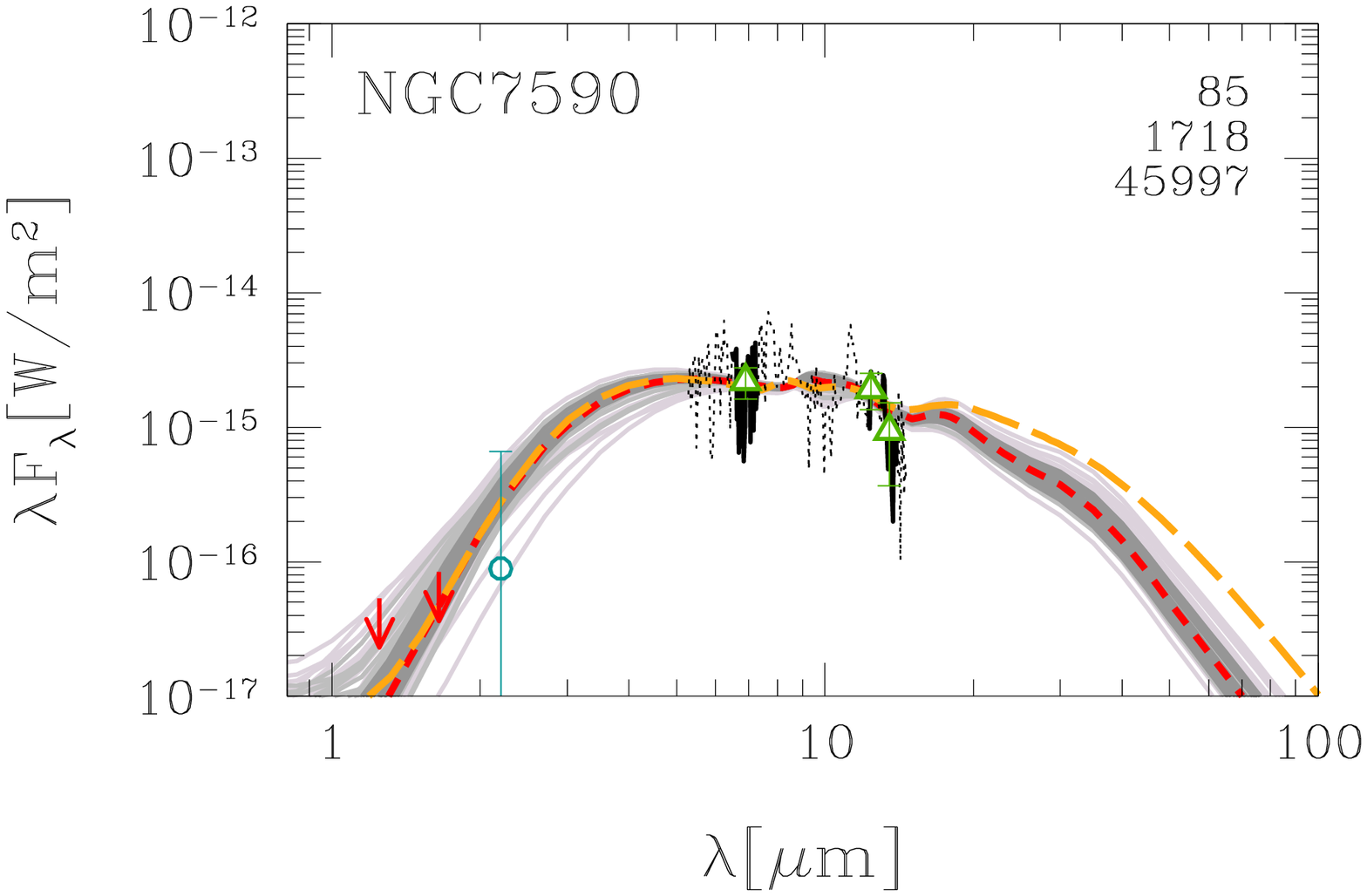}
    \includegraphics[scale=0.35,trim=50  40 45 160]{NGC7674.tor.eps}
    \includegraphics[scale=0.35,trim=50  40 45 160]{CGCG381-051.pow.eps}\\
  \figurenum{4}
  \figcaption[SEDs]{Continuation.}
\end{figure*}

\begin{turnpage}
\begin{deluxetable*}{l|rrrr|rrrr|rrrr|rrrr}
\tablecolumns{17}
\tablecaption{CLUMPY Model results \label{resultsc}}
\tablehead{
\colhead{Object} &
\multicolumn{4}{c|}{$\tau$} & 
\multicolumn{4}{c|}{$Y$} &		
\multicolumn{4}{c|}{$\angle i$} &	
\multicolumn{4}{c|}{$q$}\\
& \colhead{$mean$} & \colhead{$mod$} & \colhead{$med$} & \colhead{$67\%CL$} 
& \colhead{$mean$} & \colhead{$mod$} & \colhead{$med$} & \colhead{$67\%CL$} 
& \colhead{$mean$} & \colhead{$mod$} & \colhead{$med$} & \colhead{$67\%CL$} 
& \colhead{$mean$} & \colhead{$mod$} & \colhead{$med$} & \colhead{$67\%CL$} }
\startdata
%Object			tau_avg	  tau_mod tau_med	    Y_avg Y_mod	 Y_med		  i_avg	  i_mod	 i_med		  q_avg	  q_mod	 q_med	        
NGC\,1068\,$\wr$	&  44	&  30	&  34	&  22-58  &  20	&  20	&  20	&  12-22 &  81	&  80	&  80	&  75-85 &   0	&   0	&   0	&   0-0\\
NGC\,1144 BB		&  34	&  20	&  28	&   8-38  &  56	&  30	&  54	&   8-72 &  66	&  90	&  71	&  55-95 &   1	&   1	&   1	&   0-2\\
NGC\,1144		&  56	&  60	&  59	&  42-68  &  56	&  30	&  55	&  12-78 &  34	&   0	&  29	&   0-45 &   1	&   1	&   1	&   0-2\\
MCG\,-2-8-39 BB		& 114	& 100	& 112	&  98-122 &   5	&   5	&   5	&   2-8	 &  37	&  40	&  40	&  35-45 &   3	&   3	&   3	&   2-3\\
MCG\,-2-8-39		& 114	& 100	& 112	&  98-122 &  56	&  30	&  56	&   8-72 &  27	&  20	&  24	&  15-35 &   3	&   3	&   3	&   2-3\\
NGC\,1194		&  12	&  10	&  11	&   8-18  &   5	&   5	&   5	&   2-8	 &  40	&  40	&  40	&  35-45 &   0	&   0	&   0	&   0-0\\
NGC\,1320\,$\wr$	&  12	&  10	&  11	&   8-18  &  76	&  90	&  76	&  58-92 &  17	&  20	&  19	&  15-35 &   2	&   2	&   2	&   1-2\\
IRAS\,04385-0828	&  20	&  20	&  20	&  12-22  &   5	&   5	&   5	&   2-8	 &   0	&   0	&   0	&   0-5	 &   1	&   1	&   1	&   0-1\\
ESO\,33-G2		&  20	&  20	&  20	&  12-22  &  54	&  20	&  52	&   8-72 &  32	&  40	&  38	&  35-65 &   2	&   2	&   2	&   2-2\\
IRAS\,05189-2524	& 108	& 100	& 108	&  98-132 &  12	&  10	&  11	&   8-18 &  18	&  10	&  11	&   5-15 &   0	&   0	&   0	&   0-0\\
NGC\,3660 BB		&  91	&  60	&  86	&  58-148 &  48	&   5	&  48	&   2-68 &  56	&  90	&  62	&  35-95 &   2	&   3	&   3	&   2-3\\
NGC\,3660		& 149	& 150	& 149	& 122-150 &  56	&  90	&  56	&   8-78 &  29	&  30	&  31	&  25-45 &   2	&   2	&   2	&   2-3\\
NGC\,4388		&  20	&  20	&  20	&  12-22  &  12	&  10	&  11	&   8-18 &   3	&   0	&   2	&   0- 5 &   0	&   0	&   0	&   0-0\\
TOL\,1238-364\,$\wr$ BB	& 150	& 150	& 150	& 122-150 &  20	&  20	&  20	&  12-22 &  71	&  70	&  70	&  65-75 &   0	&   0	&   0	&   0-0\\
TOL\,1238-364		& 150	& 150	& 150	& 122-150 &  20	&  20	&  20	&  12-22 &  60	&  60	&  60	&  55-65 &   0	&   0	&   0	&   0-0\\
NGC\,4968\,$\wr$ BB	&   5	&   5	&   5	&   2-8	  &  91	&  90	&  91	&  88-98 &  90	&  90	&  90	&  75-95 &   0	&   0	&   0	&   0-0\\
NGC\,4968\,$\wr$	&   5	&   5	&   5	&   2-8	  &  90	&  90	&  90	&  82-92 &  88	&  90	&  89	&  75-95 &   0	&   0	&   0	&   0-0\\
MCG\,-3-34-64 BB	& 150	& 150	& 150	& 122-150 &  12	&  10	&  11	&   8-18 &  83	&  90	&  85	&  75-95 &   0	&   0	&   0	&   0-1\\
MCG\,-3-34-64		& 150	& 150	& 150	& 122-150 &  14	&  10	&  12	&   8-18 &  77	&  80	&  78	&  55-85 &   0	&   0	&   0	&   0-1\\
NGC\,5135\,$\wr$	&  45	&  30	&  34	&  22-58  &  50	&  50	&  50	&  42-52 &  81	&  80	&  80	&  75-85 &   0	&   0	&   0	&   0-0\\
NGC\,5506 BB		&  52	&  40	&  48	&  38-62  &  15	&  10	&  14	&   8-18 &  14	&  20	&  16	&   0-25 &   1	&   1	&   1	&   0-1\\
NGC\,5506		&  60	&  60	&  61	&  58-68  &  20	&  20	&  20	&  12-22 &   8	&  10	&   8	&   5-15 &   0	&   0	&   0	&   0-1\\
NGC\,5995 BB		&  12	&  10	&  11	&   8-18  &  54	&  90	&  45	&  12-78 &  72	&  70	&  71	&  65-75 &   2	&   2	&   2	&   2-3\\
NGC\,5995		&  12	&  10	&  11	&   8-18  &  62	&  90	&  68	&  32-92 &  86	&  90	&  87	&  75-95 &   2	&   2	&   2	&   2-3\\
IRAS\,15480-0344\,$\wr$	BB&   9	&  10	&   9	&   2-12  &  74	&  70	&  73	&  62-78 &  90	&  90	&  90	&  75-95 &   0	&   0	&   0	&   0-0\\
IRAS\,15480-0344\,$\wr$	&  45	&  30	&  34	&  22-58  &  90	&  90	&  90	&  82-92 &  90	&  90	&  90	&  75-95 &   0	&   0	&   0	&   0-1\\
NGC\,6890		&   5	&   5	&   5	&   2-8	  &  50	&  50	&  50	&  42-52 &  90	&  90	&  90	&  75-95 &   0	&   0	&   0	&   0-0\\
IC\,5063		&  80	&  80	&  81	&  78-88  &  23	&  20	&  20	&   8-22 &  76	&  80	&  77	&  55-85 &   3	&   3	&   3	&   2-3\\
NGC\,7130\,$\wr$ BB	& 150	& 150	& 150	& 122-150 &  30	&  30	&  30	&  22-32 &  90	&  90	&  90	&  75-95 &   0	&   0	&   0	&   0-0\\
NGC\,7130\,$\wr$	& 150	& 150	& 150	& 122-150 &  30	&  30	&  30	&  22-32 &  80	&  80	&  80	&  75-85 &   0	&   0	&   0	&   0-0\\
NGC\,7172		&  60	&  60	&  61	&  58-68  & 100	& 100	& 100	&  92-102&  40	&  40	&  40	&  35-45 &   0	&   0	&   0	&   0-1\\
MCG\,-3-58-7 BB		&   5	&   5	&   5	&   2-8	  &  20	&  20	&  20	&  12-22 &  90	&  90	&  90	&  75-95 &   0	&   0	&   0	&   0-0\\
MCG\,-3-58-7		&   5	&   5	&   5	&   2-8	  &  20	&  20	&  20	&  12-22 &  90	&  90	&  90	&  75-95 &   0	&   0	&   0	&   0-0\\
NGC\,7496 BB		& 150	& 150	& 150	& 122-150 &  12	&  10	&  11	&   2-12 &   7	&   0	&   5	&   0-15 &   2	&   2	&   2	&   2-3\\
%Object	tau_avg	tau_mod tau_med		  Y_avg	Y_mod	Y_med		 i_avg	i_mod	i_med	         q_avg	q_mod	q_med	        N_avg	N_mod	N_med	            sig_avgsig_modsig_med	  index_avgindex_modindex_med	   Av_avg Av_mod Av_med
NGC\,7496 PL		& 150	& 150	& 150	& 122-150 &  10	&  10	&  10	&   2-12 &  21	&  20	&  22	&  15-35 &   2	&   2	&   2	&   2-3\\
NGC\,7496		& 150	& 150	& 150	& 122-150 &   5	&   5	&   5	&   2-8	 &  25	&  20	&  25	&  15-35 &   0	&   0	&   0	&   0-0\\
NGC\,7582 BB		&  47	&  40	&  43	&  32-48  & 100	& 100	& 100	&  92-102&  30	&  30	&  30	&  25-35 &   0	&   0	&   0	&   0-1\\
NGC\,7582		&  47	&  40	&  43	&  32-48  & 100	& 100	& 100	&  92-102&  30	&  30	&  30	&  25-35 &   0	&   0	&   0	&   0-1\\
NGC\,7590		&   8	&   5	&   7	&   2-12  &  49	&   5	&  47	&   2-68 &  71	&  90	&  79	&  65-95 &   2	&   2	&   2	&   2-3\\
NGC\,7674\,$\wr$	&  45	&  30	&  34	&  22-58  &  12	&  10	&  11	&   8-18 &  70	&  70	&  70	&  65-75 &   0	&   0	&   0	&   0-1\\
CGCG\,381-051 BB	& 145	& 150	& 146	& 118-150 &  39	&   5	&  29	&   2-58 &  15	&   0	&  12	&   0-25 &   3	&   3	&   3	&   2-3\\
CGCG\,381-051		& 148	& 150	& 148	& 112-150 &  47	&   5	&  44	&   2-68 &  10	&   0	&   8	&   0-15 &   3	&   3	&   3	&   2-3\\
\enddata

\tablecomments{Model results from the fitting procedure using CLUMPY
  models. For each parameter the average ($mean$), the mode ($mod$), and the
  median ($med$) with 1-sigma confidence limits (CL) of the probability
  distribution are given. Because of the discrete nature of the parameter
  space, for very narrow probability distributions the CLs can correspond to
  the same parameter value. A $\wr$ indicates Compton Thick (CT)
  sources. Objects that were also modeled using a black-body or power-law
  component are labeled using a BB or PL.}
\end{deluxetable*}
\end{turnpage}

\begin{turnpage}
\begin{deluxetable*}{l|rrrr|rrrr|rrrr|rrr}
\tablenum{3}
\tablecolumns{16}
\tablecaption{CLUMPY Model results \label{resultsc}}
\tablehead{
\colhead{Object} &
\multicolumn{4}{c|}{$\mathcal{N}_0$} &
\multicolumn{4}{c|}{$\sigma$ (degs)} & 
\multicolumn{4}{c|}{T(K)/$\alpha$ \dag} &
\multicolumn{3}{c}{$\log A_V$(los) \ddag} \\
& \colhead{$mean$} & \colhead{$mod$} & \colhead{$med$} & \colhead{$67\%CL$} 
& \colhead{$mean$} & \colhead{$mod$} & \colhead{$med$} & \colhead{$67\%CL$} 
& \colhead{$mean$} & \colhead{$mod$} & \colhead{$med$} & \colhead{$67\%CL$} 
& \colhead{$mean$} & \colhead{$mod$} & \colhead{$med$} }
\startdata
%Object			   N_avg  N_mod	 N_med	         sig_avg  sig_mod sig_med	 Temp_avgTemp_modTemp_med    	     Av_avg Av_mod Av_med 	
NGC\,1068\,$\wr$	&    4	&   4	&   4	&   4-4	 &  27	&  30	&  28	&  22-38 & ---	& ---	& ---	& ---	   & 2.3 & 2.4	& 2.1\\
NGC\,1144 BB		&   11	&  15	&  11	&   6-14 &  49	&  60	&  51	&  38-68 &2264	&2500	&2300	&2150-2550 & 2.5 & 2.7	& 2.5\\
NGC\,1144		&   10	&  12	&  11	&   6-14 &  50	&  60	&  52	&  38-68 & --- 	& --- 	& --- 	& ---      & 2.2 & 2.8	& 2.2\\
MCG\,-2-8-39 BB		&    8	&   8	&   8	&   8-8	 &  60	&  60	&  60	&  52-68 &1818	&1800	&1803	&1750-1850 & 2.7 & 2.9	& 2.7\\
MCG\,-2-8-39		&    7	&   7	&   7	&   6-8	 &  60	&  60	&  60	&  52-68 & --- 	& --- 	& --- 	& ---      & 2.4 & 2.7	& 2.4\\
NGC\,1194		&   15	&  15	&  15	&  14-16 &  60	&  60	&  60	&  52-68 & ---	& ---	& ---	& ---	   & 2.0 & 2.3	& 2.0\\
NGC\,1320\,$\wr$	&   14	&  15	&  15	&  12-16 &  49	&  45	&  48	&  38-52 & ---	& ---	& ---	& ---	   & 1.3 & 2.0	& 1.3\\
IRAS\,04385-0828	&   15	&  15	&  15	&  14-16 &  45	&  45	&  45	&  38-52 & ---	& ---	& ---	& ---	   & 0.8 & 1.4	& 0.8\\
ESO\,33-G2		&    5	&   4	&   5	&   4-6	 &  53	&  60	&  54	&  38-68 & ---	& ---	& ---	& ---	   & 1.5 & 2.1	& 1.6\\
IRAS\,05189-2524	&    6	&   6	&   6	&   6-6	 &  60	&  60	&  60	&  52-68 & ---	& ---	& ---	& ---	   & 2.2 & 2.4	& 2.1\\
NGC\,3660 BB		&   11	&  10	&  11	&   6-12 &  48	&  60	&  54	&  38-68 &2441	&2500	&2462	&2350-2550 & 2.8 & 3.3	& 2.9\\
NGC\,3660		&   14	&  15	&  14	&  12-15 &  31	&  30	&  30	&  22-38 & --- 	& --- 	& --- 	& ---      & 1.6 & 2.8	& 1.7\\
NGC\,4388		&   11	&  11	&  11	&  10-12 &  60	&  60	&  60	&  52-68 & ---	& ---	& ---	& ---	   & 1.5 & 1.8	& 1.5\\
TOL\,1238-364\,$\wr$ BB	&    8	&   8	&   8	&   6-8	 &  15	&  15	&  15	&   8-22 &2042	&2000	&2002	&1750-2150 & 2.4 & 2.9	& 2.4\\
TOL\,1238-364		&   13	&  13	&  13	&  12-14 &  15	&  15	&  15	&   8-22 & --- 	& --- 	& --- 	& ---      & 1.6 & 2.8	& 1.6\\
NGC\,4968\,$\wr$ BB	&   15	&  15	&  15	&  14-15 &  60	&  60	&  60	&  52-68 &1214	&1200	&1203	&1050-1350 & 1.9 & 2.1	& 1.9\\
NGC\,4968\,$\wr$	&   15	&  15	&  15	&  14-15 &  60	&  60	&  60	&  52-68 & --- 	& --- 	& --- 	& ---      & 1.9 & 2.1	& 1.9\\
MCG\,-3-34-64 BB	&   14	&  15	&  14	&  14-15 &  46	&  45	&  46	&  38-52 &2293	&2500	&2336	&2150-2550 & 3.3 & 3.4	& 3.4\\
MCG\,-3-34-64		&   13	&  14	&  14	&  12-15 &  49	&  45	&  48	&  38-52 & --- 	& --- 	& --- 	& ---      & 3.3 & 3.4	& 3.3\\
NGC\,5135\,$\wr$	&   13	&  13	&  13	&  12-14 &  60	&  60	&  60	&  52-68 & ---	& ---	& ---	& ---	   & 2.8 & 2.9	& 2.7\\
NGC\,5506 BB		&   14	&  15	&  14	&  12-15 &  60	&  60	&  60	&  52-68 &1244	&1200	&1236	&1150-1350 & 2.2 & 2.6	& 2.2\\
NGC\,5506		&   15	&  15	&  15	&  14-15 &  45	&  45	&  45	&  38-52 & --- 	& --- 	& --- 	& ---      & 1.5 & 2.2	& 1.6\\
NGC\,5995 BB		&   14	&  14	&  14	&  14-14 &  59	&  60	&  60	&  52-68 &1564	&1900	&1817	&1050-1950 & 2.2 & 2.4	& 2.2\\
NGC\,5995		&   12	&  12	&  12	&  12-12 &  60	&  60	&  60	&  52-68 & --- 	& --- 	& --- 	& ---      & 2.2 & 2.4	& 2.2\\
IRAS\,15480-0344\,$\wr$	BB&  7	&   5	&   5	&   4-12 &  15	&  15	&  15	&   8-22 &2415	&2500	&2428	&2350-2550 & 1.9 & 2.2	& 1.7\\
IRAS\,15480-0344\,$\wr$	&    2	&   2	&   2	&   2-2	 &  16	&  15	&  16	&   8-22 & --- 	& --- 	& --- 	& ---      & 2.0 & 2.2	& 1.9\\
NGC\,6890		&   15	&  15	&  15	&  14-16 &  15	&  15	&  15	&   8-22 & ---	& ---	& ---	& ---	   & 1.9 & 2.1	& 1.9\\
IC\,5063		&   14	&  13	&  14	&  12-14 &  60	&  60	&  60	&  52-68 & ---	& ---	& ---	& ---	   & 3.0 & 3.1	& 3.1\\
NGC\,7130\,$\wr$ BB	&    8	&   8	&   8	&   8-8	 &  15	&  15	&  15	&   8-22 &2176	&2500	&2191	&1750-2350 & 3.1 & 3.1	& 3.1\\
NGC\,7130\,$\wr$	&   11	&  11	&  11	&  10-12 &  15	&  15	&  15	&   8-22 & --- 	& --- 	& --- 	& ---      & 3.1 & 3.3	& 3.1\\
NGC\,7172		&   15	&  15	&  15	&  14-15 &  60	&  60	&  60	&  52-68 & ---	& ---	& ---	& ---      & 2.7 & 2.9	& 2.7\\
MCG\,-3-58-7 BB		&   15	&  15	&  15	&  14-15 &  21	&  15	&  21	&   8-38 &2150	&2000	&2043	&1850-2250 & 1.9 & 2.1	& 1.9\\
MCG\,-3-58-7		&   15	&  15	&  15	&  14-15 &  15	&  15	&  15	&   8-22 & --- 	& --- 	& --- 	& ---      & 1.9 & 2.1	& 1.9\\
NGC\,7496 BB		&   15	&  15	&  15	&  14-15 &  60	&  60	&  60	&  52-68 &1865	&1800	&1835	&1750-1950 & 2.5 & 2.9	& 2.5\\
NGC\,7496 PL		&   14	&  15	&  15	&  14-16 &  60	&  60	&  60	&  52-68 &0.39	&0.39	&0.39	&0.44-0.34 & 2.8 & 3.1	& 2.8\\
NGC\,7496		&   14	&  15	&  14	&  12-15 &  45	&  45	&  45	&  38-52 & --- 	& --- 	& --- 	& ---      & 2.4 & 2.9	& 2.5\\
NGC\,7582 BB		&   15	&  15	&  15	&  14-15 &  60	&  60	&  60	&  52-68 &1696	&1700	&1698	&1650-1750 & 2.4 & 2.6	& 2.4\\
NGC\,7582		&   15	&  15	&  15	&  14-15 &  60	&  60	&  60	&  52-68 & --- 	& --- 	& --- 	& ---      & 2.4 & 2.6	& 2.4\\
NGC\,7590		&   11	&  13	&  11	&   8-14 &  40	&  60	&  42	&  22-68 & ---	& ---	& ---	& ---	   & 1.9 & 2.3	& 1.9\\
NGC\,7674\,$\wr$	&    5	&   5	&   5	&   4-6	 &  15	&  15	&  15	&   8-22 & ---	& ---	& ---	& ---	   & 1.6 & 2.3	& 1.5\\
CGCG\,381-051 BB	&   12	&  13	&  12	&  10-15 &  53	&  60	&  54	&  38-68 &2283	&2500	&2320	&2150-2550 & 2.4 & 3.0	& 2.4\\
CGCG\,381-051		&   12	&  15	&  13	&  10-15 &  35	&  30	&  33	&  22-38 & --- 	& --- 	& --- 	& ---      & 1.0 & 1.7	& 0.7\\
\enddata 

\tablecomments{Continuation from previous page. \dag\ Temperature of the
  black-body (BB) or index of the power-law (PL) secondary
  component. \ddag\ The value of $A_V$ along the line-of-sight are calculated
  following the equation found in text. Therefore no associated error
  estimates are give.}
\end{deluxetable*}
\end{turnpage}

\subsubsection{Fitting Sources with Strong Silicate Absorption}

A few of our sources present significant silicate absorption at 9.7 $\mu$m
while showing strong near-IR emission, this is, showing a 'broad' SED. The
most extreme silicate absorption features are seen in NGC\,7172 and NGC\,7582,
while NGC\,1194, IRAS\,04385-0828, NGC\,4388 and NGC\,5506 correspond to more
moderate cases.

Using $S_{10}$ to quantity the depth of the silicate feature ($S_{10} = \ln
F_{\lambda}/F_{c,\lambda}$, where $F_{\lambda}$ corresponds to the flux at the
deepest point of the absorption feature and $F_{c,\lambda}$ corresponds to the
flux in the interpolated continuum) we find that NGC\,7172 and NGC\,7582 have
$S_{10} \sim -2.0$ and $\sim -1.4$, respectively, while the other 4 sources
listed above have $S_{10}$ values between $\sim -0.7$ and $-1$.

We find that it is not possible to properly fit NGC\,5506, NGC\,7172 and
NGC\,7582 with the set of current CLUMPY models because a combination of deep
silicate absorption and SEDs with strong near-IR emission is not
available. Even though NGC\,5506 has a moderate absorption feature it has a
particularly broad SED and therefore the obtained fits are poor.

This is because CLUMPY models can only produce $S_{10} \ga -1$ for
intermediate optical depths, large number of clouds and edge-on lines of
sight. At the same time, a large number of clouds yields models with a
pronounced decrease in flux below 20$\mu$m, as can be seen in Fig.~6 of
Nenkova et al.~(2008b), therefore producing 'narrow' SEDs. This issue is seen
in the fits performed by AH11 to NGC\,5506, NGC\,7172 and NGC\,7582 (notice
that we share common photometry with AH11 for NGC\,5506 and NGC\,7172). From
their Fig.~6 it is clear how the fits to NGC\,5506 and NGC\,7582 become much
steeper in the near-IR once the spectral information from the silicate
absorption is taken into account. Only Circinus in Fig.~5 of AH11 presents a
combination of SED shape and silicate absorption strength compatible with the
models.

On the other hand, Stalevski et al.~(2012) show that their 2pC models can have
a value of $S_{10}$ as low as $-2.4$. Crucially, their diffuse torus component
contributes more significantly to the near-IR spectral range than the CLUMPY
models, producing 'broader' SEDs. However, as noted by Stalevski et al.~(2012)
this effect is more substantial for face-on orientations, while we expect that
a large fraction of our sources prefer edge-on orientations.

Levenson et al.\ (2007) have argued that a deep silicate absorption feature
can be obtained if the emission source is deeply embedded in an optically and
geometrically thick dusty structure. This is a reminiscence of the old
continuous torus models, with the consequence that broad SEDs are not
recovered. Alternatively, deep silicate absorption can be introduced by a
thick screen of cold dust located further away from the central source, as
proposed by AH11. Given the rather small expected torus sizes, it is possible
that an external screen of dust will introduce the absorption feature along
the line of sight towards us. However, as before, this screen will also
reprocess the near-IR emission into longer wavelengths, and therefore the
final SED will look narrower that the observations.

We used DUSTY (Ivezi\'c et al.~1999) to explore further this scenario using
CLUMPY torus models as the incident spectrum on a slab of cold or hot dust. We
found that a SED with strong near-IR emission is only recovered when the
temperature of the slab is as high as $\sim 1000$K and for moderate optical
depths. However, this combination of parameters essentially retains the
silicate feature of the incident spectrum ($\tau_V \sim 20$ implies
$\tau_{9.7} \sim 1$) while adding the emission from the screen of hot dust to
the near-IR range. Essentially, this corresponds to a scenario where enough
hot dust is added to the SED to increase its near-IR output without altering
significantly the torus emission in the mid-IR. This can be regarded as an
analogous model to the one proposed in \S 3.2.3 to explain the up-turn in the
near-IR emission of many of our sources. Notice that although this scenario
might seem similar to that proposed Stalevski et al.~(2012), our added
black-body component is completely arbitrary and has no physical connection to
the torus structure.

In summary, we finally conducted the fitting to sources with deep silicate
absorption and a broad SED as described previously: allowing the presence of a
black body component with temperatures in the 1000--2500 K range.

\section{Results}

\begin{figure*}
  \begin{flushleft}
    \includegraphics[scale=0.135,trim=180 0 0 -90]{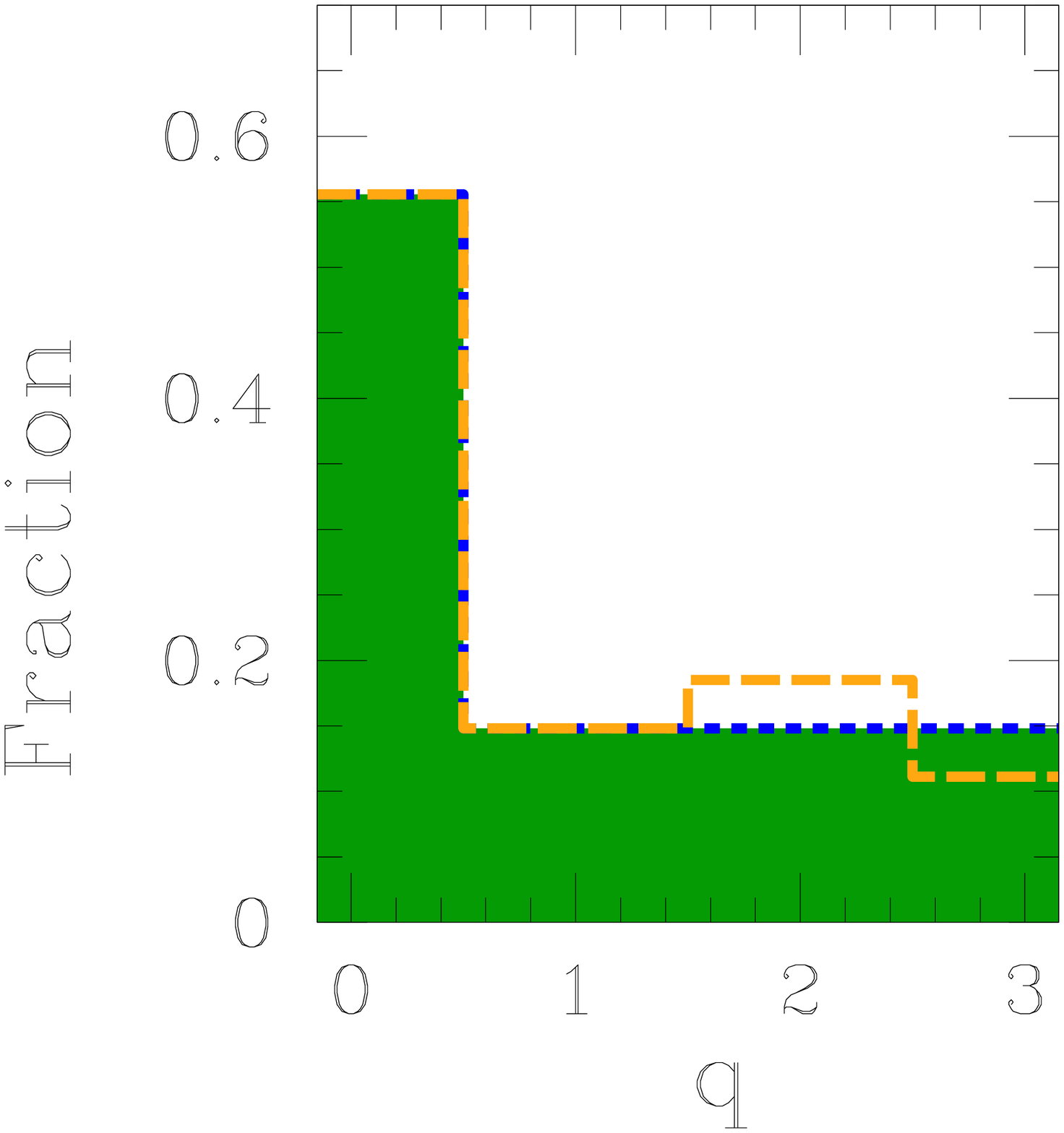}%
    \includegraphics[scale=0.135,trim=80 0 0 -90]{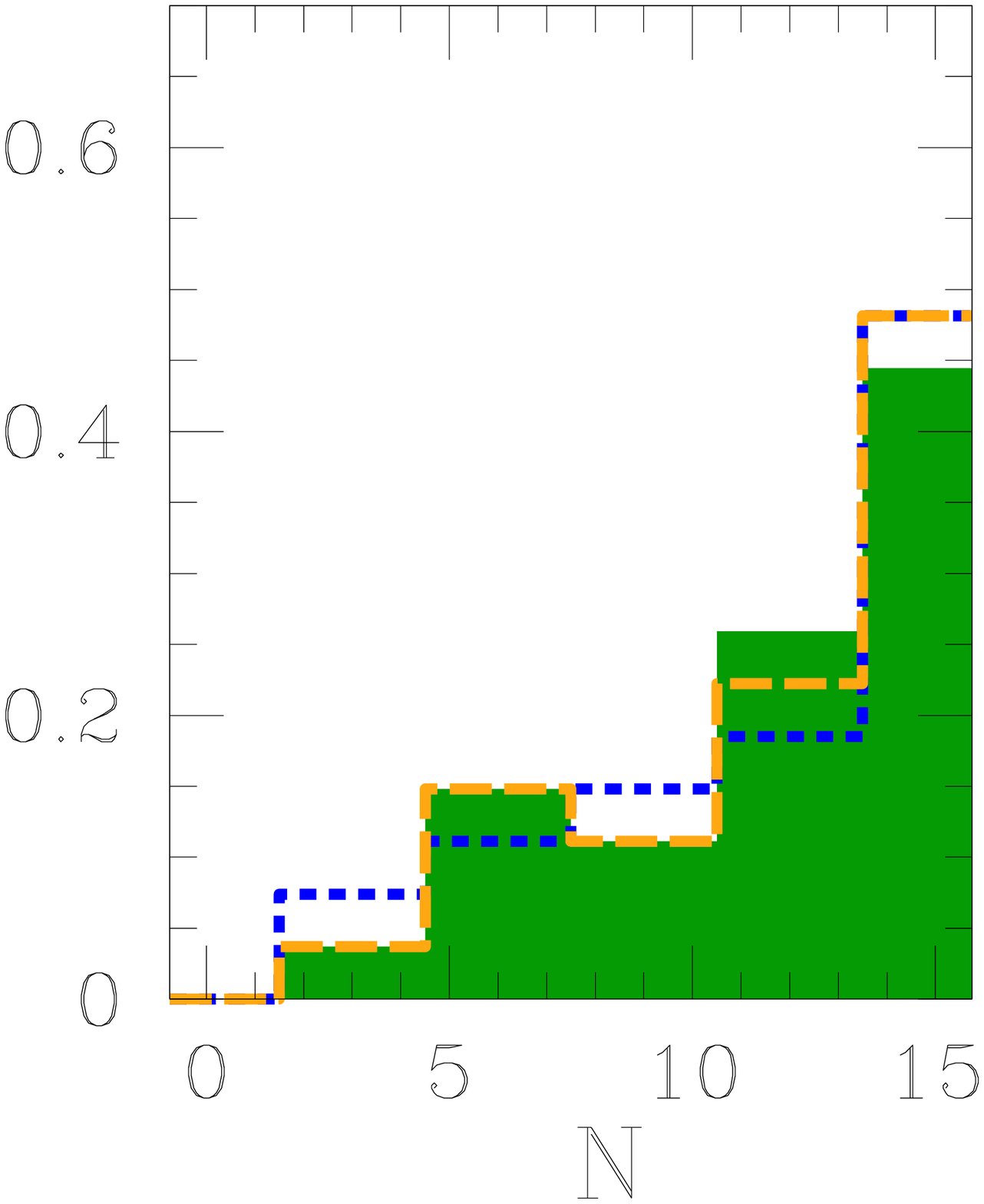}%
    \includegraphics[scale=0.135,trim=80 0 0 -90]{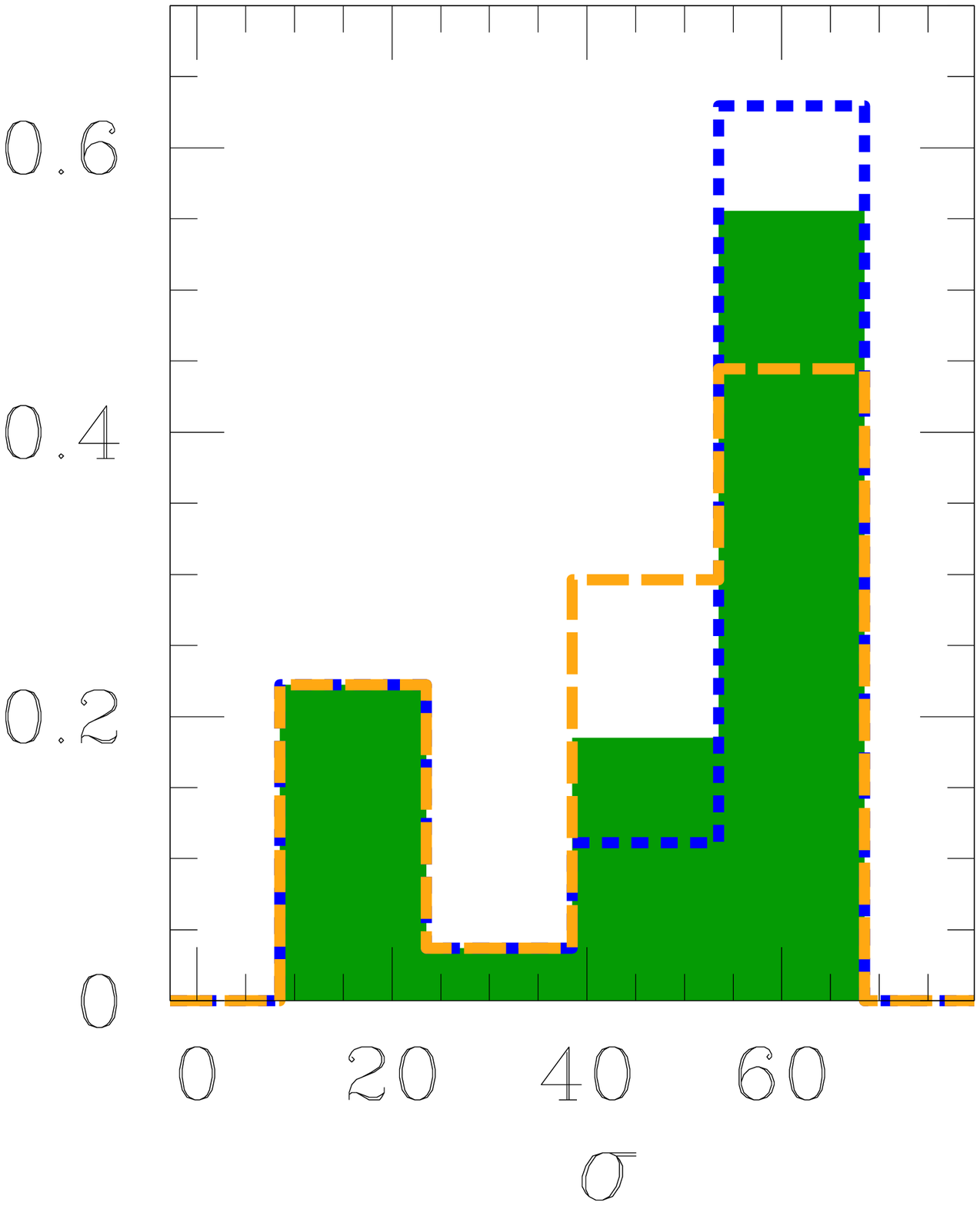}%
    \includegraphics[scale=0.135,trim=80 0 0 -90]{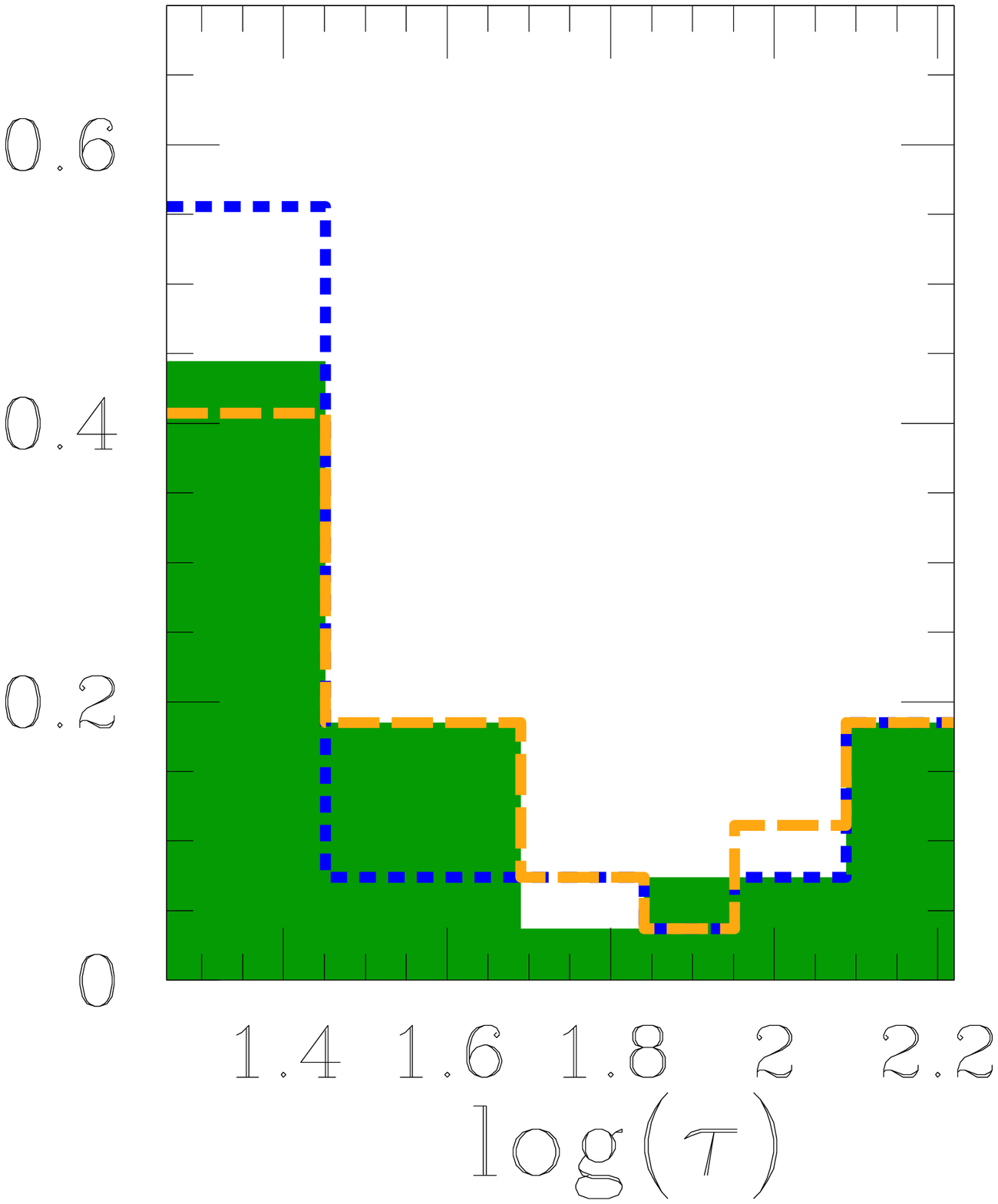}%
    \includegraphics[scale=0.135,trim=80 0 0 -90]{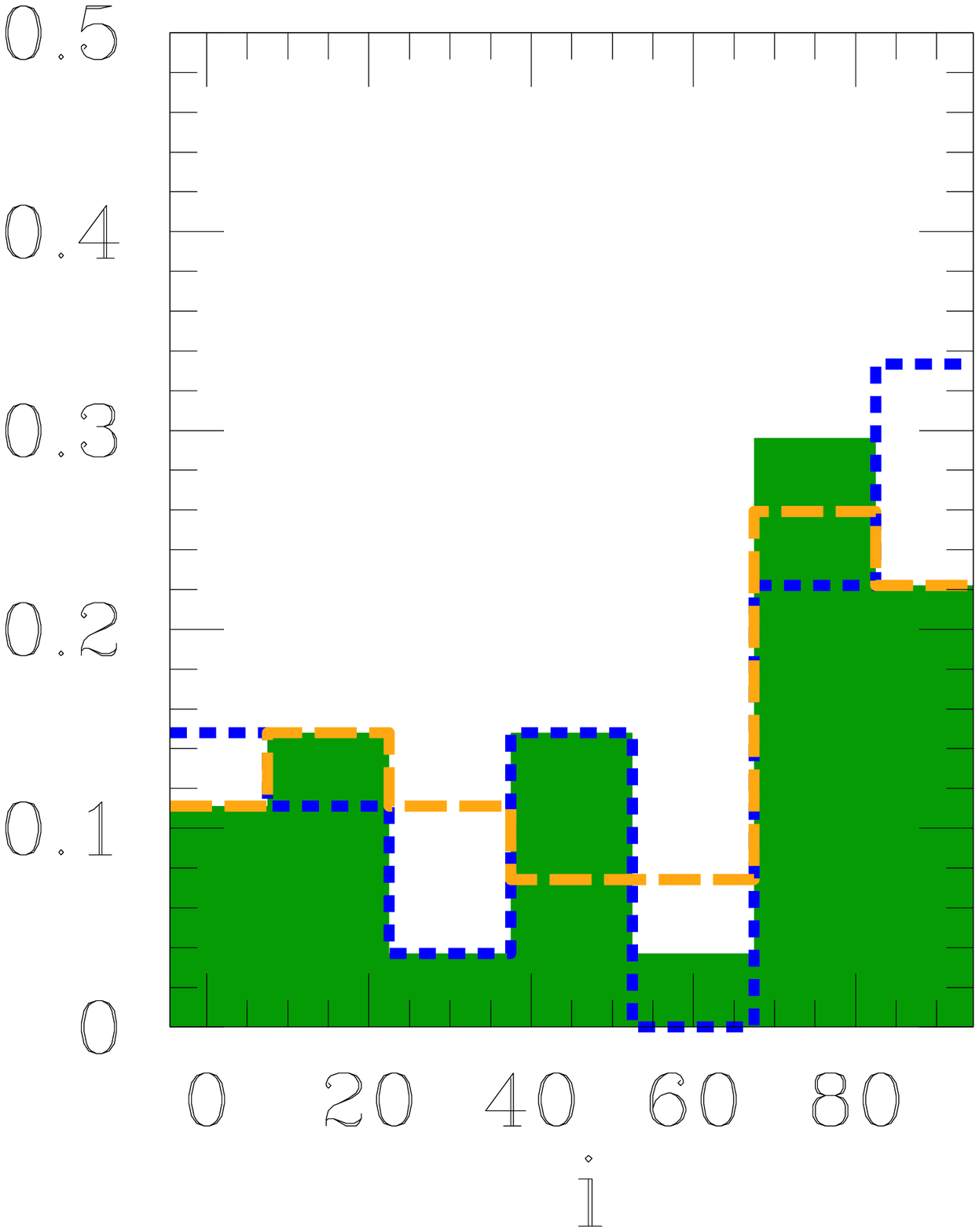}%
    \includegraphics[scale=0.135,trim=80 0 0 -90]{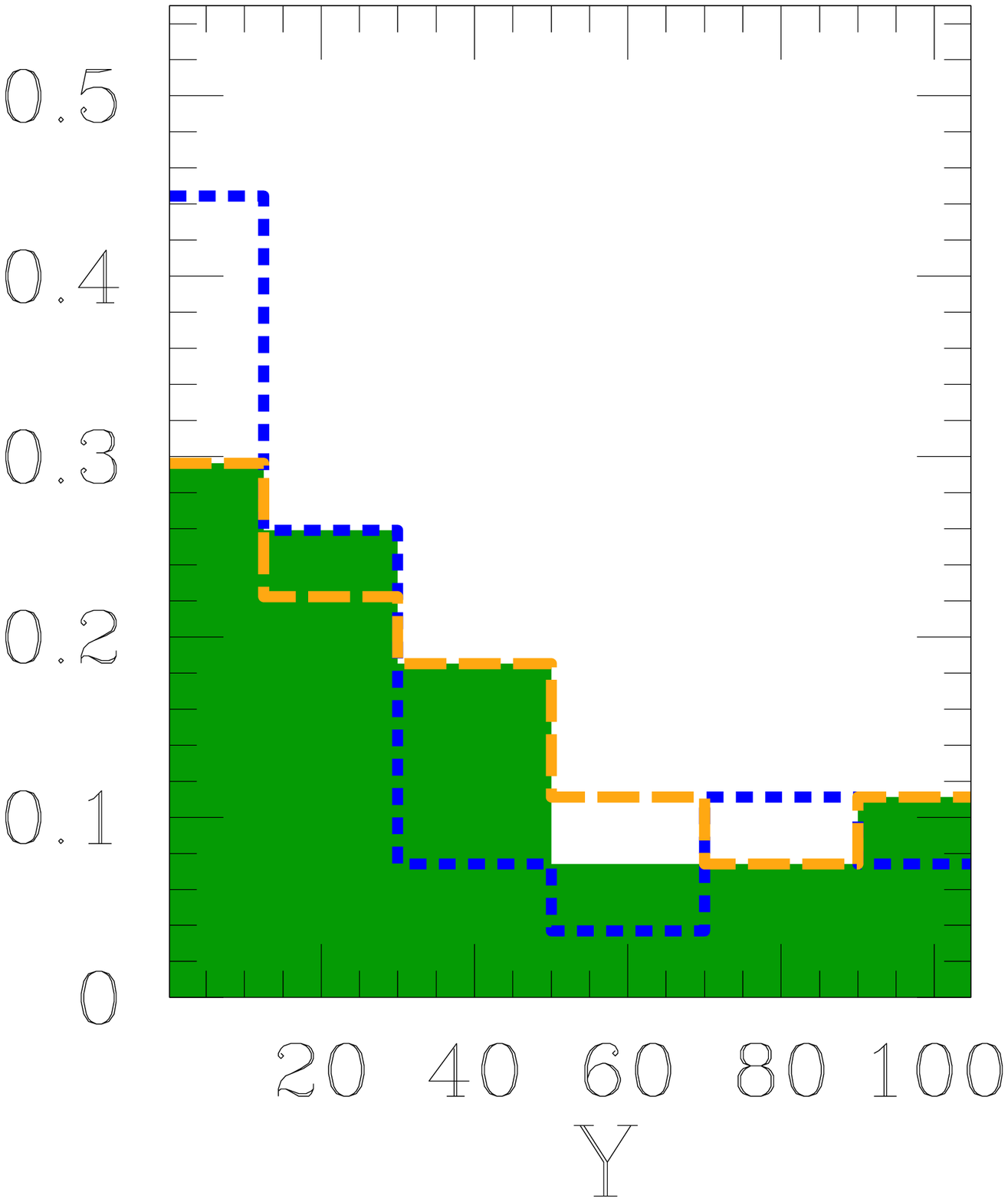}%
    \includegraphics[scale=0.135,trim=80 0 0 -90]{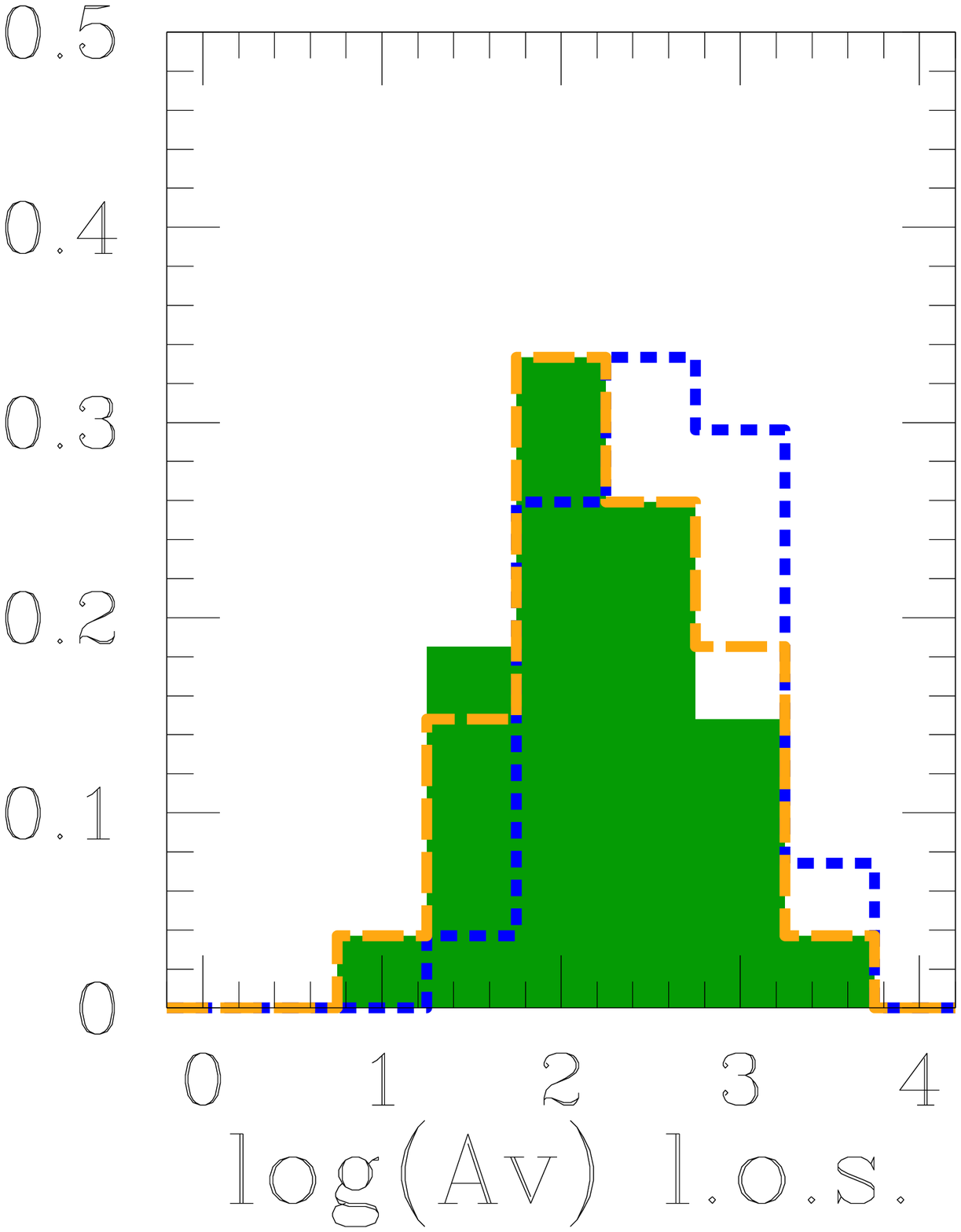}%
    \includegraphics[scale=0.135,trim=80 0 0 -90]{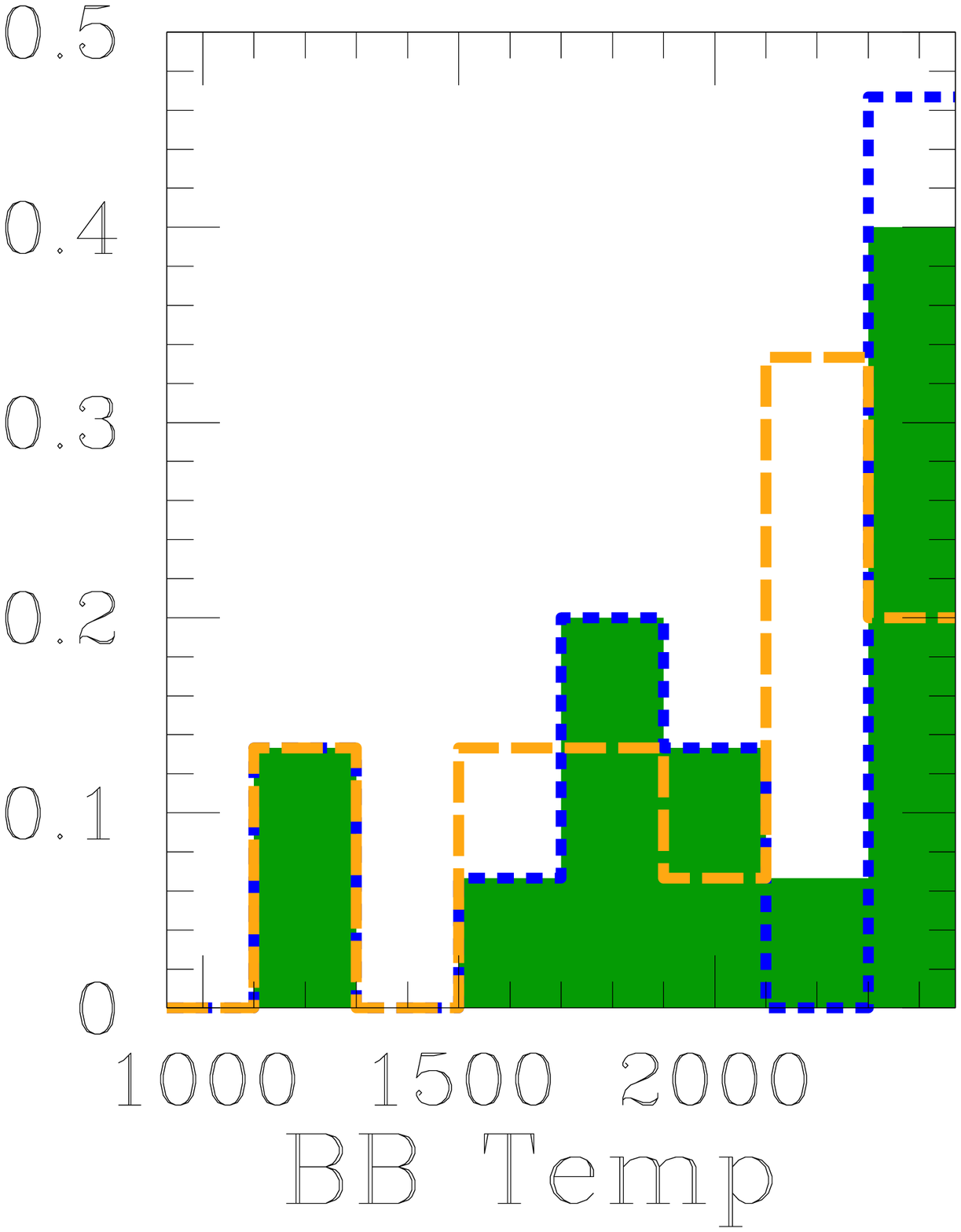}\\
    \includegraphics[scale=0.135,trim=180 0 0 -100]{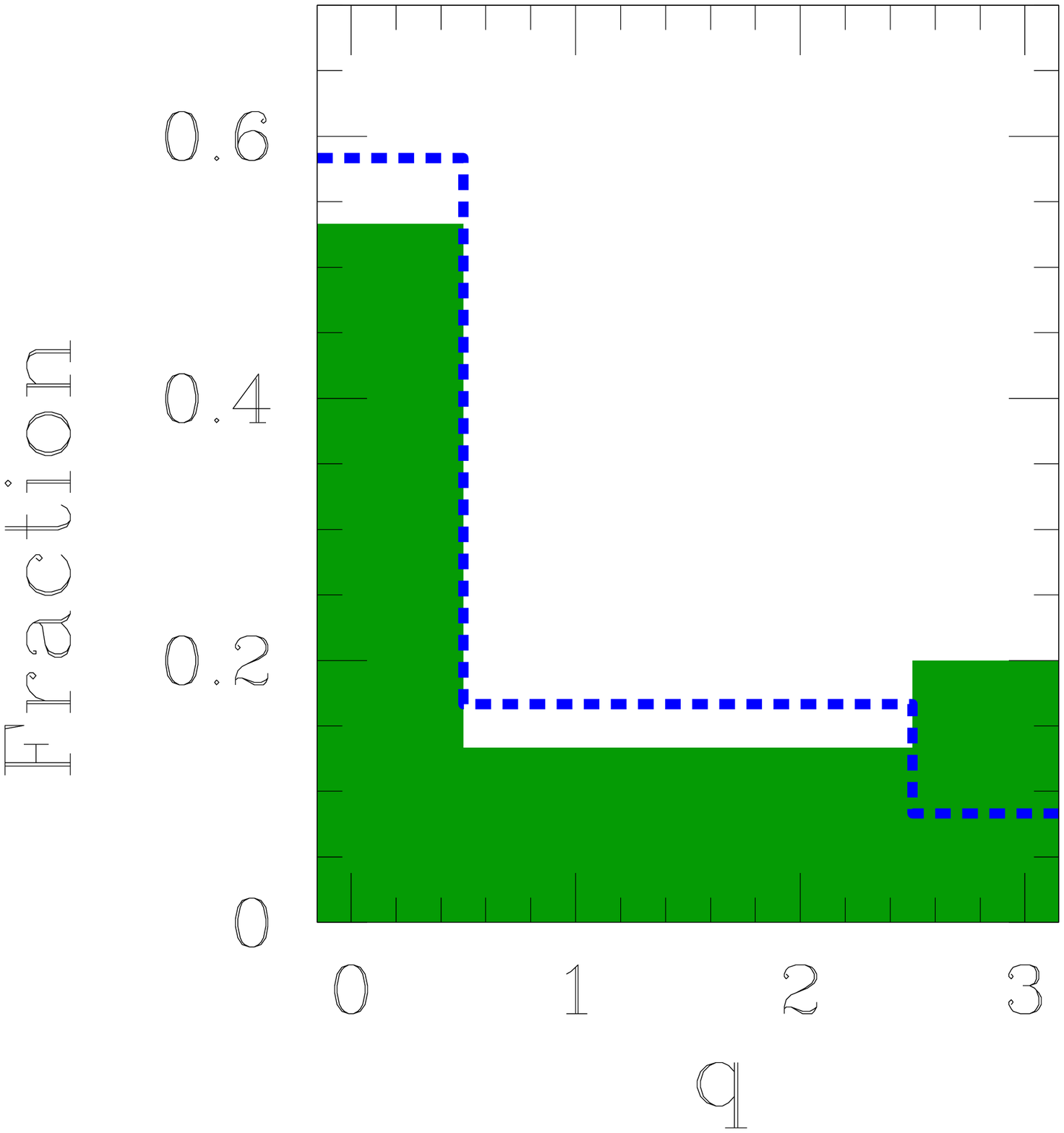}%
    \includegraphics[scale=0.135,trim=80 0 0 -100]{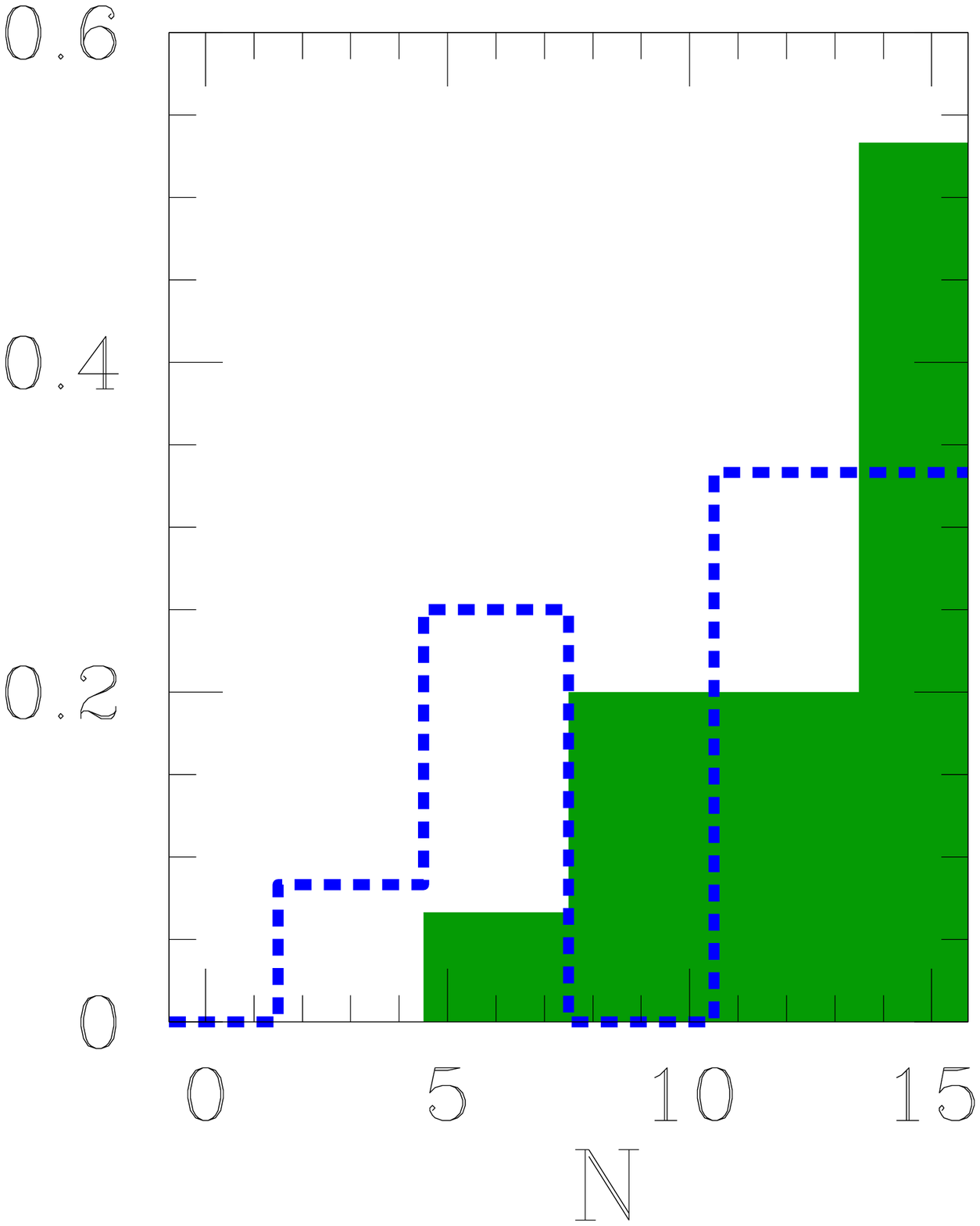}%
    \includegraphics[scale=0.135,trim=80 0 0 -100]{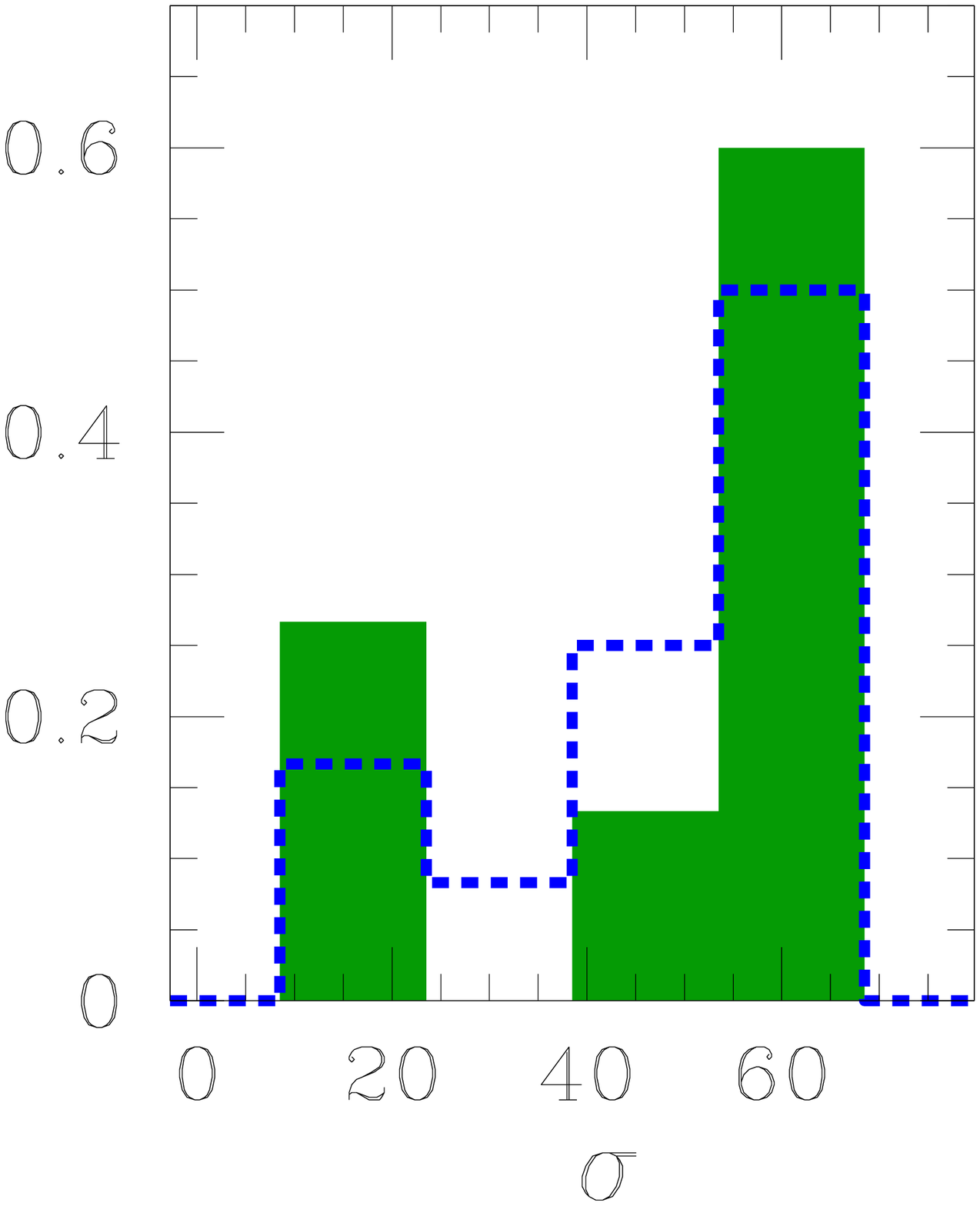}%
    \includegraphics[scale=0.135,trim=80 0 0 -100]{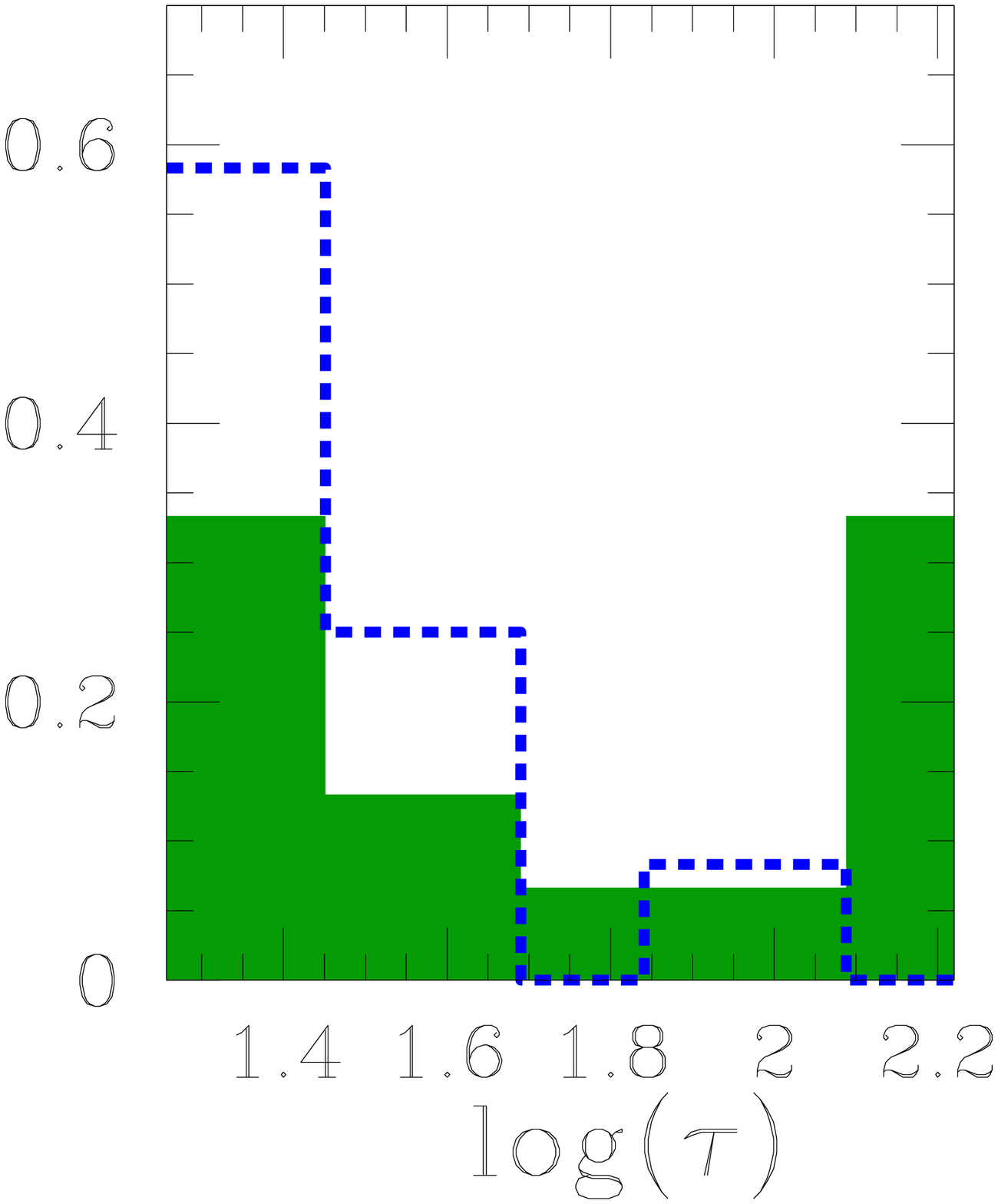}%
    \includegraphics[scale=0.135,trim=80 0 0 -100]{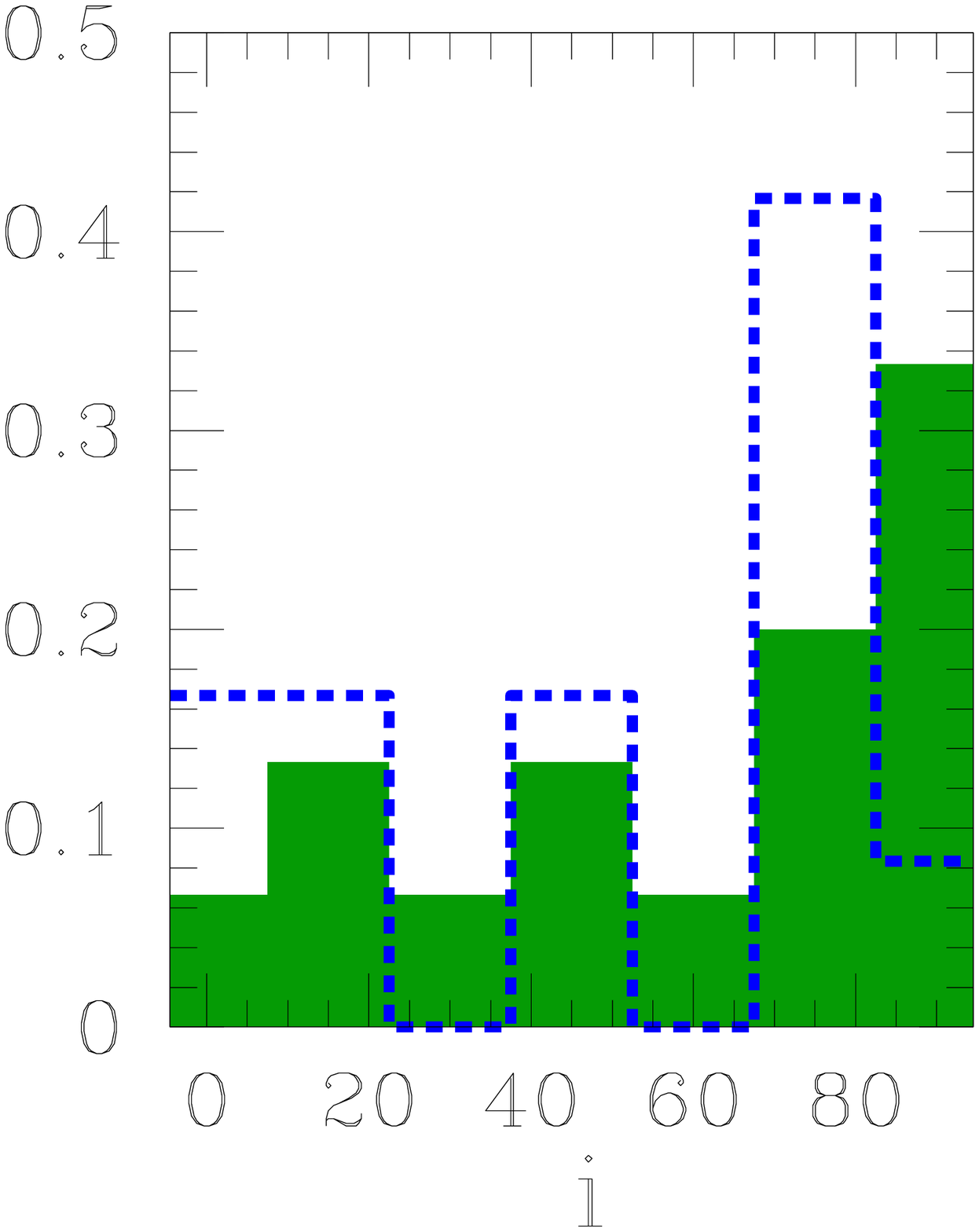}%
    \includegraphics[scale=0.135,trim=80 0 0 -100]{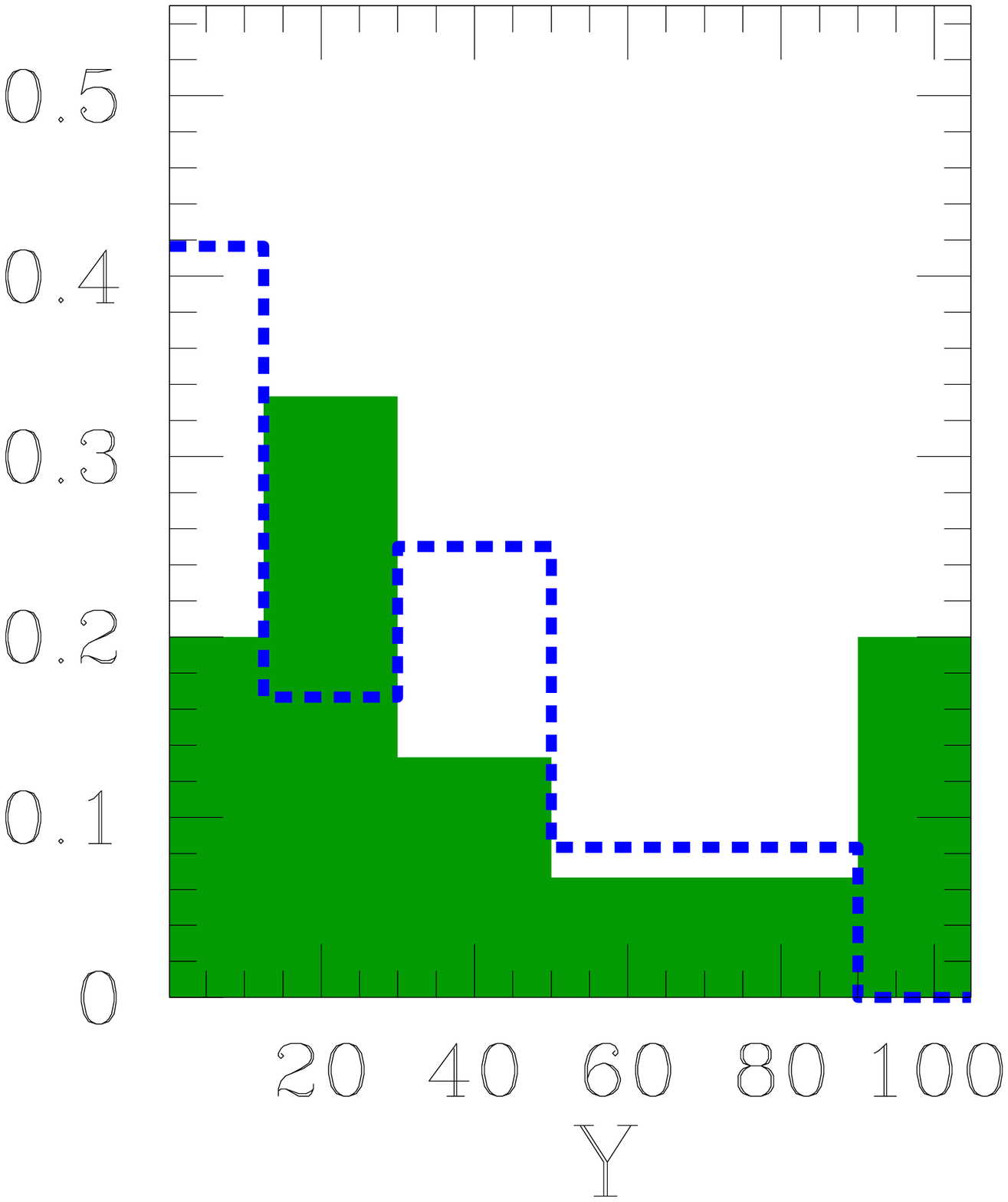}%
    \includegraphics[scale=0.135,trim=80 0 0 -100]{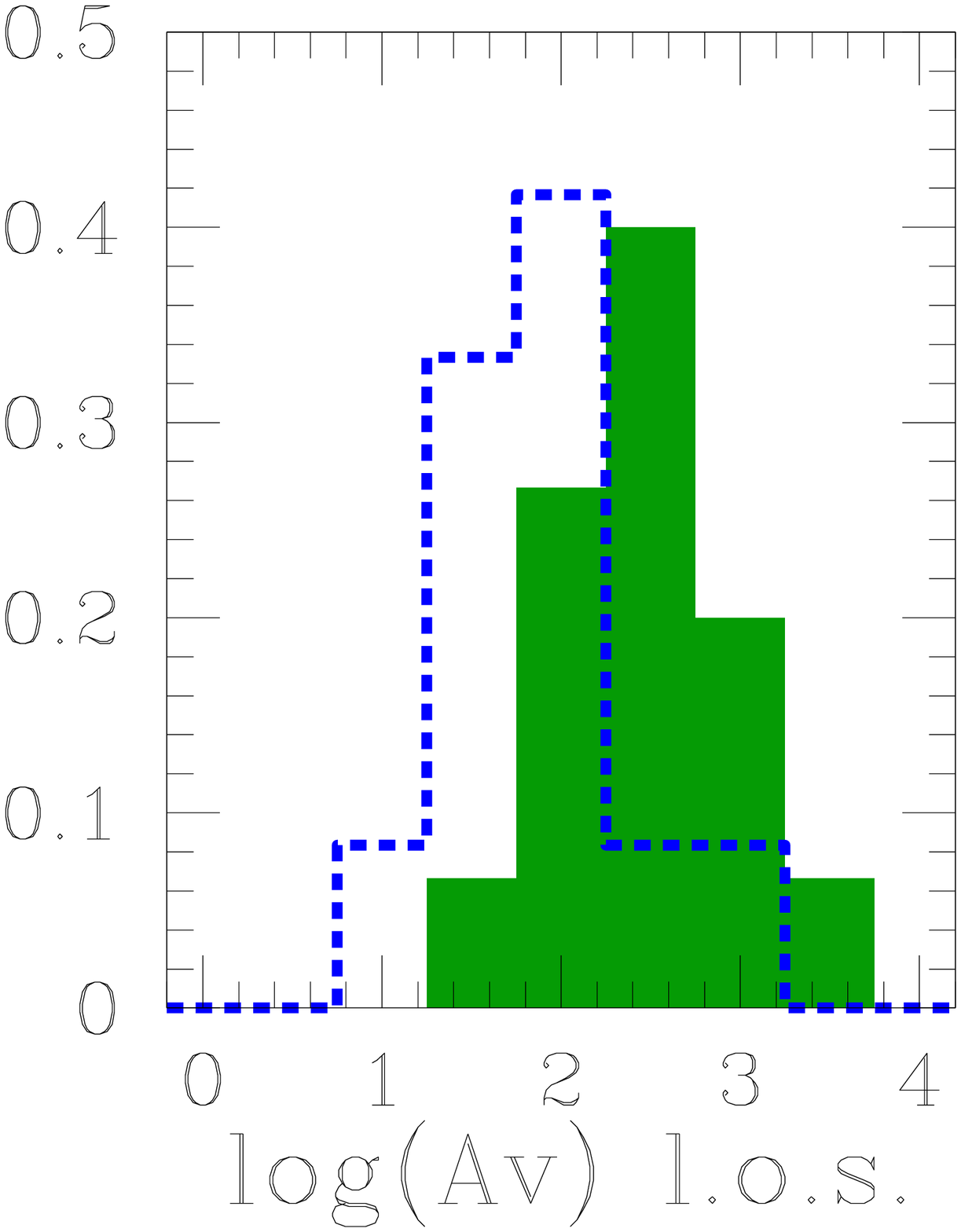}\\
    \includegraphics[scale=0.135,trim=180 0 0 -90]{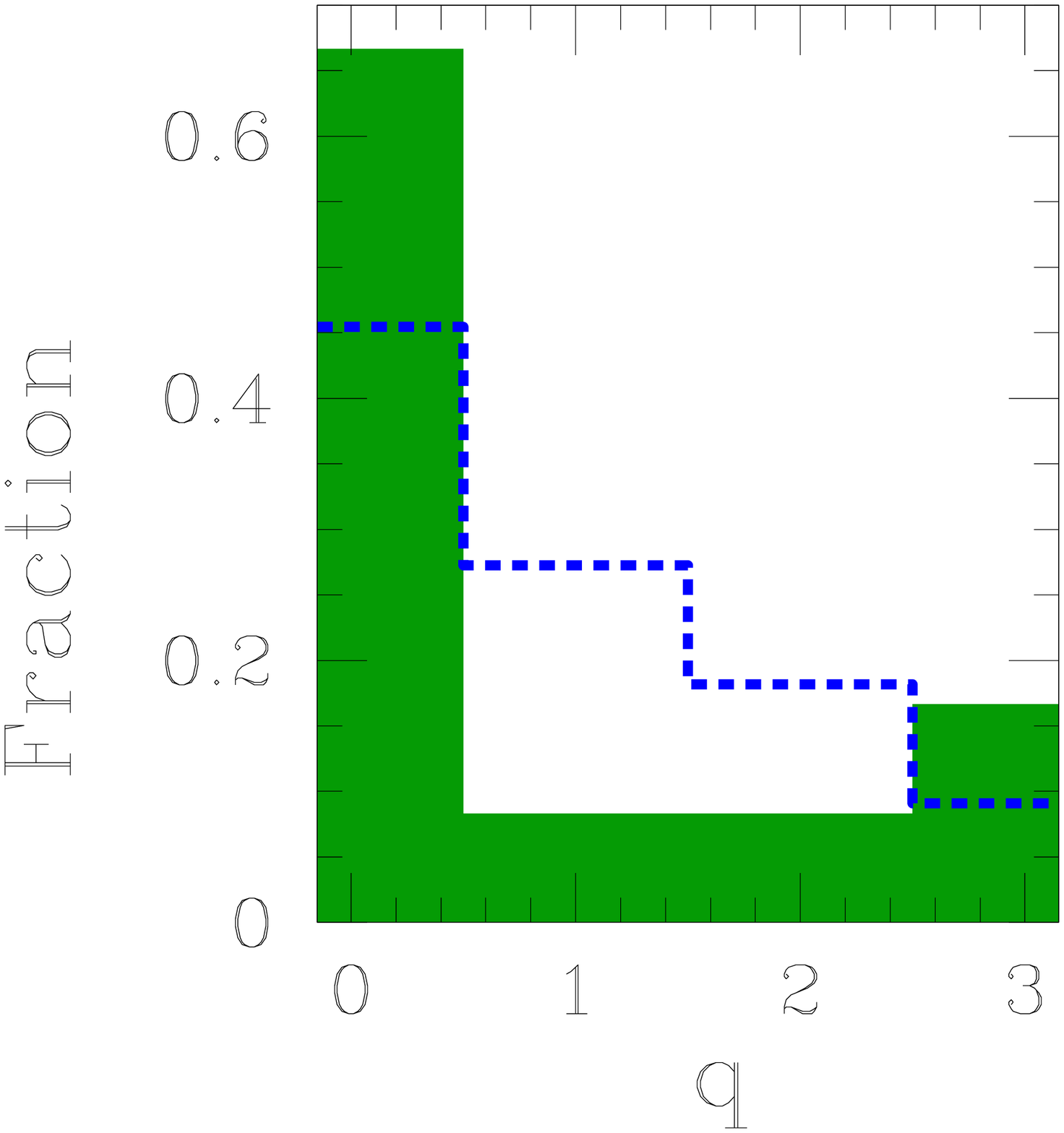}%
    \includegraphics[scale=0.135,trim=80 0 0 -90]{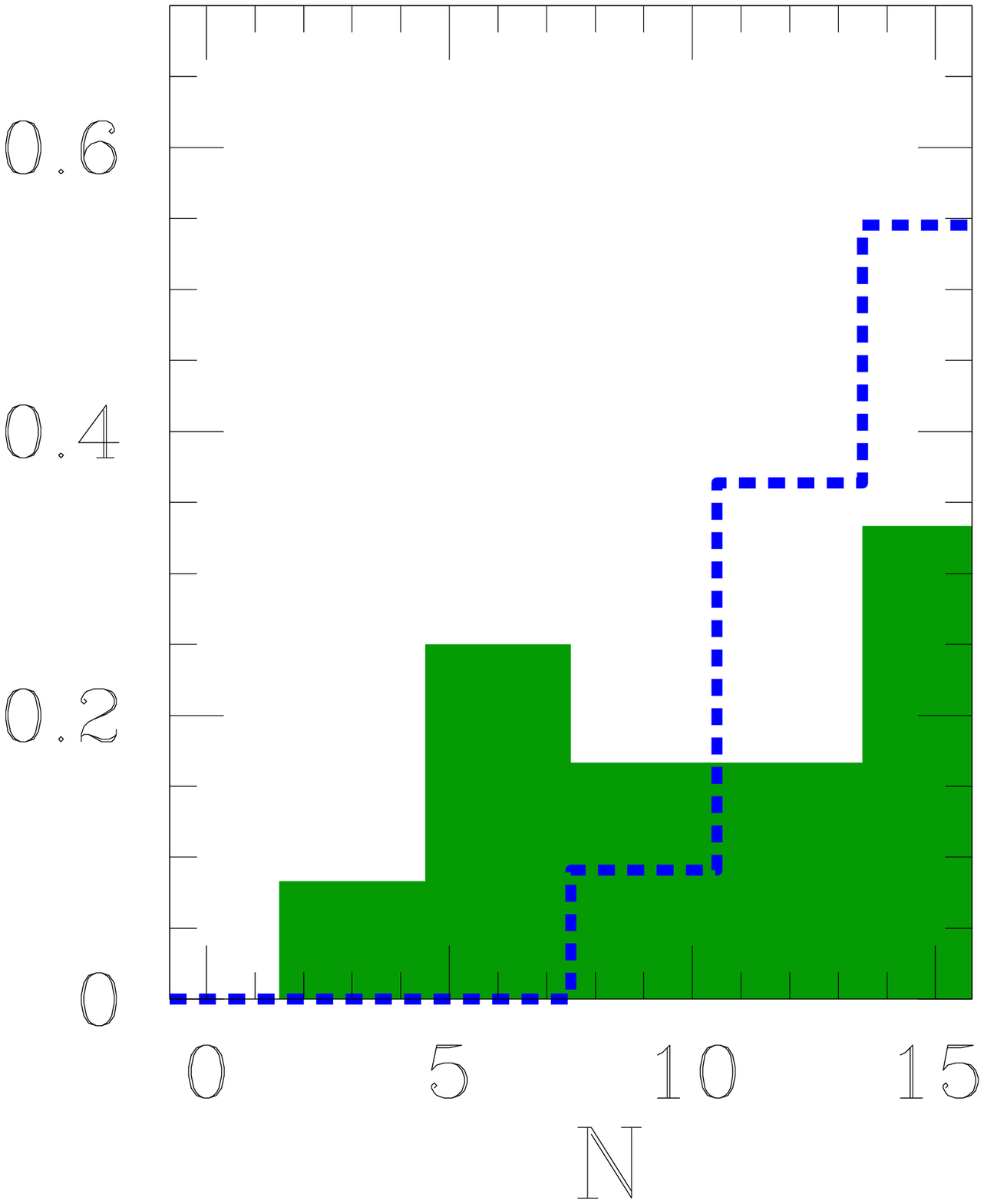}%
    \includegraphics[scale=0.135,trim=80 0 0 -90]{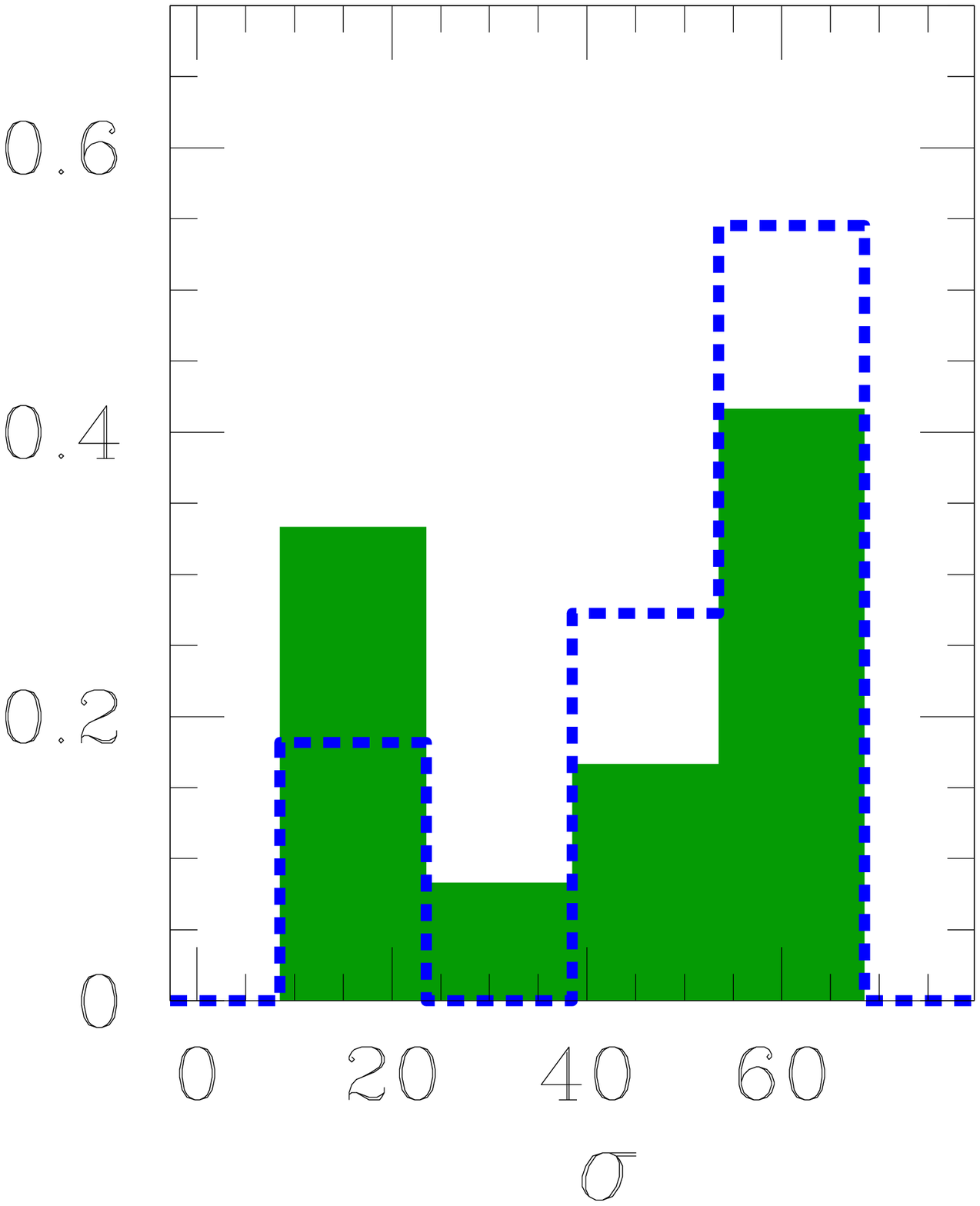}%
    \includegraphics[scale=0.135,trim=80 0 0 -90]{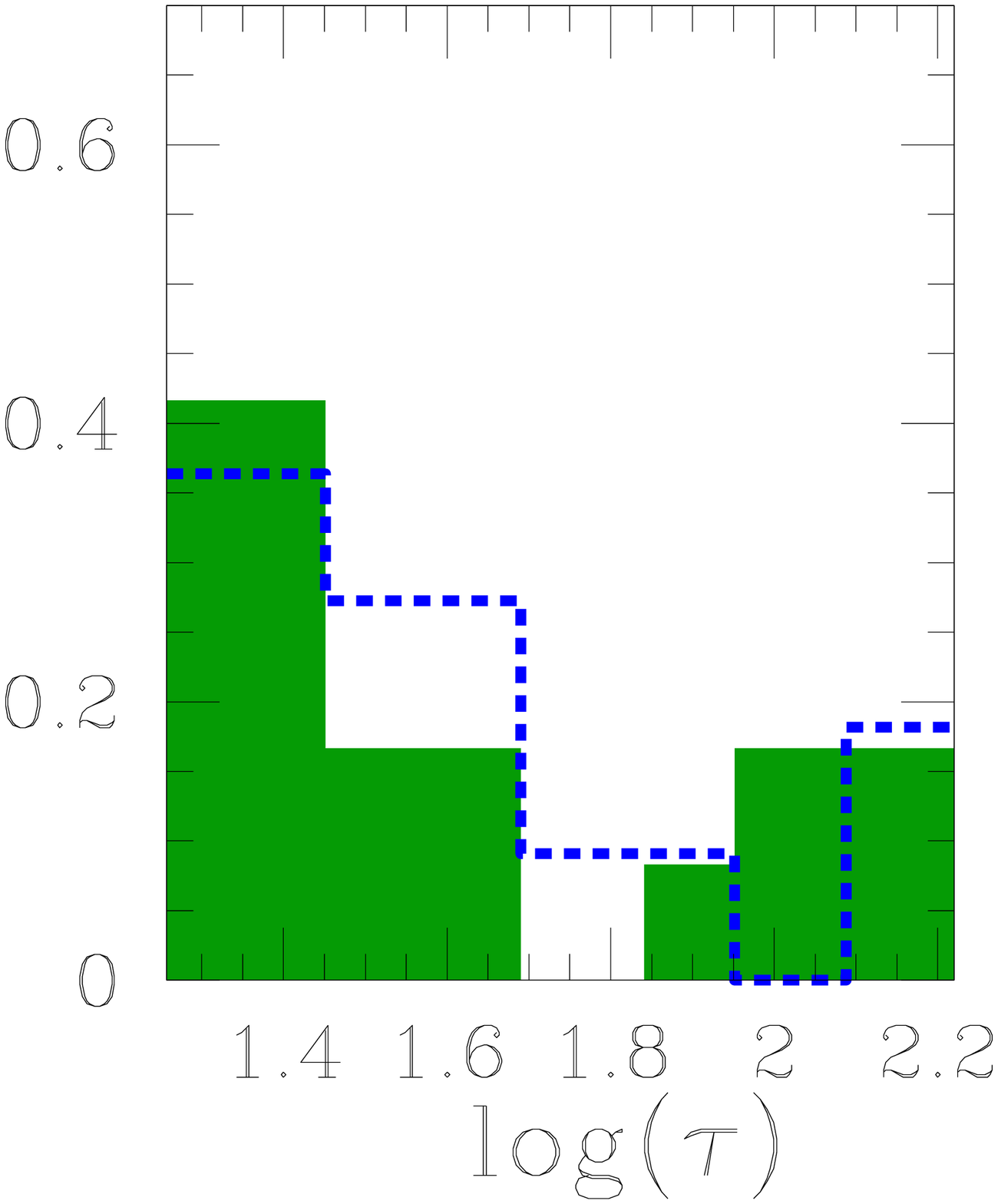}%
    \includegraphics[scale=0.135,trim=80 0 0 -90]{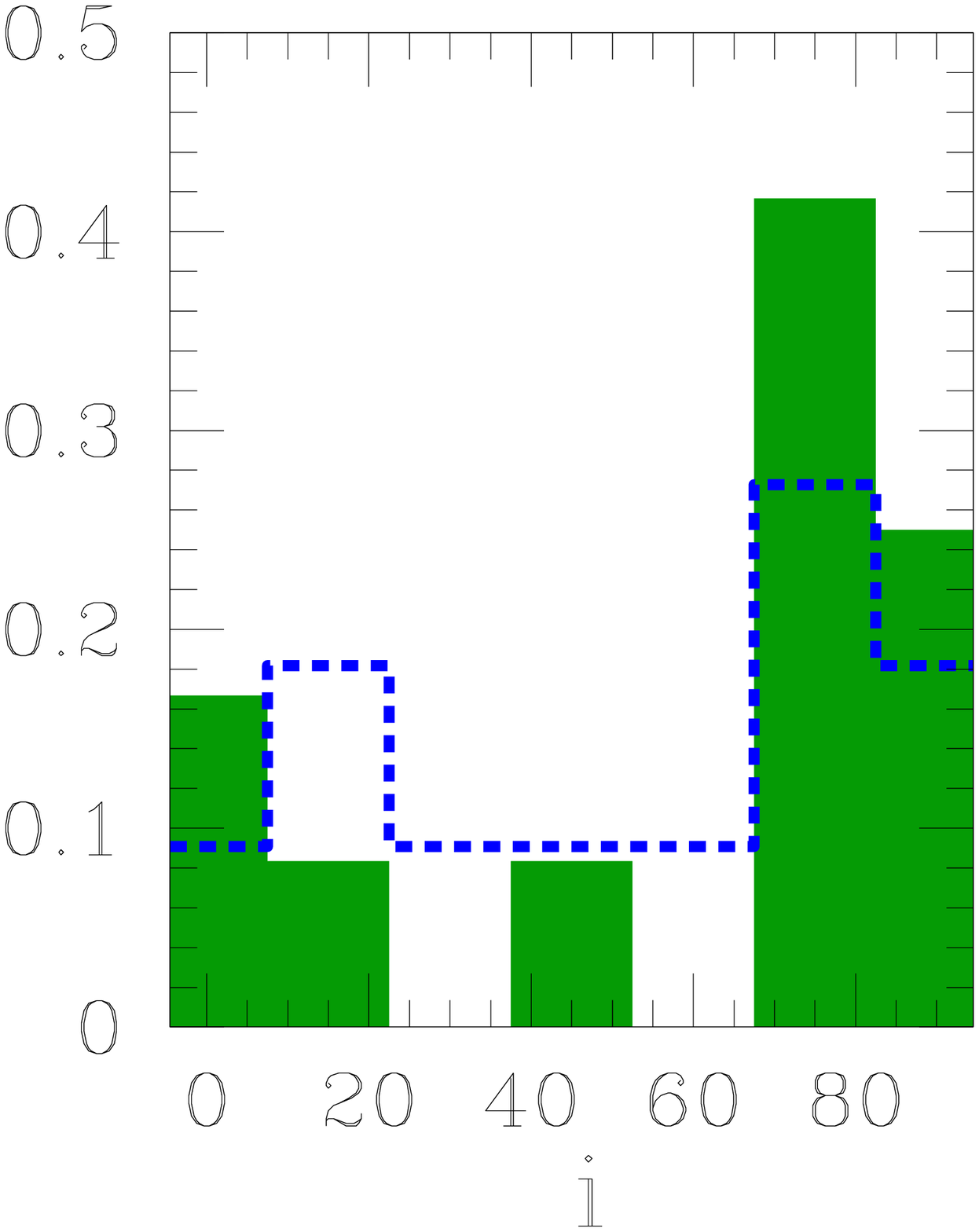}%
    \includegraphics[scale=0.135,trim=80 0 0 -90]{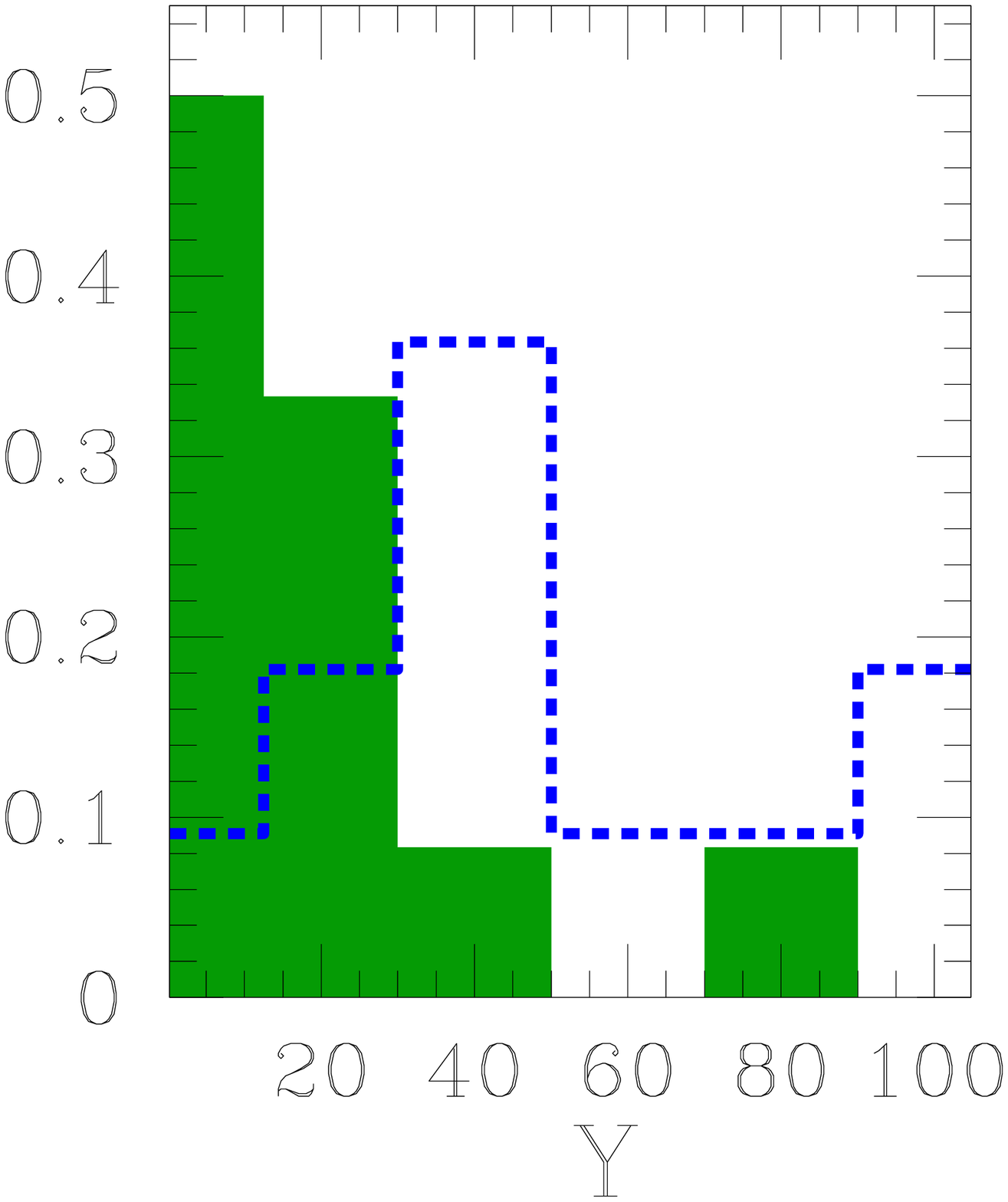}%
    \includegraphics[scale=0.135,trim=80 0 0 -90]{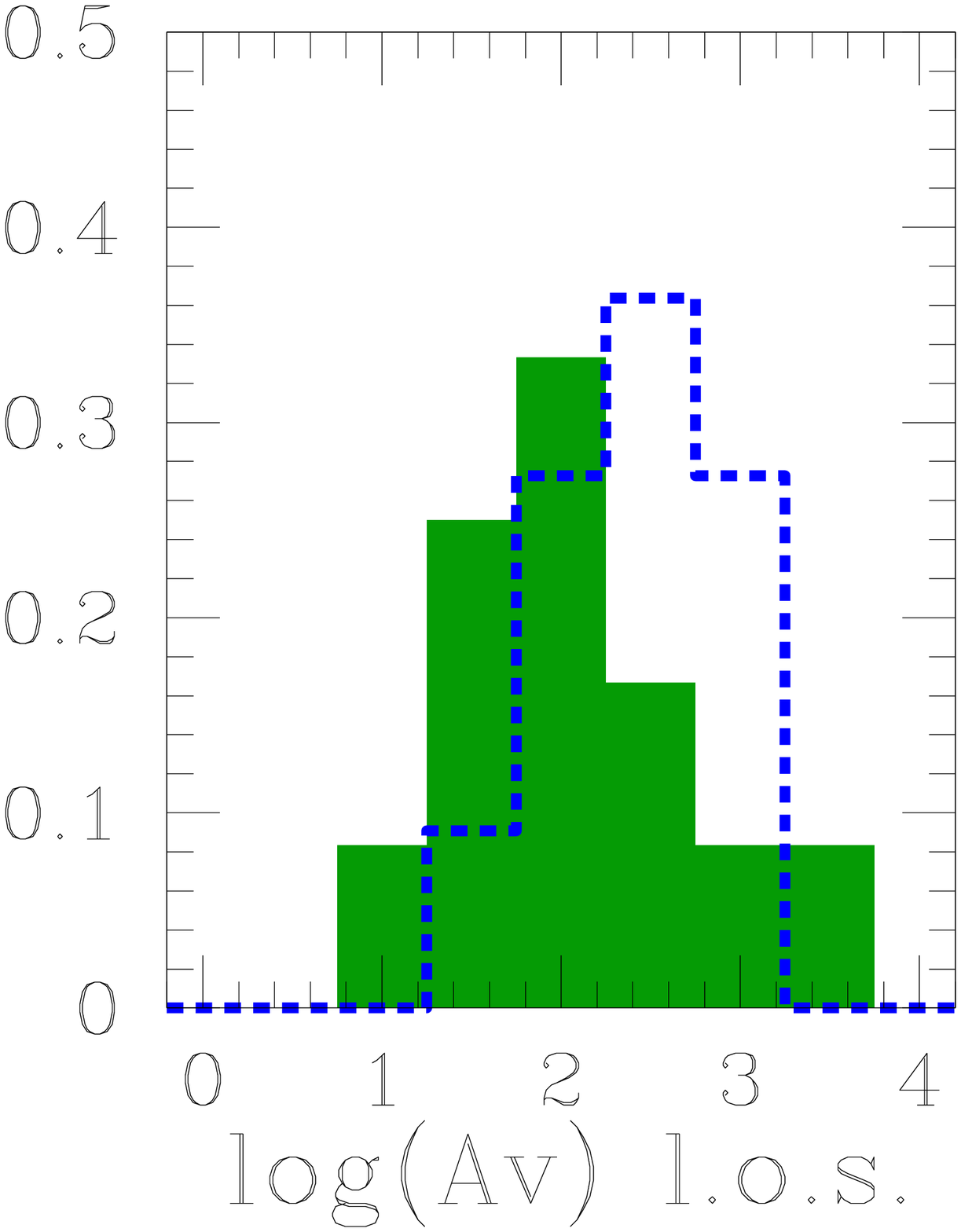}%
    \includegraphics[scale=0.135,trim=80 0 0 -90]{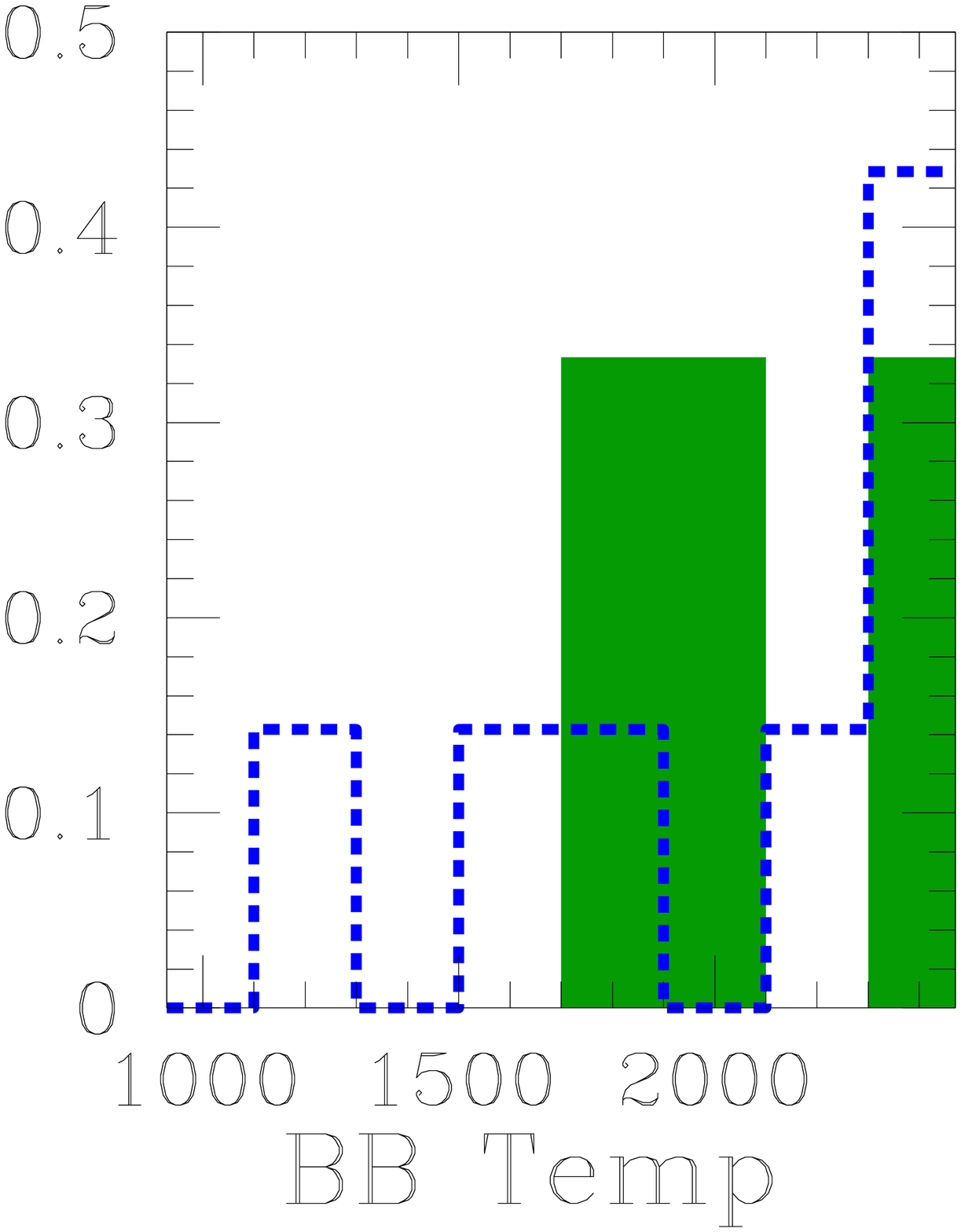}

    %trim=l b r t (left, bottom, right, top) 
  \end{flushleft}
  \figcaption{Top: Distribution of the best fitted values of $q,
    \mathcal{N}_0, \sigma, \tau, \angle i, Y$, $\log A_V$(los) and temperature
    of the additional black-body component if needed. The panels show the
    normalized histograms of the median (solid), and average and mode
    distributions (dashed lines) for the 26 objects with acceptable
    fits. Middle: Median distribution for objects that require a black-body
    component during their SED fitting (solid), compared with those that do
    not (dashed). Bottom: Median distribution for objects with a detected
    Hidden Broad Line Region (solid), compared with those without a HBLR
    (dashed).
\label{params}}
\end{figure*}

\subsection{General Results}

Fig.~\ref{seds} presents the fitting results to the SEDs of the 36 sources
found in Table 2.

The 2pC model with the lowest $\chi^2$ is shown with a long-dashed line. No
additional components were added to these models. 

The CLUMPY model with the highest probability is shown in Fig.~\ref{seds}
using a thick short-dashed line which includes the sum of any additional
component, if present. The additional component (black-body or power-law) is
separately shown using a thick dash-dotted line.

In most cases, the probability distribution of the remaining CLUMPY models
tends to be highly peaked, in the sense that the probabilities rapidly fall to
values much lower than that of the best-fit model. We have divided the
probability distribution into three ranges: top$-50\%$, $50-10\%$ and $10-1$\%
of the best fit probability. The number of models found in each of these
ranges is shown on the top-right corner of in each panel in
Fig.~\ref{seds}. At most twenty models are shown in dark-gray, medium-gray and
light-gray thin-continuous lines, for each of the probability ranges. If more
than 20 models are available for each range, then only 20 models are randomly
drawn from the pool of models available and shown. This gives a representation
of the level of departure of the model SEDs from the best-fit.

In what follows we discuss poor or inadequate fits. The mentioned sources will
not be considered when drawing conclusions from the fitting process in \S 4.2.

The need for mid-IR data to properly constrain the peak of the infrared
emission is clearly seen in the fits to IRAS\,00521-7054, ESO\,541-IG12, and
IRAS\,01475-0740, where the best models only follow the rapid raise in near-IR
flux but then flare-out for $\lambda > 10 \mu$m due to the lack of restriction
in this range. These fits are also characterized by a much less peaked model
probability distribution, as shown by the large number of models in the
probability ranges defined above. Therefore all sources without mid-IR
observations (this also includes IRAS\,00198-7926\,N and IRAS\,00198-7926\,S)
will be excluded from the subsequent analysis. Of these discarded sources, 2
have $\alpha\! <\! 1$ (see \S 3.3.3).

NGC\,34 and NGC\,5953 show peculiar SEDs because of contamination from a
starburst component, showing very strong near-IR emission. Both sources are
also characterized by {\emph{Spitzer}} spectra showing the largest PAH
residuals seen in the sample. We used a power-law to account for this
component during the fitting process, but the results are poor. These two
galaxies will not be considered in the analysis carried out below.

NGC\,4941 lacks enough data to constrain its SED and will not be analyzed any
further.

NGC\,4501 presents a very steep SED throughout the near and mid-IR, possibly
because of the presence of a beamed Synchrotron jet, as already
discussed. This is in sharp contrast with the SED determined by Kishimoto et
al.~(2009a) using the Keck interferometer, which presents a canonical shape,
rising from the near-IR to the mid-IR. In our observations, the presence of a
dominant power-law component means that the torus properties cannot be
successfully recovered from the fitting.

In all, 27 sources will be considered in the following sections, unless
otherwise noted. Of these, 14 required an extra component to account for the
near-IR excess, i.e., very close to a 50\%\ of the sample. More details are
reported in \S 4.4.

\subsection{Results from SED Fitting \& Best Fit Parameters}

\subsubsection{CLUMPY Fitting Results}

Table 3 presents the main results from the CLUMPY SED fitting procedure to the
nuclear data of 27 sources, listing the weighted mean, the median, the mode
and the 67\%\ confidence limits for the probability distribution of each torus
parameter.

Fig.~\ref{params} presents the distributions of the median (in solid), average
and mode (both in dashed) parameters for the sample. Parameters are: $q,
\mathcal{N}_0, \sigma, \tau, \angle i, Y$, $A_v$ (along the line-of-sight) and
black-body temperature when needed. In what follows we discuss the inferred
values for each parameter.

The exponent to the radial distribution of clouds shows a preference for
$q \sim 0$. This corresponds to a flat distribution with the number of
clouds showing a very weak dependence with the distance to the central Black
Hole (Nenkova et al., 2008b). However, about half of the source require larger
$q$ values.

The number of clouds along the equator shows a strongly rising distribution
towards a large number of clouds, with $\mathcal{N}_0 \ga 10$ being clearly
favored. This was also hinted by AH11. Nenkova et al.\ (2008b) showed that a
rather small number of clouds is required by clumpy models to reproduce
observed SEDs, with typically $\mathcal{N}_0 \sim 5$. The difference with our
findings could be due to our sample selection, which prefers obscured,
IR-bright objects. Whether IR-bright AGN require a larger number of clouds in
order to reproduce their properties is a tentative result from this work.

A large number of clouds ($\mathcal{N}_0>10$) pushes the results from the
modeling of clumpy media to the limits of its parameter space (M.~Elitzur,
private communication). This is because the current calculations compute the
radiation field produced by directly illuminated clouds and then solves for
those clouds found in their shadow.  The emission from clouds found in the
shadow of other clouds is not taken into account. A complete solution would
require to iterate over the whole cloud population until a converging solution
is attained. However, this is too demanding for current computer capabilities
and these iterative steps are not followed (Nenkova et al.\, 2008a). Clearly,
the larger the number of clouds, the larger the deviation between the current
model prescriptions and a complete solution, particularly for compact
geometries. For $\mathcal{N}_0 \rightarrow \infty$ the emission from a
continuous torus should be recovered.

The torus aperture angle shows a broad distribution with a peak at $\gtrsim
60$ degrees and about 70\%\ of the sources require $\sigma > 40$ degrees. This is
in agreement with AH11 who found that this is the case for Seyfert II
galaxies, while Seyfert Is might show a narrower distributions of torus
aperture angles.

The optical depth of individual clouds in the $V$ band is constrained to small
values around $\tau \lesssim 30$, although some sources require larger values.

The distribution of torus viewing angles shows that some objects are well
fitted using intermediate values of $\angle i$, but about half of the sample
requires angles of 70 to 90 degrees, reflecting the fact that our sample is
IR-selected, and therefore should not be biased against heavily obscured objects.

The distribution of torus thickness, $Y$, shows a tendency towards small
values, with the bulk of the population requiring $Y\lesssim40$. Nenkova et
al.\ (2008b) argued that given the level of isotropy observed in the IR
emission of AGN, then torus should be rather compact ($Y\lesssim20$) or
present a steep ($q\sim2$) rather than flat ($q\sim0$) cloud distribution. We
do not see this trend: sources with small and large values of $Y$ all are more
likely to require $q\sim0$.

The number of clouds along the line of sight can be determined using
the expression:

\begin{eqnarray}
\mathcal{N} (\beta) = \mathcal{N}_0 \cdot \exp (-\beta^2/\sigma^2)
\end{eqnarray}

where $\beta$ is the angle between the torus equator and the line of sight
(i.e., $\beta = 90 - \angle i$). The product of the number of clouds and the
optical depth of each cloud gives total optical depth of the torus along the
line of sight in the $V$ band:

\begin{eqnarray}
A_V = 1.1 \cdot \tau \cdot \mathcal{N} (\beta)
\end{eqnarray}

Because of the rather narrow distribution of the values of $\tau$ and
$\mathcal{N}_0$, the distribution of the total optical depth shows a strong
peak at $A_V\sim 30-300$. 

The temperature of the black-body component is an additional parameter for the
fits. The distribution of temperature values shows that most objects require a
very high temperature ($T > 1500$K) and only 2 fits require $T\sim 1200$ K. 

Crucially, the addition of a secondary component does not change the torus
parameter distributions. Individually, only a few sources show significant
changes in some parameters after the black-body component is added (see Table
3). This can also be seen collectively when comparing the histograms for
objects where an additional component was required and for those where the
torus model alone yielded a good fit (middle panels in Fig.~\ref{params}). The
distributions are almost indistinguishable.

\subsection{Previous CLUMPY Fitting Results}

\begin{deluxetable}{crrrrrr}
\tablecolumns{7}
\tablecaption{Comparison of model parameters from RA09, AH11, and this work.\label{comp_clumpy}}
\tablehead{
\colhead{Galaxy} & \multicolumn{2}{c}{RA09} & \multicolumn{2}{c}{AH11} & \multicolumn{2}{c}{This work} \\
\colhead{Parameter} & \colhead{$med$} & \colhead{$mod$} & \colhead{$med$} & \colhead{$mod$} & \colhead{$med$} & \colhead{$mod$}}
\startdata
&&&&&& \\
NGC\,1068&&&&&&\\              
$q$              & --   & --  & 2.2 & 2.0& 0  &  0\\ 
$\mathcal{N}_0$  & --   & --  & 14  & 15 & 4  & 4\\ 
$\sigma$         & --   & --  & 26  & 21 & 30  & 28\\ 
$\tau$           & --   & --  & 49  & 49 & 30  & 34\\ 
$\angle i$       & --   & --  & 88  & 89 & 80  & 81\\ 
&&&&&& \\ \hline                                      
&&&&&& \\                                             
IC\,5063&&&&&&\\               
$q$              &$<$1.5& 0.4 & 2.6 & 0.8& 3  &  3\\ 
$\mathcal{N}_0$  &$>$11 & 14  & 14  & 15 & 14  & 13\\ 
$\sigma$         &$>$57 & 75  & 60  & 47 & 60  & 60\\ 
$\tau$           &   70 & 66  & 130 & 99 & 81  & 80\\ 
$\angle i$       &$>$65 & 89  & 82  & 84 & 82  & 90\\ 
&&&&&& \\ \hline                                      
&&&&&& \\                                             
NGC\,5506&&&&&&\\    
$q$             & 2.5  & 2.7 & 0.4  & 0.3 & 0.7 & 1 \\ 
$\mathcal{N}_0$ &$<$2  & 1   & 14   & 15  & 14 & 15 \\ 
$\sigma$        & 25   & 15  & 43   & 40  & 60 & 60 \\ 
$\tau$          &$<$68 & 22  & 100  & 99  & 48 & 40 \\  
$\angle i$      & 85   & 85  & 34   & 35  & 16 & 20 \\  
&&&&&& \\ \hline                                      
&&&&&& \\                                             
NGC\,7172&&&&&&\\    
$q$             &$>$1.7&2.9 & 1.1& 1.5  & 0.5 & 0.5 \\  
$\mathcal{N}_0$ & 5    &  5 & 13 & 15    & 15 & 15\\   
$\sigma$        &$>$54 & 74 & 61 & 68    & 60 & 60\\   
$\tau$          &$<$12 & 10 & 59 & 52    & 60 & 61\\   
$\angle i$      &$>$45 & 89 & 77 & 85    & 50 & 50\\   
&&&&&& \\ \hline                                      
&&&&&& \\                                             
NGC\,7582&&&&&&\\    
$q$             &$>$2.5& 3.0 & 0.3 & 0.1 & 0.5 & 0.5 \\ 
$\mathcal{N}_0$ &$<$2  & 1   & 13 &  15  & 15 &  15 \\ 
$\sigma$        &$<$29 & 16  & 48 &  49  & 60 & 60 \\ 
$\tau$          &$<$27 & 14  & 89 &  97  & 40 & 43 \\ 
$\angle i$      & 41   & 58  & 12 &   0  & 30 & 30 \\ 
\enddata 
\tablecomments{Notice that RA09 and AH11 limited the range of $Y$ to values up
  to 30. AH11 also incorporated a foreground host absorption component during
  their fitting corresponding to $A_V = 7, 11, 8-13$ magnitudes for IC\,5063, NGC\,5506
  and NGC\,7582, respectively.}
\end{deluxetable}

\begin{deluxetable}{l|rrrrr}
\tablecolumns{6}
\tablecaption{2pC Model results \label{results2pc}}
\tablehead{
\colhead{Object} & \colhead{$\tau_{9.7}$} & \colhead{$p$} & \colhead{$q$} & \colhead{$i$} & \colhead{$\chi^2$}}
\startdata
%Object			
NGC\,1068\,$\wr$	&20	&0	&4	&50	&4\\
NGC\,1144		&20	&1	&4	&80	&3\\
MCG\,-2-8-39		&20	&0	&0	&40	&44\\
NGC\,1194		&5	&1	&4	&90	&4\\
NGC\,1320\,$\wr$	&10	&1	&6	&70	&3\\
IRAS\,04385-0828	&10	&0	&6	&80	&7\\
ESO\,33-G2		&10	&1	&4	&50	&3\\
IRAS\,05189-2524	&20	&0	&4	&40	&0.1\\
NGC\,3660		&0.1	&0	&6	&0 	&30\\
NGC\,4388		&5	&0	&2	&70	&8\\
TOL\,1238-364		&20	&0	&0	&40	&11\\
NGC\,4968\,$\wr$	&15	&0	&0	&40	&14\\
MCG\,-3-34-64		&20	&0	&0	&50	&44\\
NGC\,5135\,$\wr$	&20	&0	&0	&80	&82\\
NGC\,5506		&5	&1	&4	&90	&7\\
NGC\,5995		&10	&1	&6	&80	&5\\
IRAS\,15480-0344\,$\wr$	&20	&0	&0	&40	&13\\
NGC\,6890		&15	&0	&0	&40	&23\\
IC\,5063		&10	&0	&0	&50	&24\\
NGC\,7130\,$\wr$	&20	&0	&0	&60	&45\\
NGC\,7172		&10	&0	&0	&90	&53\\
MCG\,-3-58-7		&15	&1	&6	&60	&6\\
NGC\,7496		&20	&0	&0	&50	&460\\
NGC\,7582		&5	&0	&0	&80	&25\\
NGC\,7590		&5	&1	&6	&80	&9\\
NGC\,7674\,$\wr$	&20	&1	&4	&50	&3\\
CGCG\,381-051		&20	&0	&2	&0 	&5\\
\enddata 

\tablecomments{Model results from the fitting procedure using 2pC two-phase
  models. As before, a $\wr$ indicates Compton Thick (CT) sources.}
\end{deluxetable}

To check how robust our results are we compare our best-fit parameters with
previous work which have studied common sources and which have also used the
\citet{nenkova08a} CLUMPY models to represent the infrared SEDs of nearby
AGN. Ramos-Almeida et al.\ (2009, hereafter RA09) and AH11 presented the
modeling of Type I, Type II and intermediate Type Seyfert galaxies, of which 5
objects are in common with our sample. AH11 adds N-band ground-based
spectroscopy to the SEDs, while AR09 uses only photometric data. They both use
a Bayesian inference algorithm (Bayesclumpy, Asensio-Ramos \& Ramos-Almeida,
2009) to determine the best fit values to a given SED and the probabilistic
distribution of the inferred model parameters. Notice also that AH11 limited
the range of $Y$ to values up to 30 for all objects, while the viewing angle
was restricted to $30-50$ for NGC\,5506 and $60-90$ for NGC\,1068, while our
only restriction was the $L^{\rm bol}_{\rm OIII}$ prior described in
\S\ 3.2.2.

For three of the objects presented by RA09 and AH11, NGC\,1068, NGC\,5506, and
NGC\,7172 the photometric measurements are common with this work (except for
the Q-band measurements reported by RA09). IC\,5063 and NGC\,7582, on the
other hand, have independent photometric observations. The ground-based
spectra presented by AH11 was obtained with much higher spatial resolution
than the {\emph{Spitzer}} and {\emph{ISO}} data used here. However,
examination of the spectra shows that the ground-base data are comparable to
the space-born observations. Hence the SEDs are not significantly modified by
the different spectral data.

Table \ref{comp_clumpy} presents the median and mode of the parameter
distributions determined by RA09 and AH11 and by this work for those objects
common to the samples. Our approach gives very similar results to those found
by AH11. The only significantly different model parameters are $q$ and
$\mathcal{N}_0$ for NGC\,1068. It can be seen that the largest differences are
found between the work of RA09 and AH11, due to the inclusion of spectroscopic
information around $10\mu$m. This is clear indication of the importance of
including detailed information of the SED around the silicate absorption
feature.

\subsubsection{2pC Fitting Results}

Table \ref{results2pc} presents the results from the fitting using the 2pC
models for the 27 considered sources. The $\chi^2$ value from the best fit is
also shown. About half of the sources have $\chi^2<10$ and we label these as
acceptable fits.

Examining Fig.~\ref{seds} it becomes apparent that the 2pC models mostly fail
to reproduce those SEDs that already represented a challenge for the CLUMPY
models, namely, those with a near-IR excess and those with a strong silicate
absorption accompanied by substantial near-IR emission. We have not attempted
to obtain new fits using the current library of 2pC models adding a black-body
component because of the rather limited range of parameter values available
when compared with the CLUMPY models.

Two parameters can be directly compared between the results obtained using 2pC
and CLUMPY models: the total optical depth at the equator of the torus and the
inclination angle subtended by the observer. Neither of these comparisons
yield a proper correlation. In fact, 2pC covers an optical depth range of
$\tau_{9.7} = 0.1-20$, while CLUMPY considers values has high as $\tau_{9.7}
\ga 100$.

Fig.~\ref{comp_ang} presents the comparison between the best fit inclinations
angles. Even if only acceptable 2pC fits are taken into account, the
distribution resembles a scatter plot. Again, notice that 2pC models do not
cover intermediate angles (smaller than 40 degrees but larger than 0) and the
opening angle of the cloud distribution is fixed to 50 degrees.

Therefore, it seems that the most important drawback from the 2pC models is the
narrow range of parameters so far explored. 

\begin{figure}
  \begin{center}
    \includegraphics[scale=0.35,trim=50 50 0 0]{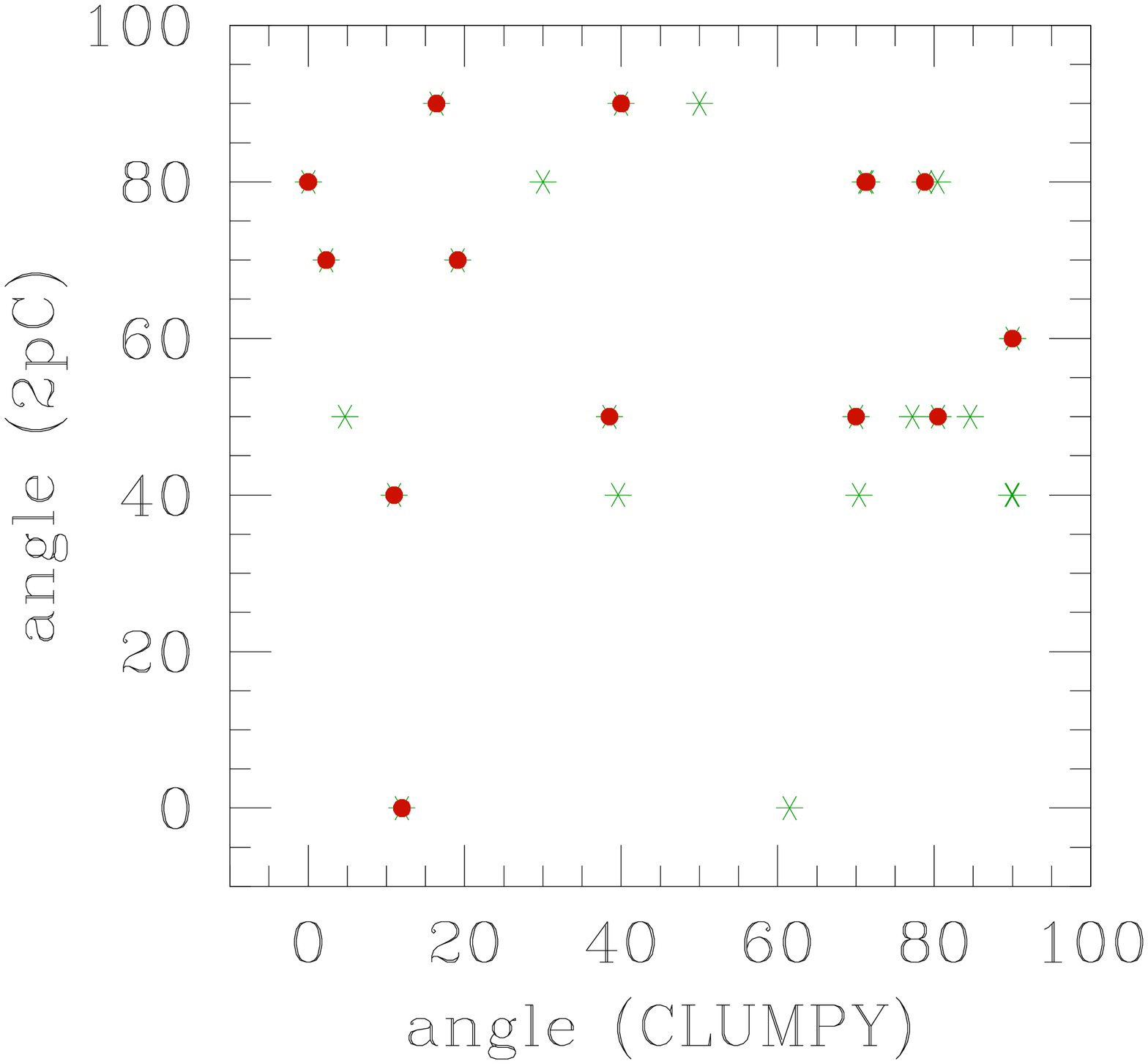}
  \end{center}
  \figcaption{Comparison between inclination angles obtained using the CLUMPY
    and 2pC models. All 27 sources are plotted with an asterisk. Sources with
    2pC fits with values of $\chi^2 < 10$ are also shown with a circle.
    \label{comp_ang}}
\end{figure}

\subsection{Sources with near-IR excess \& strong Si absorption}

As already discussed, several sources were identified in Paper I as having a
near-IR up-turn (i.e., $\alpha < 1$; see also Table \ref{gralinfo}) and their
SEDs were fitted using an additional black-body component, which considerably
improved the quality of the fits. The required temperatures are generally very
high ($T \sim 1700-2500$ K) and the emission typically peaks at around
2$\mu$m. Only 2 sources require lower temperatures ($T \sim 1200$ K):
NGC\,4968 and NGC\,5506.

Several other works have found necessary to use an additional component to
explain the excess of near-IR emission observed in luminous Type I QSOs with
respect to the CLUMPY torus models (Mor, Netzer \& Elitzur 2009; Mor \&
Trakhtenbrot, 2011; Deo et al.~2011; Landt et al.~2011). All these works find
that the black-body emission is characterized by $T \sim 1200-1400$K, while we
find that the required temperatures are much higher, with $T \sim 1700-2500$
K. These values are too high to interprete this component as emission from hot
dust and therefore its nature remains unexplained. Notice, however, that given
the difficulties in isolating the AGN near-IR emission in obscured Seyferts
compared with the dominant AGN emission in luminous QSOs might contribute to
the problem. 

Of the 12 sources with $\alpha < 1$ in Table \ref{gralinfo}, 3 correspond to
radio-loud sources and were also fit with an additional power-law component
(\S 3.3.4). IRAS\,01475-0740, however, had not enough mid-IR information to
properly restrict the fitting. NGC\,4501 seems dominated by a power-law and
the inclusion of this additional component allows for a proper fit to the SED,
as seen in Fig.~\ref{seds}. This component, however, heavily dilutes the torus
emission, and it is not clear whether the derived parameters are
representative of its intrinsic emission.  The last radio loud source,
NGC\,7496 can be successfully fit using a black-body or a power-law as
secondary component. Fig.~\ref{seds} shows the results from the fitting using
a black-body component, while Table 3 shows the best fit parameters for both
cases. It can be seen that the results are consistent with each other.

In \S 3.3.5 it was also discussed that sources presenting deep silicate
absorption were fitted using an additional black-body component to supply the
near-IR flux lacking in CLUMPY torus models that have a strong 9.7$\mu$m
absorption features, which is the case of NGC\,5506, NGC\,7582 and
NGC\,7172. 

A good fit was obtained for NGC\,5506 and NGC\,7582, as can be seen in
Fig.~\ref{seds}. In Table 3 we report results for fits with and without the
black-body component. For NGC\,7172, on the other hand, no possible
combination would reproduce both, the extremely deep 9.7$\mu$m absorption
feature and the near-IR photometry. However, the best fit 2pC model provides a
reasonable fit for this source. It corresponds to a model with a 90 degree
inclination angle and a large $\tau_{9.7}$ value.  Large $\tau_{9.7}$
corresponds in fact the regime where the two-phase medium introduces the
largest departure from the clumps-only models. The CLUMPY results are also
consistent with a large optical depth and high inclination angles.

Fig.~\ref{seds} shows that other sources require additional near-IR flux,
other than those with a index $\alpha < 1$ or strong Si absorption: NGC\,4968,
NGC\,5995, and MCG\,-3-58-7. A very good fit was found for NGC\,4968 with a
CLUMPY model and a black-body component with a temperature of $\sim 1200$
K. The fits to MCG\,-3-34-64 and MCG\,-3-58-7 are rather poor in the near-IR
region, but the fit provided by the 2pC models to MCG\,-3-58-7 is quite good
($\chi^2 = 6$). Table 3 presents the results using CLUMPY models with and
without the additional black-body component for all these sources.

It is very encouraging that the inclusion of a black-body component does not
have a significant impact on the torus parameters derived from the fitting
process using CLUMPY models. This can be seen in the middle panel of
Fig.~\ref{params}, but also when examining individual fits reported in Table
\ref{resultsc}. This is in contrast with the work of Deo et al.~(2011), who
found significant changes in the best fit parameters when introducing a
black-body component to their fits. This could be due to the Type I nature of
their sources and therefore a different level of constraining coming from the
9.7$\mu$m silicate feature in their sample of QSOs. The largest changes are
seen in parameters $q$ and $Y$. One of the most interesting findings from this
work is that most QSOs require large inclination angles and a small number of
clouds. 

In summary, to solve for the lack of sufficient near-IR emission in CLUMPY
torus models, we added an additional black-body component to $\sim 50\%$
(14/27) of our sources and obtain acceptable fits in most cases. However, the
temperatures of these components are too high to correspond to dust emission,
and therefore its true nature is unclear. 2pC models can provide better
results for some of these sources, but the limited parameter space currently
explored by these models does not allow to ascertain that the two-phase
approach is a definite solution to the lack of near-IR emission in observed
SEDs.

\section{Analysis}

In the previous section it has been established that the results obtained
using CLUMPY models are quite robust to the addition of a black-body component
in the near-IR. Because of the large parameter space explored by these models
and the well restricted results obtained for most parameters in \S 4.2.1 in what
follows we will only consider CLUMPY results for our analysis.

We have compiled ancillary data for our sample from the literature. Table
\ref{gralinfo} presents results on the detection of Hidden Broad Line Regions,
    [OIII] fluxes and luminosities, hydrogen column densities determined from
    X-ray observations and hard (2-10 keV) fluxes, radio fluxes, and Balmer
    decrements as determined in Paper I.

\subsection{The Inferred Hydrogen Columns}

One of the most powerful diagnostics to characterize the different classes of
AGN is the inferred hydrogen column density, $N_H$, as determined from the
photoelectric cut-off experimented by the power-law X-ray spectrum emitted by
the central source. The observed values of $N_H$ probes the amount of material
along the line of sight towards the active nuclei and correlate strongly with
other diagnostics to determine the Seyfert type. In fact, it has been shown
that while Type I objects suffer from little absorption, Type II systems
usually present absorbing columns of $10^{22}$ cm$^{-2}$ or more (Smith \&
Done 1996; Turnet et al.~1997; Maiolino et al.~1998; Bassani et al.~1999).

However, measurements of the column $N_H$ towards the central region obtained
through other methods, such as the ratio $H_{\alpha}/H_{\beta}$ for broad
emission lines (since narrow lines would probe the extinction affecting the
much more extended narrow emission line region), typically give smaller values
of $N_H$ than those obtained from X-ray observations (Maccacaro et al., 1982;
Reichert et al., 1985). The optical depth can also be inferred using key
features in the extinction curve, such as the 2200\AA\ 'bump' and the silicate
absorption features in the mid-IR.

Possible solutions for these differences have been postulated: a dust free
inner region (interior to the torus) could be responsible for the excess
column probed by X-rays; an anomalous $H_{\alpha}/H_{\beta}$ ratio could be
due to the collisional effects present in the high-density clouds found in the
broad line region; the line of sight probed by the X-rays could be
significantly different to that probed by other estimators; Maiolino et
al.~(2001ab) and Gaskell et al.\ (2004) argue that AGN environments might have
a different dust size distribution, either because of the presence of larger
grains or because small grains are depleted, although other works argue for
normal dust properties in AGN (Mason et al., 2004; Nenekova et al.,
2008b). X-ray absorption by dust-free material within the sublimation radius
of the torus should not be significant as this ionized gas would produced very
intense narrow emission lines which are not observed (Maiolino et al.~2001a).

Shi et al.\ (2006) have shown that there is a broad but clear correlation
between the strength of the $9.7\mu$m silicate feature and the $N_H$ columns
derived from X-ray data. In very broad terms, unabsorbed systems show
$9.7\mu$m feature in emission while absorbed systems show it in absorption. In
Fig.~\ref{comp_NHs} we compare the column density $N_H$ along the line of
sight derived from our SED fitting with those obtained from the photoelectric
cut-off from X-ray observations. Arrows show objects for which $N_H >
10^{24-25}$ cm$^{-2}$ upper limits have been determined. One system has a $N_H
> 10^{22}$ cm$^{-2}$ upper limit coming from observations in the soft
X-rays. No clear trend is found in the plot, and we are not able to reproduce
the findings of Shi et al.\ (2006). Crucially, we do not find a systematic
offset between the two measurements. We also show the combination of
parameters $\mathcal{N}_0$ and $\tau$ to yield the total Hydrogen column $N_H$
(assuming a gas-to-dust ratio $N_H / A_v = 1.79\times10^{21}$ cm$^{-2}$
mag$^{-1}$) at the torus equator and along the line-of-sight.

The presence of clumpy media around the active nucleus seems to be the best
way to interpret our results. While our SEDs probe the average conditions of
the dusty medium {\it in emission}, the X-rays probe a particular line of
sight towards the nucleus {\it in absorption}. Rapid and dramatic changes of
the X-ray $N_H$ in NGC\,1365 and NGC\,4388 (Risaliti et al., 2005; Elvis et
al., 2004) seem to validate this scenario (although Elvis et al.\ oppose it).

So, can we consider our column densities more representative of the average,
long term conditions of the physical conditions of the duty torus? Probably
yes, but these will have to be revised as new models and better observations
come along in the future. Since from the model fitting no $N_H$ values are
found outside the $10^{22.5-25}$ cm$^{-2}$, this might be a reasonable range
of columns to be adopted as representative of the average values for Type II
Seyferts.

\begin{figure}
  \begin{center}
    \includegraphics[scale=0.35,trim=0 0 0 0]{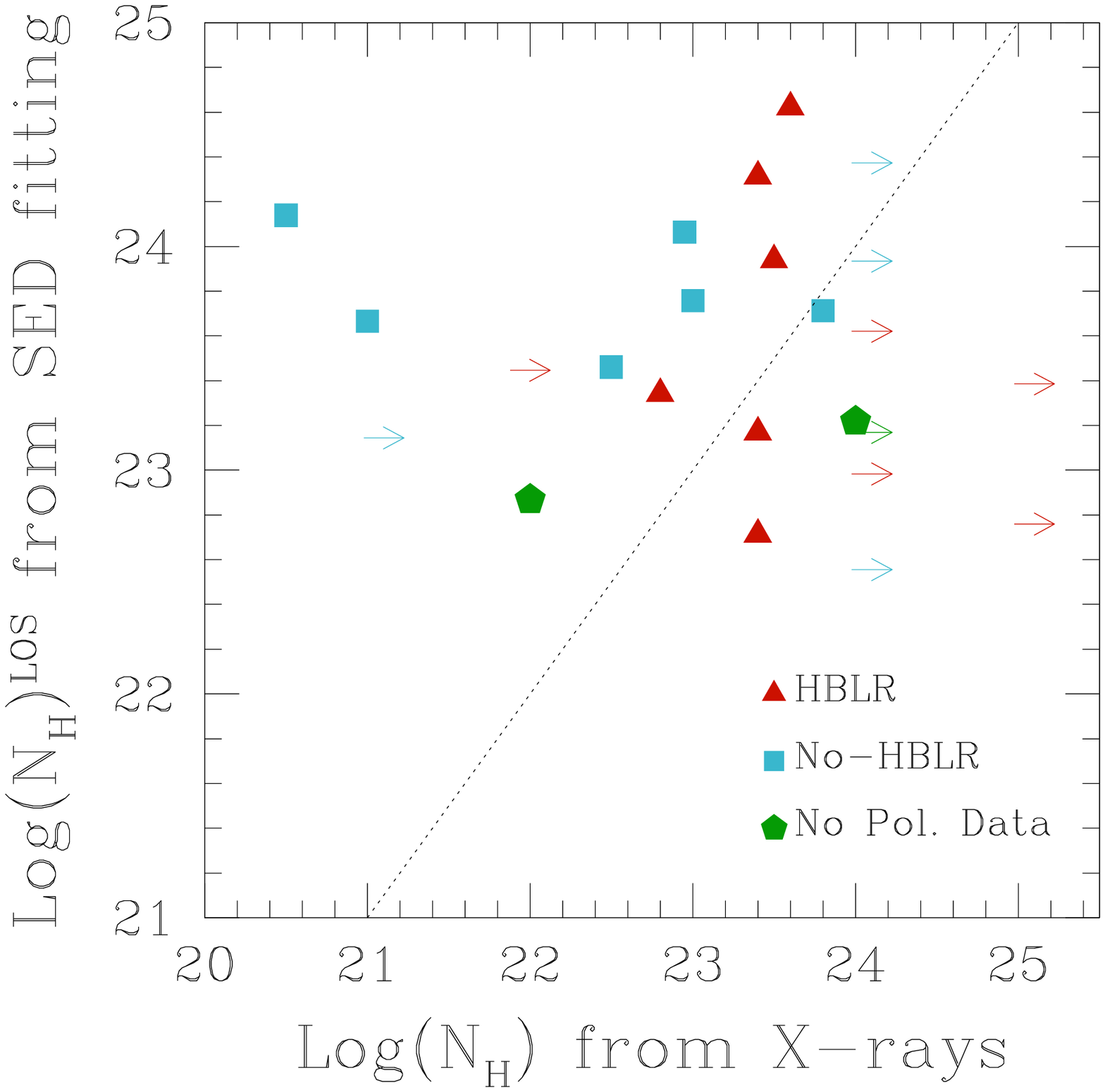}
    \includegraphics[scale=0.35,trim=0 0 0 0]{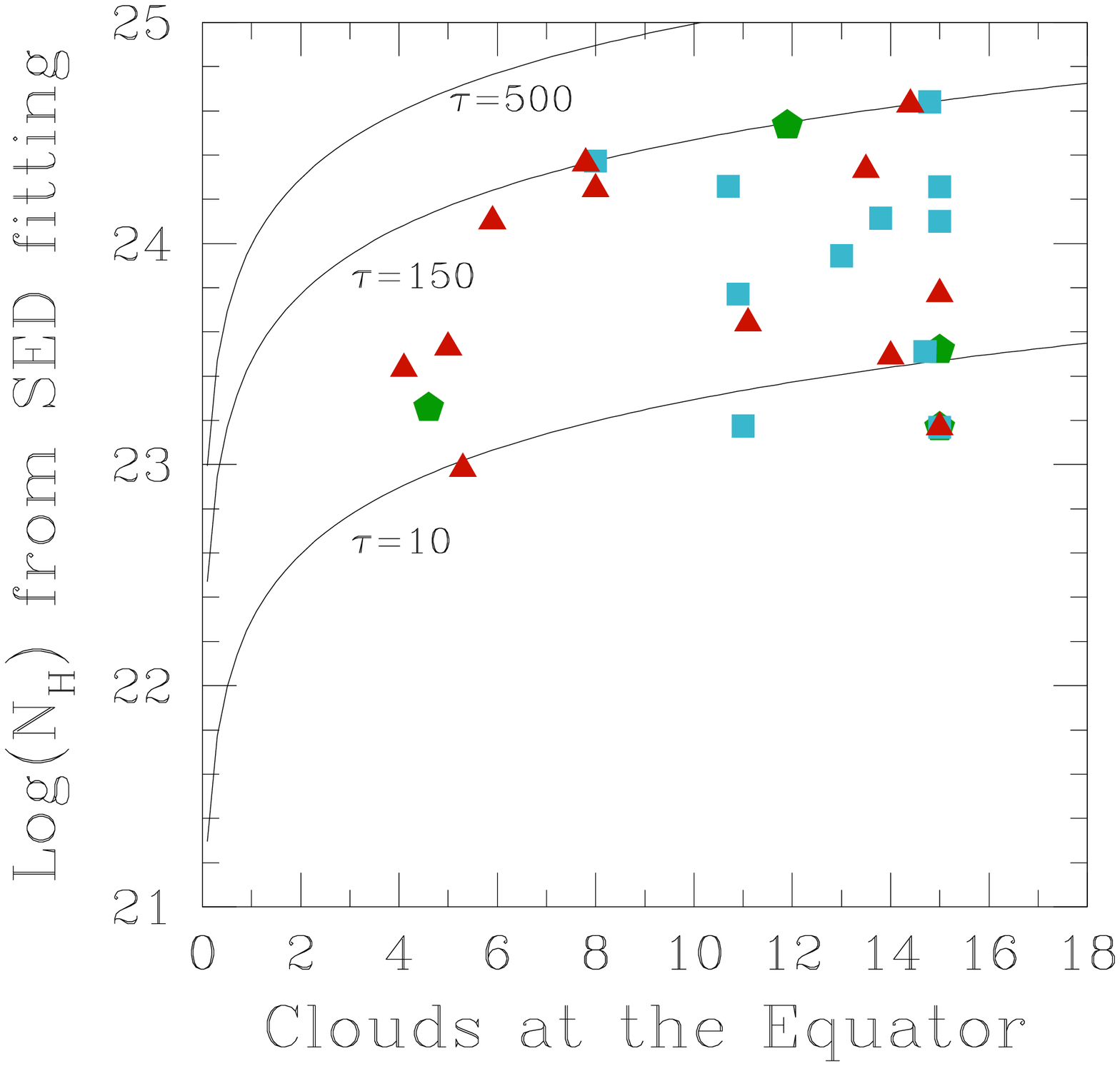}
    \includegraphics[scale=0.35,trim=0 0 0 0]{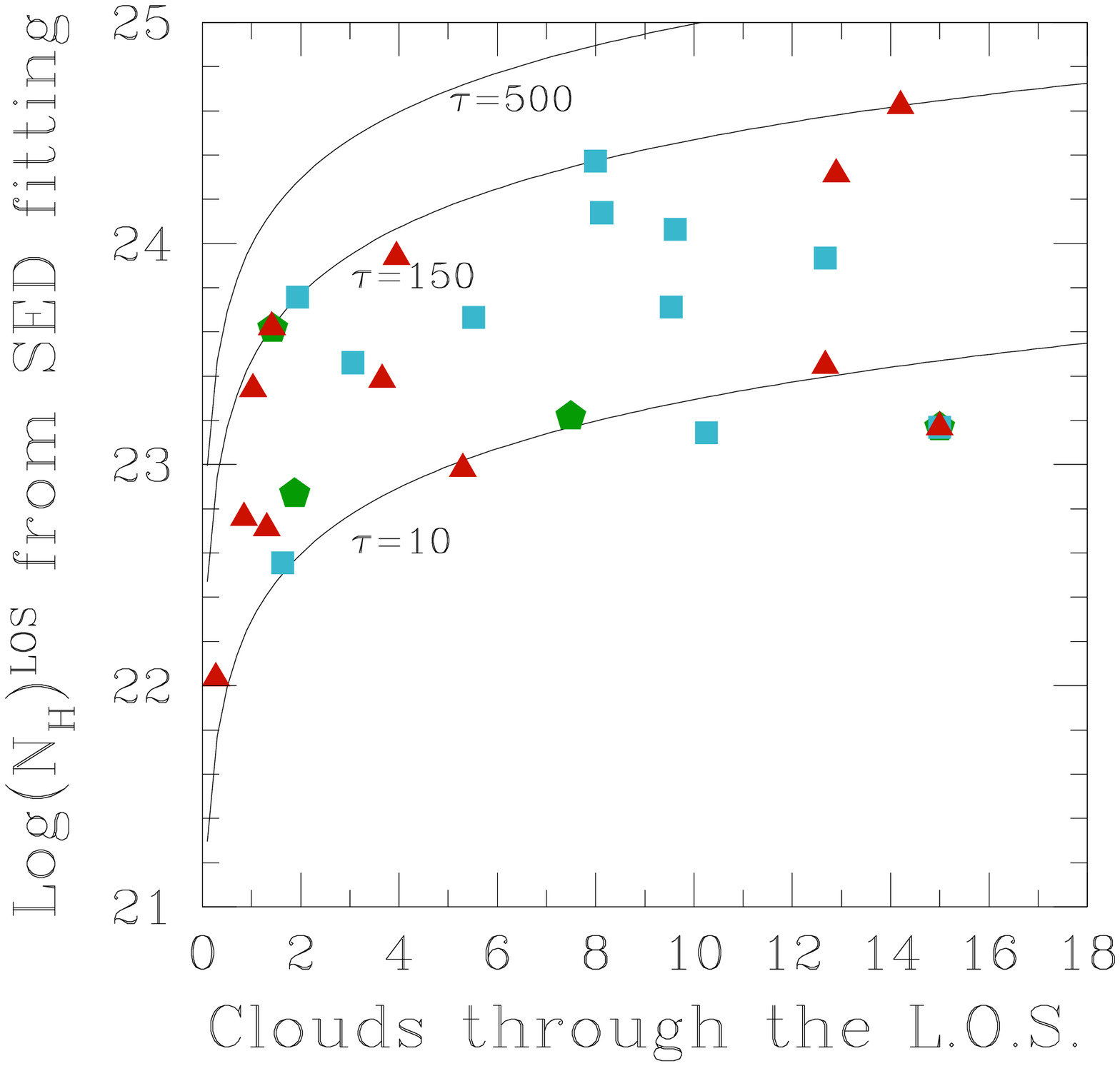}
  \end{center}
  \figcaption{Top panel: comparison of column densities along the
    line-of-sight, $N_H$ (cm$^{-2}$), derived from the CLUMPY emission models
    for the torus and derived through X-ray observations.  Middle panel:
    distribution of the number of clouds at the torus equator
    ($\mathcal{N}_0$) versus the total $N_H$ column derived from our fittings,
    also at the equator. Continuous lines correspond to the total opacity for
    different optical depths of individual clouds ($\tau = 10, 150, 500$).
    Bottom panel: same as above but this time showing the number of clouds
    ($\mathcal{N}(\beta)$) and the $N_H$ column along the line-of-sigh.
    \label{comp_NHs}}
\end{figure}

Compton Thick nuclei are defined as those where the X-ray derived $N_H$
columns are in excess of 10$^{24-25}$ cm$^{-2}$. The current compilation of
$N_H$ columns presented in Table \ref{gralinfo} shows that of the total number
of objects with measured column densities (30 out of the 39 galaxies in the
full sample), 11 are Compton-Thick. This is in line with previous findings
(Risaliti et al., 1999; Bassani et al., 1999; Bassani et al., 2006).

Out of the 11 CT sources found in our sample, 8 have acceptable SED fits: the
bona fide CT galaxy NGC\,1068, NGC\,1320, Tol\,1238-364, NGC\,4968, NGC\,5135,
F\,15480-0344, NGC\,7130 and NGC\,7674. The inferred angle that the torus axis
subtends with our line of sight are in the 70--90 degree range for 7 objects,
while for NGC\,1320 the angle found is rather small ($\sim 20$). See Table 3.

AH11 has shown that the probability for photons to scape the obscuring
material can still be low for systems with intermediate inclinations, provided
that the angle subtended by the torus is fairly large. This is the case for
most of our Seyfert II systems which present rather large values of $\sigma$.
The exceptions are Tol\,1238-364, IRAS\,15480-0344, NGC\,6890, NGC\,7130,
MCG\,-3-58-7 and NGC\,7674. Interestingly, 4 of these sources are CT
systems. It should be remembered, however, that the X-ray derived $N_H$
columns measure a very specific line-of-sight property of the central region,
as clearly validated by those objects with variable columns (Risaliti et al.,
2005; Elvis et al., 2004).

Elitzur (2012) has recently proposed that Type I and Type II nuclei are
examples of objects preferentially drawn from the two ends of the
distribution of torus covering factor, with Type II sources being examples of
particularly puffed-up tori, characterized by large values of $\mathcal{N}_0$
and $\sigma$. This is clearly supported by our results.

\subsection{The presence of Hidden Broad Line Regions.}

The presence of a Hidden Broad Line Region is the most clear indication that
at least some Type II AGN have broad emission lines. About $40\%$ of Seyfert 2
galaxies in the 12$\mu$m Sample are found to have HBLRs, in agreement with the
fraction found in optically selected Seyfert 2 samples (Tran 2003). The
fraction is closer to $50\%$ for the 27 objects with good SED
fittings. Unfortunately, there are significant differences in the sensitivity
of the spectropolarimetric studies found in the literature and since the data
are not provided in many of these works it is not possible to visually verify
the presence or absence of the HBLRs.

The reasons behind the lack of detection of a HBLR in some Type II sources,
have been a matter of heated debate (Heisler, Lumsden \& Baily, 1997;
Alexander 2001; Gu et al., 2001; Tran 2001; Lumsden \& Alexander, 2001; Gu \&
Wang; Tran 2003; Deluit 2004; Lumsden, Alexander \& Hough, 2004; Zhang \&
Wang, 2006; Shu et al., 2006; Haas et al., 2007). Some works advocate that the
non-detection of a HBLR is due to the presence of a dominant galaxy component
that dilutes the emission from the active nucleus. Others interpret the
observations as evidence for the existence of a different class of AGN where a
BLR is not present.

It could be imagined that in objects that truly lack a broad line region, no
dusty structure needs to be present either, but this does not have to be the
case. The non-HBLR objects might have an active nucleus {\em and} a torus,
lacking only the BLR. Evidence seems to support this: Haas et al.,~(2007)
looked at the mid-IR properties of a sample of Seyfert galaxies as obtained
with high spatial resolution images and found that the nuclear properties of
the 12$\mu$m/[OIII] ratio showed no distinction between sources with or
without a HBLR. This seems to suggest that despite the presence or absence of
the HBLR, hot dust is still present in the nuclear region of both types of
sources.

Besides, Tran (2003) finds that although the IRAS $25\mu$m/$60\mu$m
color and the luminosity of the AGN are well correlated with the
presence of a HBLR, the level of extinction towards the nuclei, is
similar in both types of sources.

We find some indications that sources with and without a HBLR might have
systematic differences in their infrared emission. Bottom panels in
Fig.~\ref{params} suggest that for sources with a HBLR the torus might be less
extended than for sources with an undetected HBLR. However, as already
discussed in \S\ 5.2, the parameter $Y$ is not well constrained by the fitting
process of our data.

Sources with a HBLR also might have systematically smaller line-of-sight
extinction values than sources with an undetected HBLR. Eq.~2 states that the
number of clouds along the line-of-sigh depends on the number of equatorial
clouds ($\mathcal{N}_0$), the thickness of the torus ($\sigma$) and the
inclination angle ($\angle i$). Fig.~\ref{params} shows that $\sigma$ is very
similar for sources with and without a HBLR, while $\angle i$ does not show a
statistically significant difference. The number of clouds, on the other hand,
presents a more clear differences between sources with and without a
HBLR. This can also be seen in Fig.~\ref{comp_NHs} where nuclei with a HBLR
cover a wide range of number of clouds at the torus equator, while nuclei
without a HBLR cluster at the higher end of the distribution. However,
Fig.~\ref{comp_NHs} also shows that these differences are much less clear when
looking at the number of clouds along the line-of-sight, $\mathcal{N}
(\beta)$, a parameter much closely related to the line-of-sight $A_V$
(Eq.~3). Kolmogorov-Smirnov (KS) tests in fact do not confirm that the
perceived differences are statistically significant.

If further evidence that the different distributions of the line-of-sight
$A_V$ for sources with and without a HBLR are different is found, we need to
explore some possible explanations. At face value this contradicts a strict
unification scheme, where the only difference between Type I and Type II
sources is the view-point of the observer. However, is becoming clear that a
strict unification scheme is not plausible (Elitzur 2012).

We can postulate that a larger number of clouds obscures the polarized
emission from the BLR. This would require a rather compact scattering region,
with a comparable scale height as that of the torus itself. This result is in
agreement with the analysis presented in Lumsden et al.~(2004) where it is
claimed that the fraction of Seyfert galaxies with a HBLR is larger when
looking at only Compton-thin nuclei (as determined by X-ray observations),
implying a sample of sources with less obscured central regions.

\subsection{Torus Sizes and Masses}

The CLUMPY modeling does not provide an absolute torus size, but instead, the
parameter $Y = R_{out} / R_{in}$. While most of the fits favor $Y < 50$, some
sources can have very extended torus with $Y \la 100$ (see Fig.~\ref{params}).

Clumpy torus models are characterized by clouds showing a range of
temperatures for a given distance from the central source (Nenkova et al.,
2008b; Schartmann et al., 2005). This is in sharp contrast with the
predictions from a continuous dust distributions, where a unique temperature
is found as a function of radius. Hence, the SED shapes for CLUMPY models are
not very sensitive to the $Y$ parameter or the size of the torus. Nenkova et
al.\ (2008b) show that observations below 5 $\mu$m will not be able to
distinguish between $Y$ values, irrespective of the cloud distribution, which
is determined by the $q$ parameter. At longer wavelengths some differences can
be appreciated for a flat cloud distribution ($q\sim1$), but these are only
significant for wavelengths above $15\mu$m and therefore are not well probed
by our observational SEDs (see also \S5.5).

Following the results from Suganuma et al.\ (2006) we can assume that $R_{in}$
is indeed set by the dust sublimation radius as $R_{in} = 0.4 \sqrt(L^{\rm
  bol}/10^{45})$ pc, where $L^{\rm bol}$ is in units of ergs/s. As already
explained, we have used the [OIII] luminosities listed in Table \ref{gralinfo}
as a prior for the intrinsic nuclear luminosities in our sources and found
that the torus inner radii vary between 0.05 and 1 pc.

Torus outer sizes ($R_{out}$) are presented in Fig.~\ref{massize}. As
it can be seen there is a wide distribution of $R_{out}$, but most of
them are below 5 pc in extent, with some torus being as small as 0.1
pc.

The total mass of the torus can be estimated as \citep{nenkova08a}:

\begin{eqnarray}
M_{\mathrm{torus}} &=& 4 \pi m_H \sin(\sigma) N_H^{eq} R_{in}^2 Y I_q(Y)
\end{eqnarray}

where $N_H^{eq}$ is the equatorial column density of the torus, $R_{in}$ is
calculated as the dust sublimation radius, $\sigma$ and $Y$ are parameters of
the model, and $I_q$ is 1 if $q=2$ or 3, $Y/2lnY$ if $q$=1, or $Y/3$ if
$q=0$. We find some torus masses, up to $10^7 M_\odot$, as shown in
Fig.~\ref{massize}. However, in most cases these are driven by large $Y$
values, a parameter not well constrained by the best fitting results. Still,
Siebenmorgen et al.,~(2005) reported the dust masses implied by a simple model
of the dust emission in 2 quasars, being of the order of $10^{6-7}
M_\odot$. Fritz et al.,~(2006) reported on the masses implied by the smooth
modeling of the emission of Type I and Type II Seyferts, ranging from $\sim70
- 10^7 M_\odot$. Our results show a strong peak, however, with most masses in
the $\ga 10^{4} M_\odot$ range, as also seen by AH11.

\begin{figure}
  \begin{center}
    \includegraphics[scale=0.3,trim=0 0 0 0]{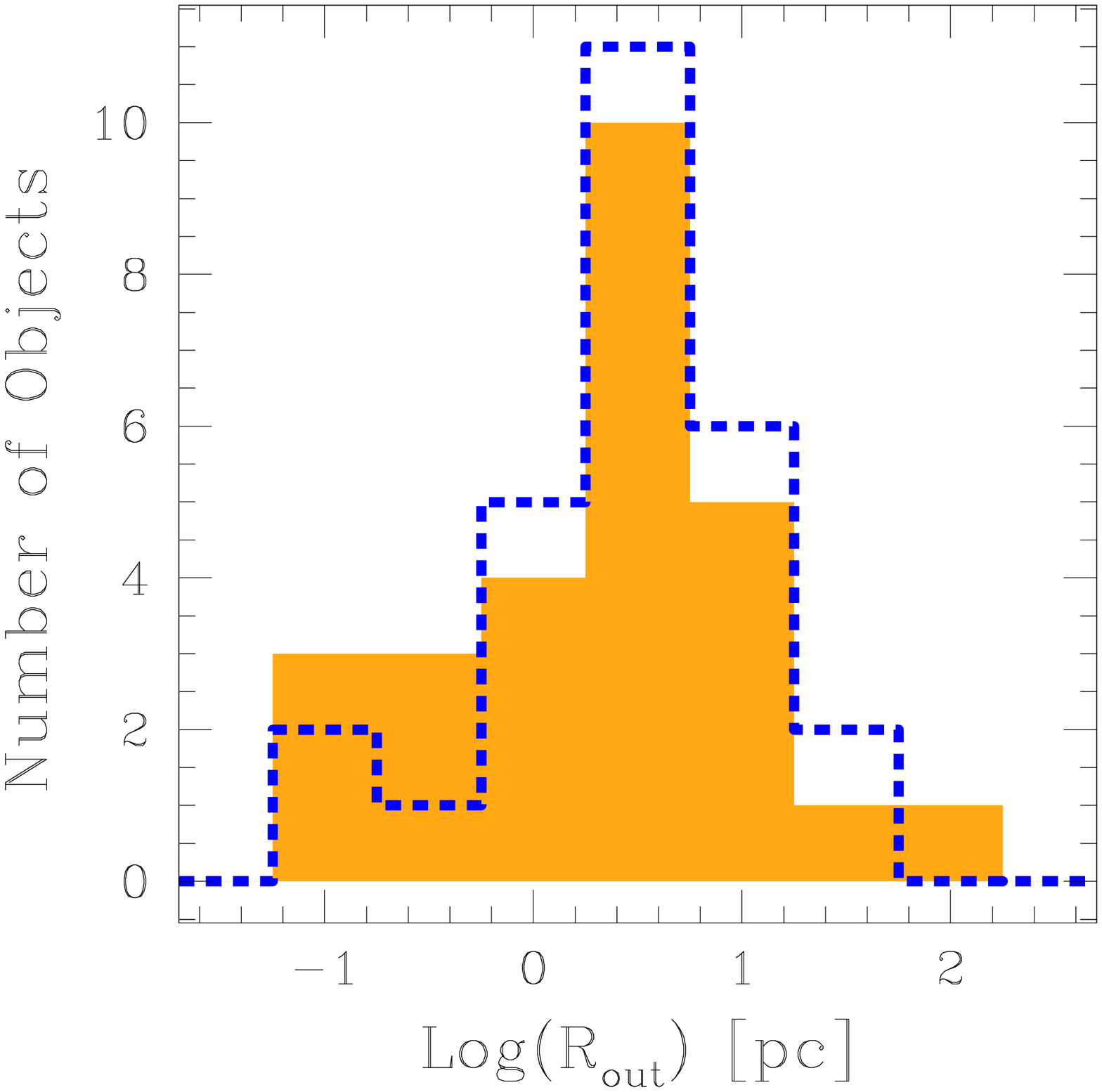}\\
    \includegraphics[scale=0.3,trim=0 0 0 0]{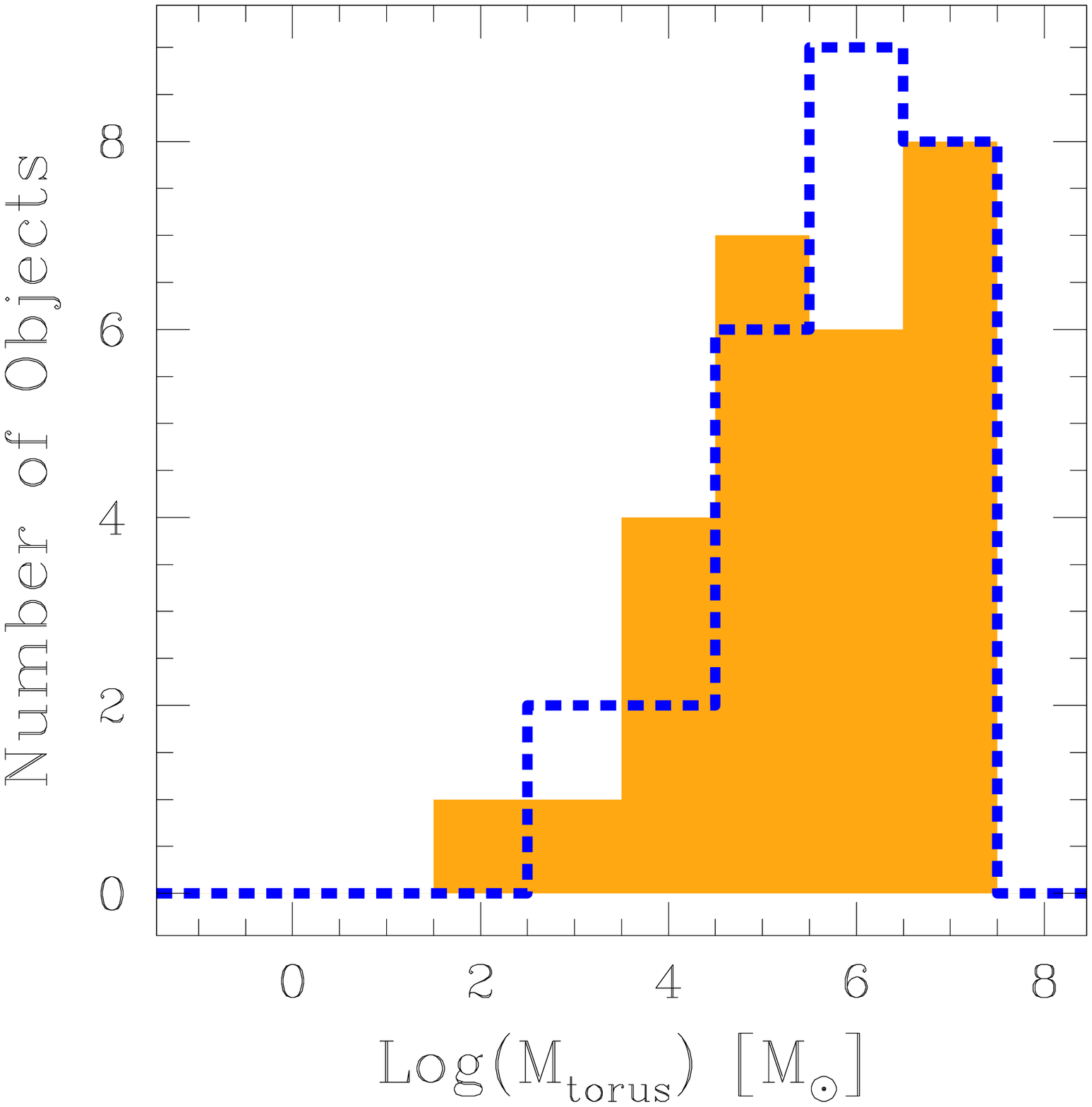}
  \end{center}
  \figcaption{Distribution of torus out radii ($R_{out}$) and masses
    ($M$). The median distribution is shown in solid color while the mode
    distribution is shown with a dashed line.\label{massize}}
\end{figure}

\subsection{Comparison with Interferometric Observations.}

Infrared interferometric observations are providing the ultimate way to
observe the putative dusty torus in AGN. Unfortunately, this technique is
limited to the brightest sources and the observations required are very
expensive and difficult to obtain. 

Until now long-based interferometry has been carried out for a couple of dozen
Type I and II AGN (Wittkowski et al., 1998; Weinberger et al., 1999; Swain et
al., 2003; Weigelt et al., 2004; Jaffe et al. 2004; Wittkowski et al., 2004;
Meisenheimer et al., 2007; Tristram et al., 2007; Beckert et al., 2008;
Kishimoto et al., 2009ab; Raban et al., 2009), with the largest sample found
in Tristram et al.\ (2009). Four objects in this last work are in common with
our sample: NGC1068, F05189-2524, NGC5506 and NGC7582, however the
observations for NGC5506 and NGC7582 did not provide useful data, while for
F05189-2524 only a very faint fringe detection was possible. Also, it has been
realized that interferometric studies yield more unambiguous results in Type I
sources (see discussion in Kishimoto et al.\ 2011). Therefore, a one to one
target comparison is not possible.

However, for most of those sources where interferometric observations
have provided restricting results, the sizes of the resolved
structures observed in the near and mid-IR are of the order of a few
parsecs (see references above). These are already in good agreement
with the results from the dust reverberation determined by Suganuma et
al.\ (2006), where the inner face of the torus is found at the dust
sublimation radius which directly depends on the luminosity of the
central source, and with the results derived in this work.

\subsection{Radio-loudness}

Ho (1999) was the first to notice that radio-loudness is a function of the AGN
bolometric luminosity, with low-luminosity objects (those below $\sim 10^{43}$
ergs/s) being more likely to have a radio-loud central source. We can see this
trend in Fig.~\ref{radio}, which includes all sources with $L^{bol}$ and $R_L$
measurements. It can be seen that the probability of being radio-loud
increases with $L^{\rm bol}$.

It is interesting to see that for those radio-loud sources with acceptable
fits (F\,04385-0828 and NGC\,7496) a very small inclination angle to the line
of sight is derived from the SED results. This is in line with an orientation
effect to be responsible for the boosting of the radio emission, although
theoretical predictions state that the scaling between the radio and the
optical output should depend weakly on the relativistic Doppler factor
\citep{falcke04}. Further data would be needed to confirm this finding.

In Fig.~\ref{radio} we also include information about objects with HBLR
detections. There seems to be a trend for more luminous nuclei to show the
presence of polarized BLRs as already noticed by other works and in line with
the hypothesis that weaker nuclear sources are out-shined by the stellar
components (see references in \S 5.2).

\begin{figure}
  \begin{center}
    \includegraphics[scale=0.35,trim=0 0 0 -50]{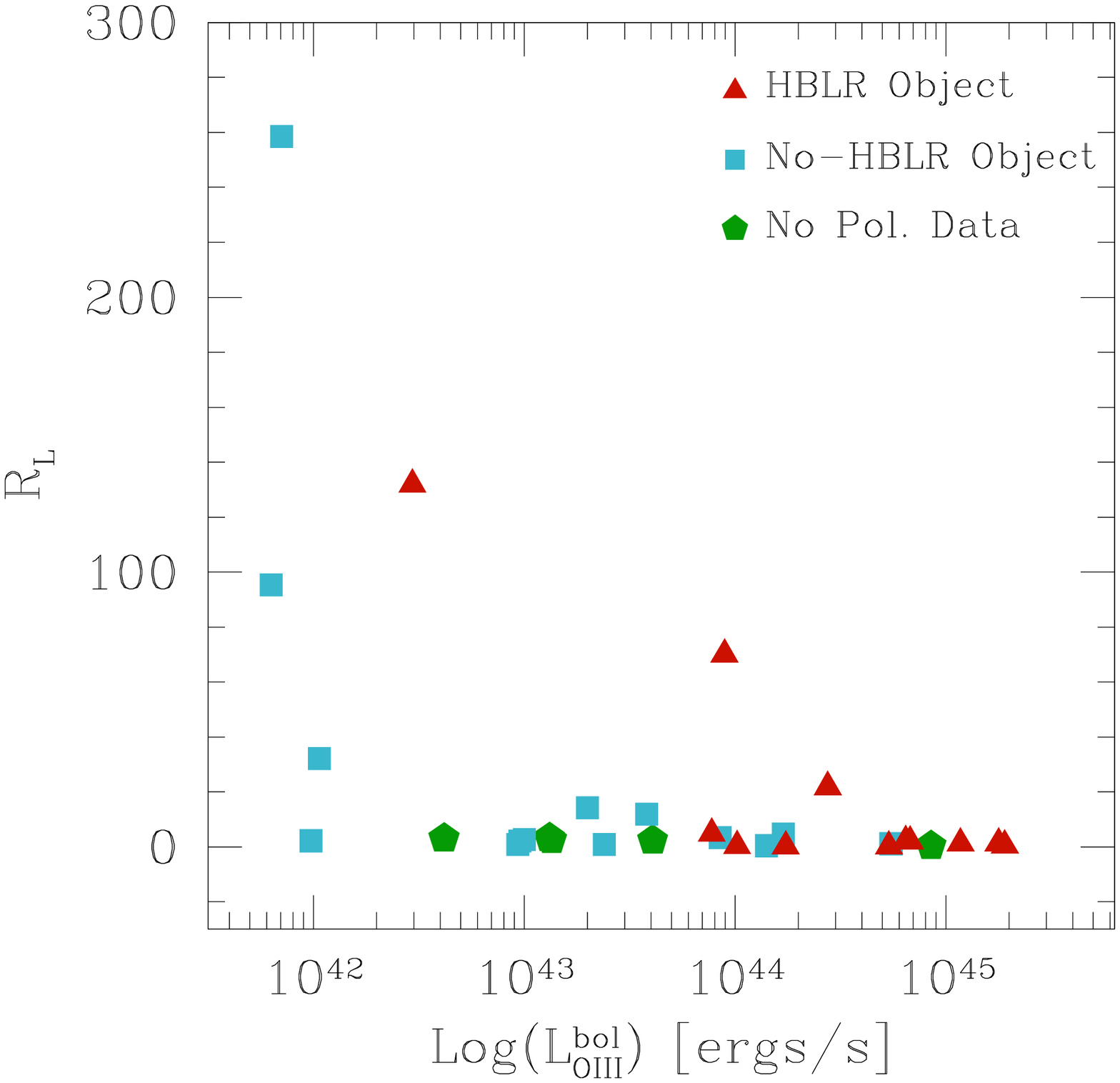}
  \end{center}
  \figcaption{Radio-loudness, as defined by the parameter $R_L$ versus
    bolometric luminosity. The presence of a Hidden Broad Line Region is also
    indicated.
\label{radio}}
\end{figure}

\subsection{Correlation with Starformation}

\begin{figure}
  \begin{center}
    \includegraphics[scale=0.35,trim=0 0 0 -50]{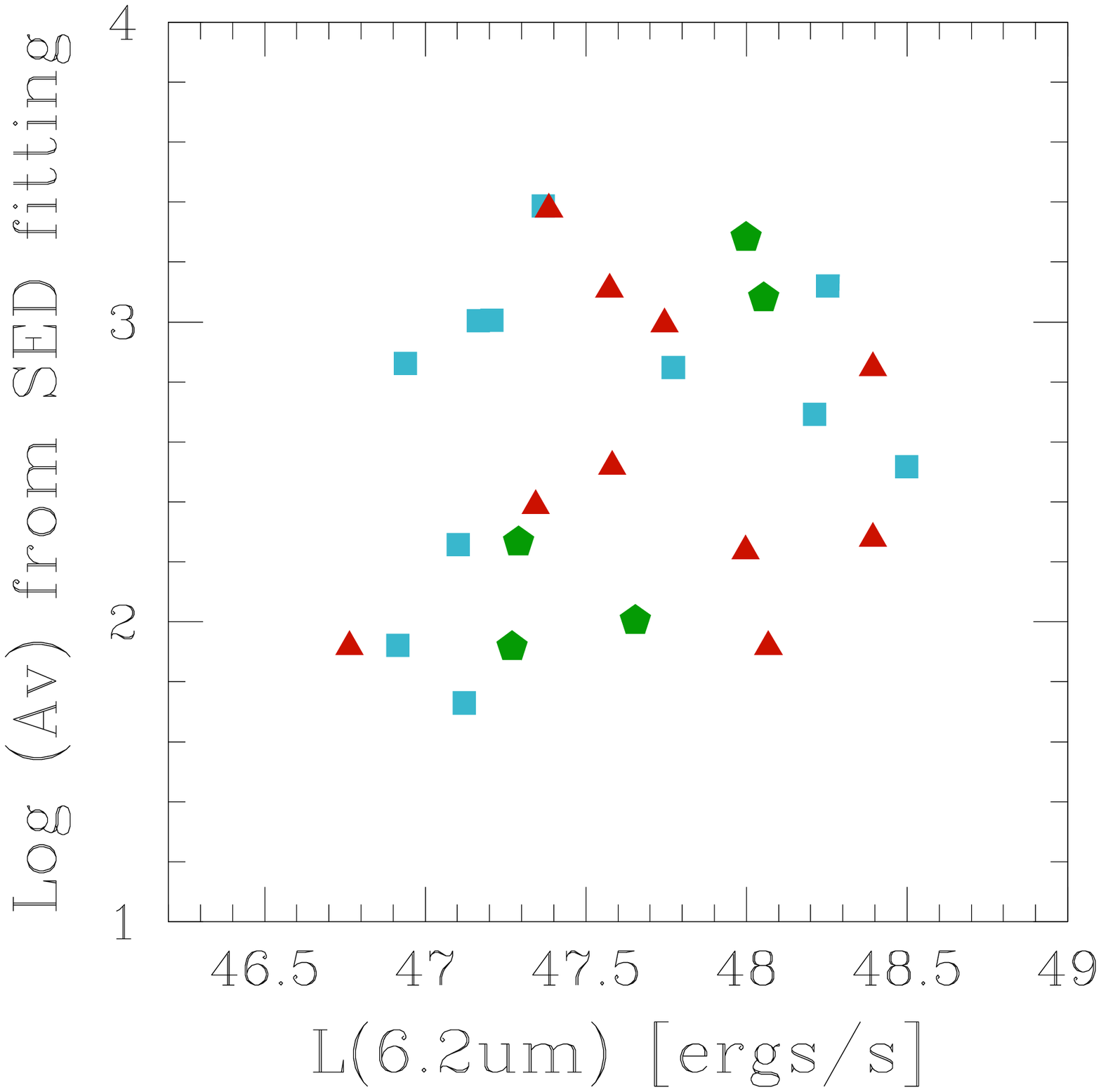}
  \end{center}
  \figcaption{Luminosity of the 6.2$\mu$m PAH feature, as a tracer of star
    forming activity, versus total obscuration at the equator of the modeled
    torus. Symbols are the same as Fig.~\ref{radio}. \label{sb_av}}
\end{figure}

One important subject to discuss is whether the presence of nuclear
obscuration is related to the level of star formation in the nucleus vicinity.
Taken at face value this scenario opposes the Unified Model, since in
principle the only difference between a Type I and Type II source is the
viewing angle. However, this is true only if all tori are exactly the same.

As already discussed, Elitzur (2012) has recently proposed that Type II
objects are examples of particularly puffed-up tori, with larger
$\mathcal{N}_0$ and $\sigma$ values. In turn this could result in a larger
mass of cold dust, because of larger shadowing from the central source which
allows for dust to cool more efficiently at large radii.

In Paper I we have used diagnostic diagrams to determine the presence of
star-formation in the nuclear region. Essentially, there is no correlation
between the level of starformation and the parameters that control the
thickness of the torus, $\mathcal{N}_0$ and $\sigma$.

We can also look at the more extended star-formation using the observed
luminosity of the 6.2$\mu$m PAH feature. No evidence for such correlation is
found between the strength of the PAH feature and the amount of extinction
determined from the SED fitting, as can be seen in Fig.~\ref{sb_av}.

\section{Summary and Conclusions}

We have performed the fitting of the near and mid-IR SEDs of a sample of 36
Seyfert II galaxies using CLUMPY and 2pC models developed by Nenkova et
al.\ (2002, 2008ab, 2010) and Stalevski et al.~(2012). Adequate fits were
reported for 27 sources.\\

Our conclusions are the following:

\begin{itemize}

\item Observations in the mid-IR, and in particular N-band spectroscopy of the
  sources, are crucial to perform an adequate fitting to the SEDs.

\item The use of the bolometric luminosity of the source as a prior during the
  fitting is also an important tool to constrain the best-fit results.

\item We find that the best-fit parameters for the CLUMPY models correspond to
  $\mathcal{N}_0 \ga 5$, $\sigma \ga 40$, $\tau \la 25$, $\angle i \ga 40$, $Y
  \la 50$ and $A_v^{\rm los} \sim 100-300$. These values translate into
  typical torus sizes and masses of $0.1-5.0$ pc and $10^{4-6} M_\odot$.

\item About half of the objects require an additional black-body component in
  the near-IR range to provide an adequate fit to the SEDs. Most of best
  fitted temperatures are very high ($T \sim 1700-2500$ K) and cannot
  correspond to the emission of hot dust.

\item 2pC models can sometimes provide a better fit to sources that require
  additional near-IR flux.

\item We find weak evidence that nuclei with HBLRs present lower levels of
  extinction than those without a HBLR.

\item Tentatively, we find that radio-loud objects are those with a very small
  inclination angle with respect to the line of sight.

\item We find no correlation between the torus properties and the presence of
  star-formation.

\end{itemize}

\acknowledgments

PL wishes to acknowledge everybody at the Department of Astronomy at the
University of Yale for their hospitality during a 6 month sabbatical
period. Also, to Moshe Elitzur and Robert Nikutta from the Department of
Astronomy at the Kentucky University, home of the Clumpy code and models, for
all their help and guidance. Finally, we are grateful of Patrick Roche and
Emeric Le Floc'h for sharing their spectroscopic observations with us.

PL is grateful of financial support by Fondecyt grant No 1080603.  LV
gratefully acknowledges fellowship support by project MECESUP UCH0118, and
partial support from Fondecyt project 1080603. This publication was also
financed by the ALMA-Conicyt Fund, allocated to the project N° 31060003. AAH
acknowledges support from the Spanish Plan Nacional de Astronom\'ia y
Astrof\'isica under grant AYA2009-05705-E.

We are happy to acknowledge the help from the anonymous referee. Finally, this
work could have not been possible without the help of A.~Cooke.


\begin{thebibliography}{}

\bibitem[Alexander(2001)]{alexander01} Alexander, D.~M.\ 2001, 
\mnras, 320, L15 

\bibitem[Alonso-Herrero et al.(1996)]{ah96} Alonso-Herrero, 
A., Ward, M.~J., \& Kotilainen, J.~K.\ 1996, \mnras, 278, 902 

\bibitem[Alonso-Herrero et al.(2003)]{ah03} Alonso-Herrero, 
A., Quillen, A.~C., Rieke, G.~H., Ivanov, V.~D., 
\& Efstathiou, A.\ 2003, \aj, 126, 81 

\bibitem[Alonso-Herrero et al.(2011)]{AH11} Alonso-Herrero, 
A., Ramos Almeida, C., Mason, R., et al.\ 2011, \apj, 736, 82 

%BLR en flujo polarizado
\bibitem[Antonucci \& Miller(1985)]{antonucci85} Antonucci, 
R.~R.~J., \& Miller, J.~S.\ 1985, \apj, 297, 621

\bibitem[Antonucci(1993)]{antonucci93} Antonucci, R.\ 1993, \araa, 31,
  473 

\bibitem[Asensio Ramos \& Ramos Almeida(2009)]{asensio09} Asensio Ramos, A., \& Ramos
  Almeida, C.\ 2009, \apj, 696, 2075 

\bibitem[Barger et al.(2005)]{barger05} Barger, A.~J., Cowie, 
L.~L., Mushotzky, R.~F., et al.\ 2005, \aj, 129, 578 

\bibitem[Bassani et al.(2006)]{bassani06} Bassani, L., Molina, 
M., Malizia, A., et al.\ 2006, \apjl, 636, L65 

\bibitem[Bassani et al.(1999)]{bassani99} Bassani, L., Dadina, 
M., Maiolino, R., et al.\ 1999, \apjs, 121, 473 

\bibitem[Beckert et al.(2008)]{beckert08} Beckert, T., Driebe, T., H{\"o}nig,
  S.~F., \& Weigelt, G.\ 2008, \aap, 486, L17

\bibitem[Bonning et al.(2009)]{bonning09} Bonning, E.~W., Bailyn, 
C., Urry, C.~M., et al.\ 2009, \apjl, 697, L81 

\bibitem[Brightman \& Nandra(2008)]{brightman08} Brightman, M., \& Nandra,
  K.\ 2008, \mnras, 390, 1241

\bibitem[Buchanan et al.(2006)]{buchnan06} Buchanan, C.~L., 
Kastner, J.~H., Forrest, W.~J., et al.\ 2006, \aj, 132, 1890 

\bibitem[Chiaberge et al.(1999)]{chiaberge99} Chiaberge, M., Capetti, A., \&
  Celotti, A.\ 1999, \aap, 349, 77

\bibitem[C{\^o}t{\'e} et al.(2006)]{cote06} C{\^o}t{\'e}, P., 
Piatek, S., Ferrarese, L., et al.\ 2006, \apjs, 165, 57 

\bibitem[Deo et al.(2007)]{deo07} Deo, R.~P., Crenshaw, 
D.~M., Kraemer, S.~B., et al.\ 2007, \apj, 671, 124 

\bibitem[Deo et al.(2011)]{deo11} Deo, R.~P., Richards, 
G.~T., Nikutta, R., et al.\ 2011, \apj, 729, 108 

\bibitem[D{\'{\i}}az-Santos et al.(2010)]{diaz10} 
D{\'{\i}}az-Santos, T., Charmandaris, V., Armus, L., et al.\ 2010, \apj, 
723, 993 

%modelo continuo de toro
\bibitem[Efstathiou \& Rowan-Robinson(1995)]{efstathiou95} Efstathiou, 
A., \& Rowan-Robinson, M.\ 1995, \mnras, 273, 649 

\bibitem[Elvis et al.(2004)]{elvis04} Elvis, M., Risaliti, G., 
Nicastro, F., et al.\ 2004, \apjl, 615, L25 

%estructuras en las SEDs
\bibitem[Elvis et al.(1994)]{elvis94} Elvis, M., et al.\ 1994, 
\apjs, 95, 1 

\bibitem[Falcke et al.(2004)]{falcke04} Falcke, H., K{\"o}rding, E., \& Markoff,
  S.\ 2004, \aap, 414, 895 

%observaciones de los conos de ionizacion
\bibitem[Falcke et al.(1998)]{falcke98} Falcke, H., Wilson, 
A.~S., \& Simpson, C.\ 1998, \apj, 502, 199 

\bibitem[Fritz et al.(2006)]{fritz06} Fritz, J., Franceschini, 
A., \& Hatziminaoglou, E.\ 2006, \mnras, 366, 767 

\bibitem[Gaskell et al.(2004)]{gaskell04} Gaskell, C.~M., Goosmann, R.~W.,
  Antonucci, R.~R.~J., \& Whysong, D.~H.\ 2004, \apj, 616, 147

\bibitem[Gallimore et al.(2006)]{2006AJ....132..546G} Gallimore, J.~F., 
Axon, D.~J., O'Dea, C.~P., Baum, S.~A., \& Pedlar, A.\ 2006, \aj, 132, 546 

\bibitem[Gilli et al.(2007)]{gilli07} Gilli, R., Comastri, A., \& Hasinger,
  G.\ 2007, \aap, 463, 79

%modelo continuo de toro
\bibitem[Granato \& Danese(1994)]{granato94} Granato, G.~L., \& Danese, 
L.\ 1994, \mnras, 268, 235

\bibitem[Greenhill et al.(2008)]{greenhill08} Greenhill, L.~J., 
Tilak, A., \& Madejski, G.\ 2008, \apjl, 686, L13 

\bibitem[Gregory et al.(1994)]{gregory94} Gregory, P.~C., 
Vavasour, J.~D., Scott, W.~K., \& Condon, J.~J.\ 1994, \apjs, 90, 173 

\bibitem[Gu et al.(2001)]{gu01} Gu, Q., Maiolino, R., \& Dultzin-Hacyan,
  D.\ 2001, \aap, 366, 765

\bibitem[Haas et al.(2007)]{haas07} Haas, M., Siebenmorgen, R., Pantin, E., et
  al.\ 2007, \aap, 473, 369

\bibitem[Heisler et al.(1997)]{heisler97} Heisler, C.~A., 
Lumsden, S.~L., \& Bailey, J.~A.\ 1997, \nat, 385, 700 

\bibitem[Hewitt \& Burbidge(1991)]{hewitt91} Hewitt, A., \&
  Burbidge, G.\ 1991, \apjs, 75, 297

\bibitem[Ho(1999)]{ho99} Ho, L.~C.\ 1999, \apj, 516, 672 

\bibitem[Hopkins et al.(2007)]{hopkins07} Hopkins, P.~F., Richards, G.~T., \&
  Hernquist, L.\ 2007, \apj, 654, 731

\bibitem[Ivezi\'c et al.(1999)]{ivezic97} Ivezic, Z., Nenkova, M., \& Elitzur,
  M.\ 1999, arXiv:astro-ph/9910475

%imagen de HST del toro en NGC4261
\bibitem[Jaffe et al.(1993)]{jaffe93} Jaffe, W., Ford, H.~C., 
Ferrarese, L., van den Bosch, F., \& O'Connell, R.~W.\ 1993, \nat, 364, 213 

%primero en interferometria IR de NGC1068
\bibitem[Jaffe et al.(2004)]{jaffe04} Jaffe, W., Meisenheimer, K.,
  R{\"o}ttgering, H., Leinert, C., \& Richichi, A.\ 2004, The
  Interplay Among Black Holes, Stars and ISM in Galactic Nuclei, 222,
  37

\bibitem[Kellermann et al.(1989)]{kellermann89} Kellermann, K.~I., 
Sramek, R., Schmidt, M., Shaffer, D.~B., \& Green, R.\ 1989, \aj, 98, 1195 

\bibitem[Kishimoto et al.(2011)]{kishimoto11} Kishimoto, M., H{\"o}nig, S.~F.,
  Antonucci, R., et al.\ 2011, \aap, 536, A78

\bibitem[Kishimoto et al.(2009)]{kishimoto09a} Kishimoto, M., H{\"o}nig, S.~F.,
  Tristram, K.~R.~W., \& Weigelt, G.\ 2009, \aap, 493, L57

\bibitem[Kishimoto et al.(2009)]{kishimoto09b} Kishimoto, M., H{\"o}nig,
  S.~F., Antonucci, R., et al.\ 2009, \aap, 507, L57

\bibitem[Kotilainen et al.(1992)]{kotilainen92} Kotilainen, J.~K., 
Ward, M.~J., Boisson, C., Depoy, D.~L., 
\& Smith, M.~G.\ 1992, \mnras, 256, 149 

\bibitem[LaMassa et al.(2010)]{lamassa10} LaMassa, S.~M., 
Heckman, T.~M., Ptak, A., et al.\ 2010, \apj, 720, 786 

\bibitem[Lamastra et al.(2009)]{lamastra09} Lamastra, A., Bianchi, S., Matt, G., et
  al.\ 2009, \aap, 504, 73 

\bibitem[Landt et al.(2011)]{landt11} Landt, H., Elvis, M., 
Ward, M.~J., et al.\ 2011, \mnras, 414, 218 

\bibitem[Lawrence \& Elvis(1982)]{lawrence82} Lawrence, A., \& Elvis, 
M.\ 1982, \apj, 256, 410 

\bibitem[Le Floc'h et al.(2001)]{lefloch01} Le Floc'h, E., Mirabel,
  I.~F., Laurent, O., et al.\ 2001, \aap, 367, 487

\bibitem[Levenson et al.(2007)]{levenson07} Levenson, N.~A., 
Sirocky, M.~M., Hao, L., et al.\ 2007, \apjl, 654, L45 

\bibitem[Lumsden et al.(2004)]{lumsden04} Lumsden, S.~L., Alexander, D.~M., \&
  Hough, J.~H.\ 2004, \mnras, 348, 1451

\bibitem[Lumsden \& Alexander(2001)]{lumsden01} Lumsden, S.~L., \& Alexander,
  D.~M.\ 2001, \mnras, 328, L32

\bibitem[Maccacaro et al.(1982)]{maccacro82} Maccacaro, T., 
Perola, G.~C., \& Elvis, M.\ 1982, \apj, 257, 47 

\bibitem[Maiolino et al.(2001)]{maiolino01a} Maiolino, R., Marconi, A., \&
  Oliva, E.\ 2001, \aap, 365, 37

\bibitem[Maiolino et al.(2001)]{maiolino01b} Maiolino, R., Marconi, A.,
  Salvati, M., et al.\ 2001, \aap, 365, 28

\bibitem[Maiolino et al.(1998)]{maiolino98} Maiolino, R., Salvati, M.,
  Bassani, L., et al.\ 1998, \aap, 338, 781

\bibitem[Marconi et al.(2004)]{marconi04} Marconi, A., Risaliti, 
G., Gilli, R., et al.\ 2004, \mnras, 351, 169 

\bibitem[Malkan et al.(1998)]{malkan98} Malkan, M.~A., Gorjian, 
V., \& Tam, R.\ 1998, \apjs, 117, 25

\bibitem[Mason et al.(2004)]{mason04} Mason, R.~E., Wright, G., 
Pendleton, Y., \& Adamson, A.\ 2004, \apj, 613, 770 

\bibitem[Mass-Hesse et al.(1993)]{mass-hesse93} Mass-Hesse, J.~M., 
Rodriguez-Pascual, P.~M., Mirabel, I.~F.,
\& Sanz Fernandez de Cordoba, L.\ 1993, First Light in the Universe.~Stars or QSO's?, 397 

\bibitem[Meisenheimer et al.(2007)]{meisenheimer07} Meisenheimer, K.,
  Tristram, K.~R.~W., Jaffe, W., et al.\ 2007, \aap, 471, 453

\bibitem[Mel{\'e}ndez et al.(2008)]{melendez08} Mel{\'e}ndez, M., 
Kraemer, S.~B., Armentrout, B.~K., et al.\ 2008, \apj, 682, 94 

\bibitem[Mor \& Trakhtenbrot(2011)]{mor11} Mor, R., \& Trakhtenbrot, B.\ 2011, \apjl, 737, L36 

\bibitem[Mor et al.(2009)]{mor09} Mor, R., Netzer, H., \& Elitzur, M.\ 2009,
  \apj, 705, 298

\bibitem[Mullaney et al.(2011)]{mullaney11} Mullaney, J.~R., 
Alexander, D.~M., Goulding, A.~D., 
\& Hickox, R.~C.\ 2011, \mnras, 414, 1082 

%primer modelo clumpy del toro
\bibitem[Nenkova et al.(2002)]{nenkova02} Nenkova, M., 
Ivezi{\'c}, {\v Z}., \& Elitzur, M.\ 2002, \apjl, 570, L9 

\bibitem[Nenkova et al.(2008a)]{nenkova08a} Nenkova, M., Sirocky, 
M.~M., Ivezi{\'c}, {\v Z}., \& Elitzur, M.\ 2008a, \apj, 685, 147 

\bibitem[Nenkova et al.(2008b)]{nenkova08b} Nenkova, M., Sirocky, 
M.~M., Nikutta, R., Ivezi{\'c}, {\v Z}., \& Elitzur, M.\ 2008b, \apj, 685, 160 

\bibitem[Nenkova et al.(2010)]{nenkova10} Nenkova, M., Sirocky, 
M.~M., Nikutta, R., Ivezi{\'c}, {\v Z}., 
\& Elitzur, M.\ 2010, \apj, 723, 1827 

\bibitem[Noguchi et al.(2010)]{noguchi10} Noguchi, K., Terashima, 
Y., Ishino, Y., et al.\ 2010, \apj, 711, 144 

\bibitem[Panessa and Bassani(2002)]{panessa02} Panessa, F., \& Bassani,
  L.\ 2002, \aa 394, 435

%primer intento por modelar el toro
\bibitem[Pier \& Krolik(1992)]{pier92} Pier, E.~A., \& Krolik, 
J.~H.\ 1992, \apj, 401, 99 

\bibitem[Pier \& Krolik(1993)]{pier93} Pier, E.~A., \& Krolik, J.~H.\ 1993,
  \apj, 418, 673

\bibitem[Prieto et al.(2001)]{prieto01} Prieto, M.,
  Aguerri, J.~A.~L., Varela, A.~M., \& Mu{\~n}oz-Tu{\~n}{\'o}n,
  C.\ 2001, \aap, 367, 405

\bibitem[Prieto et al.(2002)]{prieto02} Prieto, M.~A., Reunanen, J.,
  \& Kotilainen, J.~K.\ 2002, \apjl, 571, L7

\bibitem[Prieto et al.(2010)]{prieto10} Prieto, M.~A., Reunanen, 
J., Tristram, K.~R.~W., et al.\ 2010, \mnras, 402, 724 

\bibitem[Quillen et al.(2001)]{quillen01} Quillen, A.~C., 
McDonald, C., Alonso-Herrero, A., et al.\ 2001, \apj, 547, 129 

%interferometria IR de AGN
\bibitem[Raban et al.(2009)]{raban09} Raban, D., Jaffe, W.,
  R{\"o}ttgering, H., Meisenheimer, K., \& Tristram, K.~R.~W.\ 2009,
  \mnras, 394, 1325

\bibitem[Ramos Almeida et al.(2009)]{RM09} Ramos Almeida, C., Levenson, N.~A.,
  Rodr{\'{\i}}guez Espinosa, J.~M., et al.\ 2009, \apj, 702, 1127

\bibitem[Reichert et al.(1985)]{reichert85} Reichert, G.~A., 
Mushotzky, R.~F., Holt, S.~S., \& Petre, R.\ 1985, \apj, 296, 69 

\bibitem[Risaliti et al.(2005)]{risaliti05} Risaliti, G., Elvis, 
M., Fabbiano, G., Baldi, A., \& Zezas, A.\ 2005, \apjl, 623, L93 

%distribucion de las columnas de densidad entre Sy1s y Sys2
\bibitem[Risaliti et al.(1999)]{risaliti99} Risaliti, G., 
Maiolino, R., \& Salvati, M.\ 1999, \apj, 522, 157 

\bibitem[Roche et al.(2007)]{roche07} Roche, P.~F., Packham, 
C., Aitken, D.~K., \& Mason, R.~E.\ 2007, \mnras, 375, 99 

%extended 12 micron galaxy sample
\bibitem[Rush et al.(1993)]{rush93} Rush, B., Malkan, M.~A., 
\& Spinoglio, L.\ 1993, \apjs, 89, 1 

\bibitem[Sazonov et al.(2007)]{sazonov07} Sazonov, S., Revnivtsev, M., Krivonos, R.,
  Churazov, E., \& Sunyaev, R.\ 2007, \aap, 462, 57 

\bibitem[Schartmann et al.(2005)]{schartmann05} Schartmann, M., Meisenheimer,
  K., Camenzind, M., Wolf, S., \& Henning, T.\ 2005, \aap, 437, 861

%disco de acrecion
\bibitem[Shakura \& Sunyaev(1976)]{shakura76} Shakura, N.~I., \& 
Sunyaev, R.~A.\ 1976, \mnras, 175, 613

\bibitem[Shi et al.(2006)]{shi06} Shi, Y., Rieke, G.~H., 
Hines, D.~C., et al.\ 2006, \apj, 653, 127 

\bibitem[Shu et al.(2007)]{shu07} Shu, X.~W., Wang, J.~X., 
Jiang, P., Fan, L.~L., \& Wang, T.~G.\ 2007, \apj, 657, 167 

\bibitem[Siebenmorgen et al.(2005)]{siebenmorgen05} Siebenmorgen, R.,
  Haas, M., Kr{\"u}gel, E., \& Schulz, B.\ 2005, \aap, 436, L5

%Spitzer star-formation templates
\bibitem[Smith et al.(2007)]{smith07} Smith, J.~D.~T., et al.\ 
2007, \apj, 656, 770

\bibitem[Smith \& Done(1996)]{smith96} Smith, D.~A., \& Done, C.\ 1996, \mnras, 280, 355 

\bibitem[Stalevski et al.(2012)]{stalevski12} Stalevski, M., Fritz, 
J., Baes, M., Nakos, T., \& Popovi{\'c}, L.~{\v C}.\ 2012, \mnras, 420, 2756 

\bibitem[Suganuma et al.(2006)]{suganuma06} Suganuma, M., Yoshii, 
Y., Kobayashi, Y., et al.\ 2006, \apj, 639, 46 

\bibitem[Swain et al.(2003)]{swain03} Swain, M., Vasisht, G., 
Akeson, R., et al.\ 2003, \apjl, 596, L163 

\bibitem[Taranova \& Shenavrin(2006)]{taranova06} Taranova, O.~G., \&
  Shenavrin, V.~I.\ 2006, Astronomy Letters, 32, 439

\bibitem[Thean et al.(2000)]{2000MNRAS.314..573T} Thean, A., Pedlar, A., 
Kukula, M.~J., Baum, S.~A., \& O'Dea, C.~P.\ 2000, \mnras, 314, 573 

\bibitem[Thompson et al.(2009)]{thompson09} Thompson, G.~D., 
Levenson, N.~A., Uddin, S.~A., \& Sirocky, M.~M.\ 2009, \apj, 697, 182 

\bibitem[Tran(2003)]{tran03} Tran, H.~D.\ 2003, \apj, 583, 632 

\bibitem[Tran(2001)]{tran03} Tran, H.~D.\ 2001, \apjl, 554, L19

%interferometria IR de AGN
\bibitem[Tristram et al.(2007)]{tristram07} Tristram, K.~R.~W., et
  al.\ 2007, \aap, 474, 837

\bibitem[Tristram et al.(2009)]{tristram09} Tristram, K.~R.~W., Raban, D.,
  Meisenheimer, K., et al.\ 2009, \aap, 502, 67

%interferometria IR de AGN
\bibitem[Tristram et al.(2009)]{tristram09} Tristram, K.~R.~W., et
  al.\ 2009, \aap, 502, 67

\bibitem[Turner et al.(1997)]{turner97} Turner, T.~J., George, 
I.~M., Nandra, K., \& Mushotzky, R.~F.\ 1997, \apjs, 113, 23 

\bibitem[Veron-Cetty \& Veron(1991)]{veron-cetty91} Veron-Cetty,
  M.-P., \& Veron, P.\ 1991, European Southern Observatory Scientific
  Report, 10, 1

\bibitem[Videla et al.(2012)]{videla12} Videla, L., Lira, P., Andrews, H.,
  Alonso-Herrero, A., Alexander, D.~M., \& Ward, M.\ 2012, \apjs, xxx, xxx

\bibitem[Walcher et al.(2006)]{walcher06} Walcher, C.~J., 
B{\"o}ker, T., Charlot, S., et al.\ 2006, \apj, 649, 692 

\bibitem[Weigelt et al.(2004)]{weigelt} Weigelt, G., Wittkowski, M., Balega,
  Y.~Y., et al.\ 2004, \aap, 425, 77

\bibitem[Weinberger et al.(1999)]{weinberger99} Weinberger, A.~J., 
Neugebauer, G., \& Matthews, K.\ 1999, \aj, 117, 2748 

\bibitem[Winter et al.(2007)]{winter07} Winter, L.~M., 
Mushotzky, R.~F., \& Reynolds, C.~S.\ 2007, \apj, 655, 163 

\bibitem[Wittkowski et al.(1998)]{wittkowski98} Wittkowski, M., Balega, Y.,
  Beckert, T., et al.\ 1998, \aap, 329, L45

\bibitem[Wittkowski et al.(2004)]{wittkowski04} Wittkowski, M., Kervella, P.,
  Arsenault, R., et al.\ 2004, \aap, 418, L39

%espectros de spitzer incluidos en las SEDs
\bibitem[Wu et al.(2009)]{wu09} Wu, Y., Charmandaris, V., Huang, J.,
  Spinoglio, L., \& Tommasin, S.\ 2009, \apj, 701, 658

\bibitem[Young et al.(1996)]{young96} Young, S., Hough, J.~H., 
Efstathiou, A., et al.\ 1996, \mnras, 281, 1206 


\bibitem[Zhang \& Wang(2006)]{zhang06} Zhang, E.-P., \& Wang, J.-M.\ 2006,
  \apj, 653, 137

\end{thebibliography}
\end{document}